\documentclass[aps,prl,reprint,superscriptaddress,amssymb,amsmath, longbibliography]{revtex4-1}
 \pdfoutput=1 
\usepackage{graphicx}
\usepackage{dcolumn}
\usepackage{bm, color}
\usepackage[none]{hyphenat} 
\usepackage{bbm}
\usepackage[colorlinks]{hyperref}
\usepackage{braket}
\usepackage{tabularx}
\usepackage{float}

\newcommand{\bk}{\mathbf k}
\newcommand{\bq}{\mathbf q}

\newcommand{\beginsupplement}{%
        \setcounter{table}{0}
        \renewcommand{\thetable}{S\arabic{table}}%
        \setcounter{figure}{0}
        \renewcommand{\thefigure}{S\arabic{figure}}%
     }

\begin{document}

\title{A Charge-Density-Wave Topological Semimetal}

\author{Wujun Shi}
\thanks{These authors contributed equally to this work.}
\affiliation{Max Planck Institute for Chemical Physics of Solids, D-01187 Dresden, Germany}
\affiliation{School of Physical Science and Technology, ShanghaiTech University, Shanghai 200031, China}

\author{Benjamin J. Wieder}
\thanks{These authors contributed equally to this work.}
\affiliation{Department of Physics, Princeton University, Princeton, New Jersey 08544, USA}

\author{Holger L. Meyerheim}
\thanks{These authors contributed equally to this work.}
\affiliation{Max Planck Institute of Microstructure Physics, Weinberg 2, 06120 Halle (Saale), Germany}

\author{Yan Sun}
\affiliation{Max Planck Institute for Chemical Physics of Solids, D-01187 Dresden, Germany}

\author{Yang Zhang}
\affiliation{Max Planck Institute for Chemical Physics of Solids, D-01187 Dresden, Germany}
\affiliation{Leibniz Institute for Solid State and Materials Research, 01069 Dresden, Germany}

\author{Yiwei Li}
\affiliation{Department of Physics, University of Oxford, Oxford OX1 3PU, United Kindom}

\author{Lei Shen}
\affiliation{States Key Laboratory of Low Dimensional Quantum Physics, Department of Physics and Collaborative Innovation Center of Quantum Matter, Tsinghua University, Beijing, China}

\author{Yanpeng Qi}
\affiliation{School of Physical Science and Technology, ShanghaiTech University, Shanghai 200031, China}

\author{Lexian Yang}
\affiliation{States Key Laboratory of Low Dimensional Quantum Physics, Department of Physics and Collaborative Innovation Center of Quantum Matter, Tsinghua University, Beijing, China}

\author{Jagannath Jena}
\affiliation{Max Planck Institute of Microstructure Physics, Weinberg 2, 06120 Halle (Saale), Germany}

\author{Peter Werner}
\affiliation{Max Planck Institute of Microstructure Physics, Weinberg 2, 06120 Halle (Saale), Germany}

\author{Klaus Koepernik}
\affiliation{Leibniz Institute for Solid State and Materials Research, 01069 Dresden, Germany}

\author{Stuart Parkin}
\affiliation{Max Planck Institute of Microstructure Physics, Weinberg 2, 06120 Halle (Saale), Germany}

\author{Yulin Chen}
\affiliation{School of Physical Science and Technology, ShanghaiTech University, Shanghai 200031, China}
\affiliation{Department of Physics, University of Oxford, Oxford OX1 3PU, United Kindom}
\affiliation{States Key Laboratory of Low Dimensional Quantum Physics, Department of Physics and Collaborative Innovation Center of Quantum Matter, Tsinghua University, Beijing, China}

\author{Claudia Felser}
\affiliation{Max Planck Institute for Chemical Physics of Solids, D-01187 Dresden, Germany}

\author{B. Andrei Bernevig}
\email{bernevig@princeton.edu}
\affiliation{Department of Physics, Princeton University, Princeton, New Jersey 08544, USA}

\author{Zhijun Wang}
\email{wzj@iphy.ac.cn}
\affiliation{Beijing National Laboratory for Condensed Matter Physics, and Institute of Physics, Chinese Academy of Sciences, Beijing 100190, China}
\affiliation{University of Chinese Academy of Sciences, Beijing 100049, China}

\date{\today}
\begin{abstract}
{\bf Topological physics and strong electron-electron correlations in quantum materials are typically studied independently.  However, there have been rapid recent developments in quantum materials in which topological phase transitions emerge when the single-particle band structure is modified by strong interactions.  We here demonstrate that the room-temperature phase of (TaSe$_4$)$_2$I is a Weyl semimetal with 24 pairs of Weyl nodes.  Owing to its quasi-1D structure, (TaSe$_4$)$_2$I hosts an established CDW instability just below room temperature.  Using X-ray diffraction, angle-resolved photoemission spectroscopy, and first-principles calculations, we find that the CDW in (TaSe$_4$)$_2$I couples the bulk Weyl points and opens a band gap.  The correlation-driven topological phase transition in (TaSe$_4$)$_2$I provides a route towards observing condensed-matter realizations of axion electrodynamics in the gapped regime, topological chiral response effects in the semimetallic phase, and represents an avenue for exploring the interplay of correlations and topology in a solid-state material.}
\end{abstract}

\maketitle


Conventional solid-state Weyl semimetals~\cite{Wan2011,Weng2015, Huang2015, Lv2015TaAs, Xu2015TaAs,Xu2015NbAs, Wang2015MoTe2, Sun2015MoTe2, Soluyanov:2015WSM2, jiang2017signature, Vafek2014,WangHeusler2016} are 3D materials whose bulk Fermi pockets derive from linearly-dispersing, point-like nodal degeneracies.  Unlike in other solid-state semimetals with lower dimensionality or higher symmetry~\cite{armitage2017weyl,Wang:2013is,DDP,NewFermions}, the Fermi pockets of 3D Weyl semimetals carry integer-valued topological (chiral) charges, reflecting that the nodal points are sources and sinks of Berry curvature~\cite{Wan2011}.  Because the low-energy spectra of the nodal points in Weyl semimetals resemble the Weyl equation in high-energy physics, the nodal points have become known as condensed-matter Weyl points (WPs)~\cite{Wan2011,Weng2015, Huang2015, Lv2015TaAs, Xu2015TaAs,Xu2015NbAs, Wang2015MoTe2, Sun2015MoTe2, Soluyanov:2015WSM2, jiang2017signature, Vafek2014,WangHeusler2016}.

\begin{figure*}[!t]
\centering
\includegraphics[width=\textwidth]{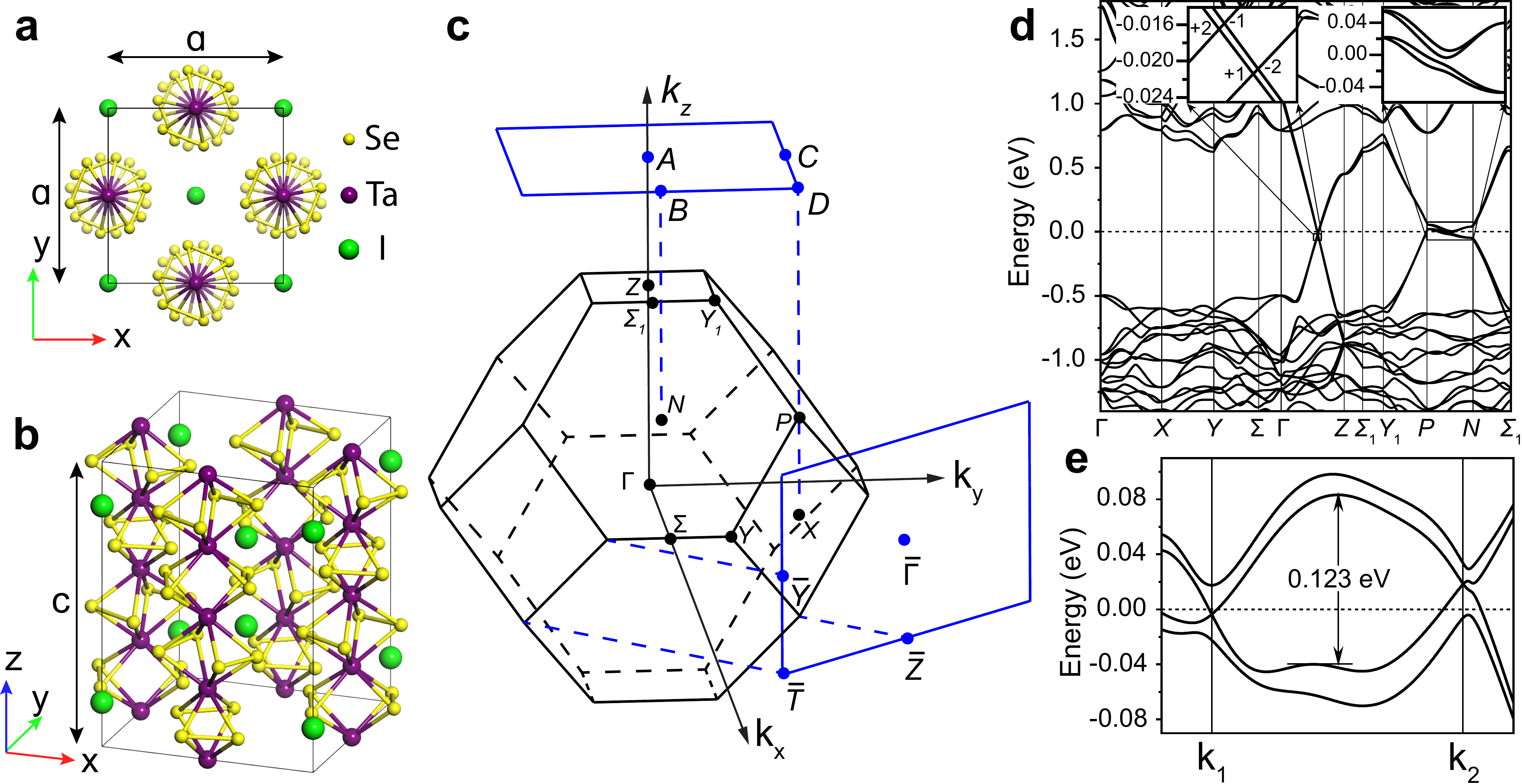}
\caption{(Color online) The crystal structure, 3D bulk and 2D surface Brillouin zones (BZs), and electronic structure of (TaSe$_4$)$_2$I.  The crystal structure is shown from both (a) top [$(001)$] and (b) tilted side perspectives, where throughout this work, surface terminations are labeled using the conventional-cell lattice vectors $(a,a,c)$.  Because (TaSe$_4$)$_2$I crystallizes in the body-centered tetragonal space group 97 ($I422$)~\cite{Ta2Se8IPrepare}, then its primitive cell contains half as many atoms as the conventional cell shown in (a,b) [SI~I].  (c) The bulk BZ and its projections onto the conventional-cell (001)- and (110)-surfaces.  (d) Electronic band structure of (TaSe$_4$)$_2$I with spin-orbit coupling.  Bands along $\Gamma Z$ cross at the Fermi energy ($E_{F}$) near the halfway points $k_{z}=\pm \pi/c$ -- a remnant of a filling-enforced nodal plane that is present in the band structure of a crystal of decoupled TaSe$_4$ chains (SI~I).  Away from the high-symmetry BZ lines, there are 48 Weyl points (WPs) lying within the energy range $-10$ meV $<(E-E_F)<15$ meV (Fig.~\ref{fig:wps}), which we designate the ``Fermi-surface WPs'' (FSWPs).  The momentum-space coordinates and chiral charges of all 48 FSWPs are provided in SI~A, and characteristic band dispersions for each FSWP are provided in SI~B.  Further below $E_{F}$, there are also eight $C_{4z}$-enforced chiral fermions (WPs) along $\Gamma Z$.  As shown in the left inset panel of (d), moving along $\Gamma Z$ in increasing $k_{z}$, four of the eight $C_{4z}$-enforced WPs exhibit the compensating chiral charges $+2$, $-1$, $+1$, and $-2$ [their time-reversal $(\mathcal{T}$) partners along $\Gamma Z$ with negative values of $k_{z}$ also exhibit the same charges, because $\mathcal{T}$ does not invert the chiral charge of a WP~\cite{Wan2011,KramersWeyl}].  In the right inset panel of (d), we show that bands along $P N$ are gapped at $E_{F}$.  (e) Bands between the $W^{+}_{1}$ and $W^{-}_{4}$ FSWPs (Fig.~\ref{fig:wps}) are separated by a large nontrivial energy window of 0.123~eV, or on the order of 1400 K.}
\label{fig:Structure}
\newpage
\end{figure*}

In Weyl semimetals, the surface projections of WPs of opposite chirality are connected by topological surface Fermi arcs~\cite{Wan2011}.  These surface Fermi arcs have emerged as the primary experimental means of confirming the presence of bulk WPs, and their signatures have been observed in angle-resolved photoemission spectroscopy (ARPES) experiments~\cite{Weng2015, Huang2015, Lv2015TaAs, Xu2015TaAs,Xu2015NbAs, Wang2015MoTe2, Sun2015MoTe2, Soluyanov:2015WSM2, jiang2017signature,AlPtObserve,CoSiObserveJapan,CoSiObserveHasan,CoSiObserveChina,PdGaObserve} and scanning tunneling microscopy probes of quasiparticle interference~\cite{Inoue2016,Zheng2016}.  Researchers have also proposed bulk probes of chiral topology in Weyl semimetals, including the intrinsic spin Hall effect~\cite{Sun2016}, the anomalous Hall effect~\cite{Burkov:2011de, Xu2011}, and the quantized circular photogalvanic effect~\cite{de2017quantized}.

Most interestingly, while the above response effects can be understood from the perspective of single-particle physics, researchers have also proposed more exotic response effects in Weyl semimetals with significant electron-electron interactions.  For example, in several theoretical proposals~\cite{W2,PhysRevLett.120.067003,LiYi-sp}, attractive electron-electron interactions have been shown to drive a Weyl semimetal into a topological superconductor.  Even in the absence of superconductivity, interactions can still drive a Weyl semimetal into an (incommensurate) charge-density wave (CDW) phase in which the CDW wavevector(s) ``nest'' bulk WPs.  If the nested WPs carry the same chiral charges, the CDW may access a gapless topological phase with monopole harmonic order~\cite{W5}, and if the nested WPs carry opposite chiral charges, the CDW may access a gapped phase in which dynamical CDW angle defects bind chiral modes as a result of effective axion electrodynamics~\cite{ShouchengCDW,TaylorCDW}.  Though there has been tremendous recent interest in measuring unconventional superconductivity and axionic response effects, the relative dearth of candidate Weyl semimetals with interacting instabilities has hindered the confirmation of these theoretical proposals.

In this work, we bridge the gap between noninteracting and correlated Weyl semimetals by employing first-principles calculations and experimental probes to demonstrate that quasi-1D (TaSe$_4$)$_2$I crystals~\cite{Ta2Se8IPrepare} are in fact Weyl semimetals whose WPs become coupled and gapped by the onset of a CDW.  Though the high-temperature phase of (TaSe$_4$)$_2$I has previously been highlighted for exhibiting linear crossings near the Fermi energy ($E_{F}$)~\cite{PhysRevLett.110.236401, OtherARPES} and Kramers-Weyl fermions far below $E_{F}$~\cite{KramersWeyl}, our work represents the first recognition that (TaSe$_4$)$_2$I hosts topological chiral fermions at $E_{F}$, and the first reported link between the bulk WPs and the CDW wavevectors.  Furthermore, because (TaSe$_4$)$_2$I crystallizes in chiral space group (SG) 97 ($I422$), it hosts WPs with opposite chiral charges at different energies~\cite{de2017quantized,KramersWeyl,chang2017large}, and therefore provides a promising platform for the observation of bulk probes of topological chirality~\cite{de2017quantized}.  In (TaSe$_4$)$_2$I, we find that all of the Fermi pockets originate from Bloch states that lie within a small energy range of the nodes of 48 WPs, which we designate as the ``Fermi-surface WPs''  (FSWPs).  The 48 FSWPs specifically lie within 15~meV~of $E_{F}$: 16 $C=+1$ FSWPs lie $\sim 9$~meV below $E_{F}$, and the remaining 32 lie above.  The net $+16$ chiral charge of the WPs below $E_{F}$ is by far the largest value predicted to date in a real material.

Previous experiments have shown that (TaSe$_4$)$_2$I transitions into an incommensurate, gapped CDW phase when cooled just below room temperature~\cite{Cava1986,260k,CDWTaSeIDFT}.  Using first-principles calculations, we compute the high-density electronic susceptibility and FSWP nesting vectors [see Sections A and G of the Supplementary Information (SI~A and SI~G, respectively)].  We then performed ARPES and X-ray diffraction (XRD) experiments on (TaSe$_4$)$_2$I samples to determine the gap, symmetry, and modulation vectors of the CDW phase (see SI~H).  Our theoretical and experimental analyses imply that (TaSe$_4$)$_2$I is the first known material to host a correlation-driven topological semimetal-insulator phase transition.

{\bf Structure.}
(TaSe$_4$)$_2$I (Inorganic Crystal Structure Database~\cite{ICSD} No. 35190, further details available at~\url{https://topologicalquantumchemistry.org/#/detail/35190}~\cite{QuantumChemistry,AndreiMaterials}) crystallizes in a quasi-1D, body-centered tetragonal chiral structure in SG 97 ($I422$)~\cite{Ta2Se8IPrepare}.  The conventional cell of (TaSe$_4$)$_2$I contains two TaSe$_4$ chains aligned along the $c$-axis and four iodine atoms separating the chains [Fig.~\ref{fig:Structure}(a,b)].  Each chain is formed of four alternating layers of Ta atoms and rectangles with four Se atoms on each corner, for a total of $4$ Ta atoms and $16$ Se atoms per chain.  Within the conventional cell, there are two chains, implying a total chemical formula of (TaSe$_4$)$_8$I$_4$ per conventional cell [Fig.~\ref{fig:Structure}(a,b)].  When decoupled, each chain exhibits exotic ``non-crystallographic'' screw symmetries, which we further detail in SI~I.  Because the crystal structure of (TaSe$_4$)$_2$I is only generated by (body-centered) lattice translations and the proper rotation symmetries, $C_{4z}$ and $C_{2x}$, where $C_{ni}$ is a rotation by $360^{\circ}/n$ about the $i$-axis, then it is structurally chiral~\cite{KramersWeyl}.  Additionally, because (TaSe$_4$)$_2$I is nonmagnetic, then its spectrum respects time-reversal ($\mathcal{T}$) symmetry.  Because the monopole chiral charges (Chern numbers) of chiral fermions are left invariant under proper rotations and $\mathcal{T}$~\cite{Vafek2014}, then WPs in (TaSe$_4$)$_2$I with opposite chiral charges are free to lie at different energies.  Though an energy offset between oppositely charged chiral fermions has been predicted in Kramers-Weyl~\cite{KramersWeyl} and unconventional-fermion semimetals~\cite{chang2017large,liu2019symmetryenforced}, (TaSe$_4$)$_2$I presents a rare example of this energy offset in a conventional Weyl semimetal.

{\bf Band Structure and Fermi Surface.}
Owing to its quasi-1D crystal structure, (TaSe$_4$)$_2$I exhibits a strongly anisotropic electronic structure.  In Fig.~\ref{fig:Structure}(d), we show the band structure of (TaSe$_4$)$_2$I calculated along high-symmetry lines in the first BZ [Fig.~\ref{fig:Structure}(c)].  We correspondingly observe weak dispersion in the $k_{z}=\pi/c$ plane along $PN$, and observe much stronger dispersion along $\Gamma Z$, as $k_{z}$ is reciprocal to the chain translation direction $c$ ($z$).  We observe that there is a $1$ eV gap, in the $k_z=0,2\pi/c$ planes, whereas there is no band gap in the vicinity of the $k_{z}=\pm \pi/c$ planes.  In fact, we find that the entire Fermi surface of (TaSe$_4$)$_2$I in SG 97 is localized near $k_{z}=\pm \pi/c$.  This is surprising, because in SG 97, generic points in the $k_{z}=\pm \pi/c$ planes are not fixed by symmetry, as they would be in a primitive tetragonal structure with a periodicity of $c$ in the $z$ direction (see SI~A for further WP symmetry analysis).  In SI~I, we show that the localization of the Fermi surface can be understood by recognizing that (TaSe$_4$)$_2$I is formed from weakly coupled screw-symmetric chains, which individually do exhibit symmetry- and filling-enforced nodal degeneracies near $k_{z}=\pi/c$.

\begin{figure}[!t]
\centering
\includegraphics[width=0.45\textwidth]{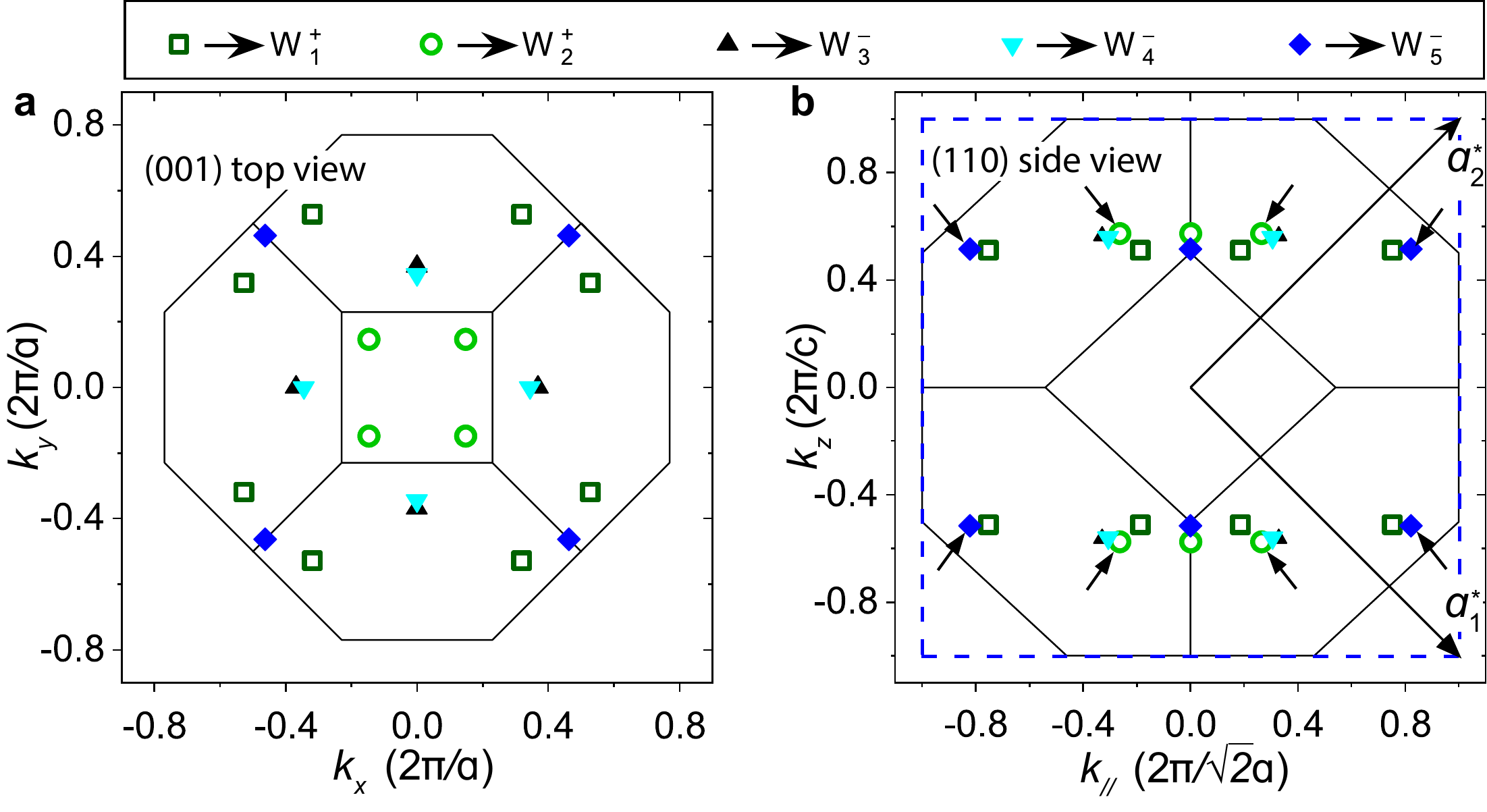}
\caption{(Color online) The distribution of the FSWPs in the first bulk BZ, viewed from (a) the top [$(001)$] (a) and (b) side [$(110)$] surfaces.  (b) The FSWPs are concentrated in the vicinity of the $k_{z}=\pm \pi/c$ planes.  The black lines in (a,b) represent the boundaries of 2D slices of the bulk BZ [Fig.~\ref{fig:Structure}(c)].  In (b), $a^{*}_{1,2}$ indicate the $(110)$-surface BZ primitive reciprocal lattice vectors (SI~C and SI~E), and the dashed blue box indicates the boundary of the second surface BZ (SI~C).  In (a), because of the bulk crystal symmetries each symbol represents a pair of WPs with the same chiral charge lying at the opposite momenta $\pm k_{z}$ (SI~A).  All of the symbols in (a) consequently represent FSWP projections with chiral charge $|C|=2$.  In (b), the relationship between the bulk WPs and symmetries is more complicated.  We therefore introduce arrows to indicate which symbols in (b) correspond to the surface projections of only a single bulk FSWP (with charge $|C|=1$); in (b), like in (a), the symbols without arrows indicate the projections of two bulk FSWPs with the same chiral charge (for a net charge of $|C|=2$).  The coordinates of the FSWPs are provided in SI~A.}
\label{fig:wps}
\newpage
\end{figure}

\begin{figure*}[!t]
\centering
\includegraphics[width=\textwidth]{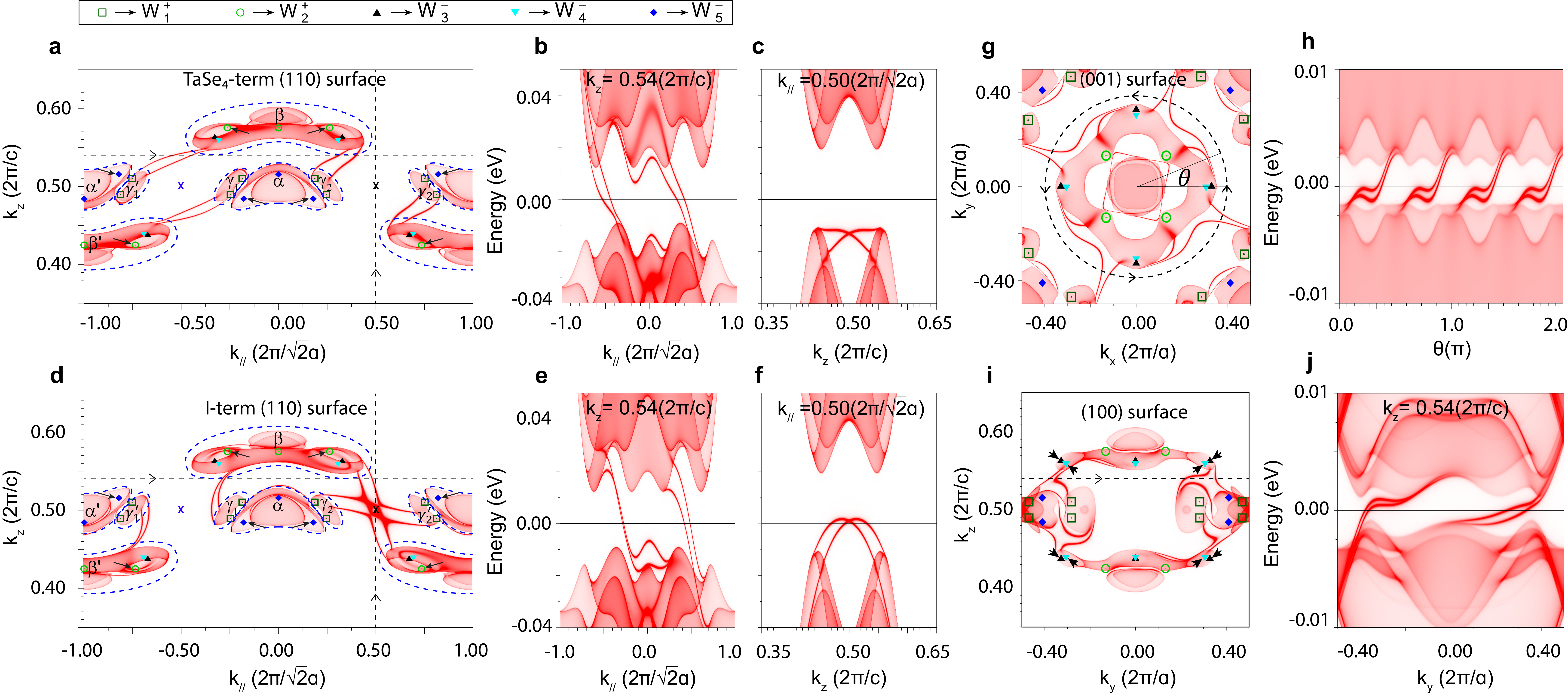}
\caption{(Color online)  The surface states of (TaSe$_4$)$_2$I terminated in (a-d) the experimentally favored~\cite{PhysRevLett.110.236401} $(110)$-direction, (g,h) the $(001)$-direction, and (i,j) the $(100)$-direction, employing the labeling established in Fig.~\ref{fig:wps}.  (a,d) The $(110)$-surface can be terminated with either TaSe$_4$ chains (Se-term) or I atoms (I-term.).  For both terminations of the $(110)$-surface, the projections of the bulk states form four islands (and their time-reversal partners), which we enclose with blue dashed lines and label $\alpha\ (\alpha')$, $\beta\ (\beta')$, $\gamma_1\ (\gamma_1')$, and $\gamma_2\ (\gamma_2')$, respectively [see SI~A and~C for the definition of $k_{\parallel}$, the distribution of states within the full $(110)$-surface BZ, and an examination of the surface projections of bulk states].  The two $\times$ symbols in (a,d) represent surface TRIM points.  The horizontal and vertical dashed lines in (a,d) indicate two cuts through the surface BZ at $k_z=0.54(2\pi/c)$ and $k_{\parallel}=0.50(2\pi/\sqrt{2}a)$, respectively.  For both the Se- and I-terminations, (b,e) the horizontal cut at $k_{z}=0.54(2\pi/c)$ exhibits Chern number $C=-4$, while (c,f) the vertical cut at $k_{\parallel}=0.50(2\pi/\sqrt{2}a)$ exhibits a trivial Chern number ($C=0$).  The trivially connected surface states in (c,f) both intersect in twofold linear crossings that are protected by $\mathcal{T}$-symmetry.  In (d,f), the I-term. surface atoms pull the trivial crossing in (c) towards the Fermi energy, and drive the four surface Fermi arcs at $k_{\parallel}> 0$ to merge in a surface Lifshitz critical point between a topological Fermi-arc connectivity linking the $\beta$ and $\gamma_{2}'$ islands and a connectivity linking the $\beta$ and $\gamma_{2}$ islands.  Additionally, in (d), the Fermi arcs that previously connected $\gamma_{1}$ and $\beta'$ (and their time-reversal partners) in (a) instead connect $\gamma_{1}$ to $\beta$ (and $\gamma_{1}'$ to $\beta'$).  Topological Fermi arcs are also present on (g,h) the $(001)$- and (i,j) $(100)$-surfaces.  In particular, like in the RhSi family~\cite{chang2017large,CoSi,NewFermions,KramersWeyl,AlPtObserve,CoSiObserveJapan,CoSiObserveHasan,CoSiObserveChina,PdGaObserve} (g) the $(001)$-surface Fermi arcs span the entire BZ, and (h) the projected Fermi pockets at $k_{x,y}=0,\pi/a$ exhibit large Chern numbers ($|C|=8$).}
\label{fig:FS}
\newpage
\end{figure*}

\vspace{0.2in}
{\bf Weyl Point Distribution.}
In 3D (TaSe$_4$)$_2$I, the entire Fermi surface is formed from topological bands connected to bulk chiral fermions (WPs).  Specifically, because (TaSe$_4$)$_2$I crystals in SG 97 ($I422$) are symmorphic, chiral, and exhibit non-negligible spin-orbit coupling (SOC), then all of their bulk degeneracies are necessarily point-like, and carry nontrivial chiral charges, as explicitly shown in~\cite{KramersWeyl} and discussed in SI~A.  In the electronic structure of (TaSe$_4$)$_2$I calculated from first principles, we observe 48 WPs within $15$~meV of $E_{F}$ (Fig.~\ref{fig:wps}), which we designate the FSWPs, as well as eight $C_{4z}$-enforced chiral fermions (two pairs of conventional WPs and two pairs of double-Weyl points~\cite{Xu2011,Huang02022016}) along $\Gamma Z$ lying between $16$ -- $20$~meV below $E_{F}$ [shown in the left inset panel of Fig.~\ref{fig:Structure}(d)].  In SI~I, we detail the origin of the $C_{4z}$-enforced WPs in terms of the symmetry eigenvalues and band connectivity of isolated TaSe$_{4}$ chains.  Because the $C_{4z}$-enforced WPs below $E_{F}$ are only weakly separated [$\Delta k_{z}\sim 0.002(\frac{2\pi}{c})$] and carry a net-zero chiral charge within each narrow grouping, then they are not likely to contribute experimentally detectable Fermi-arc surface states.  Additionally, because all eight enforced chiral fermions are fully occupied and carry compensating chiral charges, then they do not contribute to bulk response or transport effects at intrinsic filling.  Therefore, we will neglect the $C_{4z}$-enforced WPs below $E_{F}$ in further discussions of the chiral fermions in (TaSe$_4$)$_2$I.

In Fig.~\ref{fig:wps}(a,b), we show $(001)$-surface (top) and $(110)$-surface (side) views of the bulk FSWPs in the first BZ, respectively.  The solid lines indicate the projected boundary of the first bulk BZ, and the differently shaped symbols each denote one set of symmetry-related FSWPs.  Note that while the FSWPs are distributed over a wide range in $k_{x,y}$ [Fig.~\ref{fig:wps}(a)], they all lie within a close vicinity of the $k_z=\pm \pi/c$ planes [Fig.~\ref{fig:wps}(b)].  As shown in SI~I, this distribution of nodal points reflects that a crystal of decoupled TaSe$_4$ chains and iodine atoms is a filling-enforced semimetal with $4_{2}$-screw- and $\mathcal{T}$- symmetry-enforced nodal surfaces that lie in the vicinity of the $k_{z}=\pm\pi/c$ planes.

{\bf Surface States.}
Weyl semimetals most notably exhibit characteristic topological Fermi-arc surface states.  To confirm the presence of topological surface Fermi arcs in (TaSe$_4$)$_2$I, we calculate the surface states with surface Green's functions as detailed in the Methods section.  (TaSe$_4$)$_2$I is known to experimentally cleave on the conventional-cell $(110)$-surface, due to the weak van der Waals interactions between the TaSe$_4$ chains~\cite{PhysRevLett.110.236401,Ta2Se8IPrepare}.  In Fig.~\ref{fig:FS}, we show the calculated surface states of (TaSe$_4$)$_2$I on the experimentally favorable $(110)$-surface [panels (a-f)], as well as on the $(001)$- and $(100)$-surfaces [panels (g,h) and (i,j), respectively].

The conventional-cell $(110)$-surface projections of the bulk Fermi surface of (TaSe$_4$)$_2$I form four time-reversal pairs of separated islands in each surface BZ.  In Fig.~\ref{fig:FS}(a,b), we enclose the islands with dashed blue lines, and label the islands (and their time-reversal partners) $\alpha\ (\alpha')$, $\beta\ (\beta')$, $\gamma_1\ (\gamma_1')$, and $\gamma_2\ (\gamma_2')$ [see SI~C for additional details].  Each island is formed from the projected bulk Fermi pockets of the FSWPs, and can thus carry a total chiral charge.  From the surface projections of the FSWPs, we infer that $\alpha$, $\beta$, $\gamma_1$, and $\gamma_2$ respectively carry the net chiral charges $-4$, $-4$, $+4$, and $+4$.  Because the chiral charge of a WP does not change sign under $\mathcal{T}$~\cite{Wan2011}, then $\alpha'$, $\beta'$, $\gamma_1'$, and $\gamma_2'$ also exhibit the same net charges of $-4$, $-4$, $+4$, and $+4$, respectively.

\begin{figure*}[t]
\centering
\includegraphics[width=0.97\textwidth]{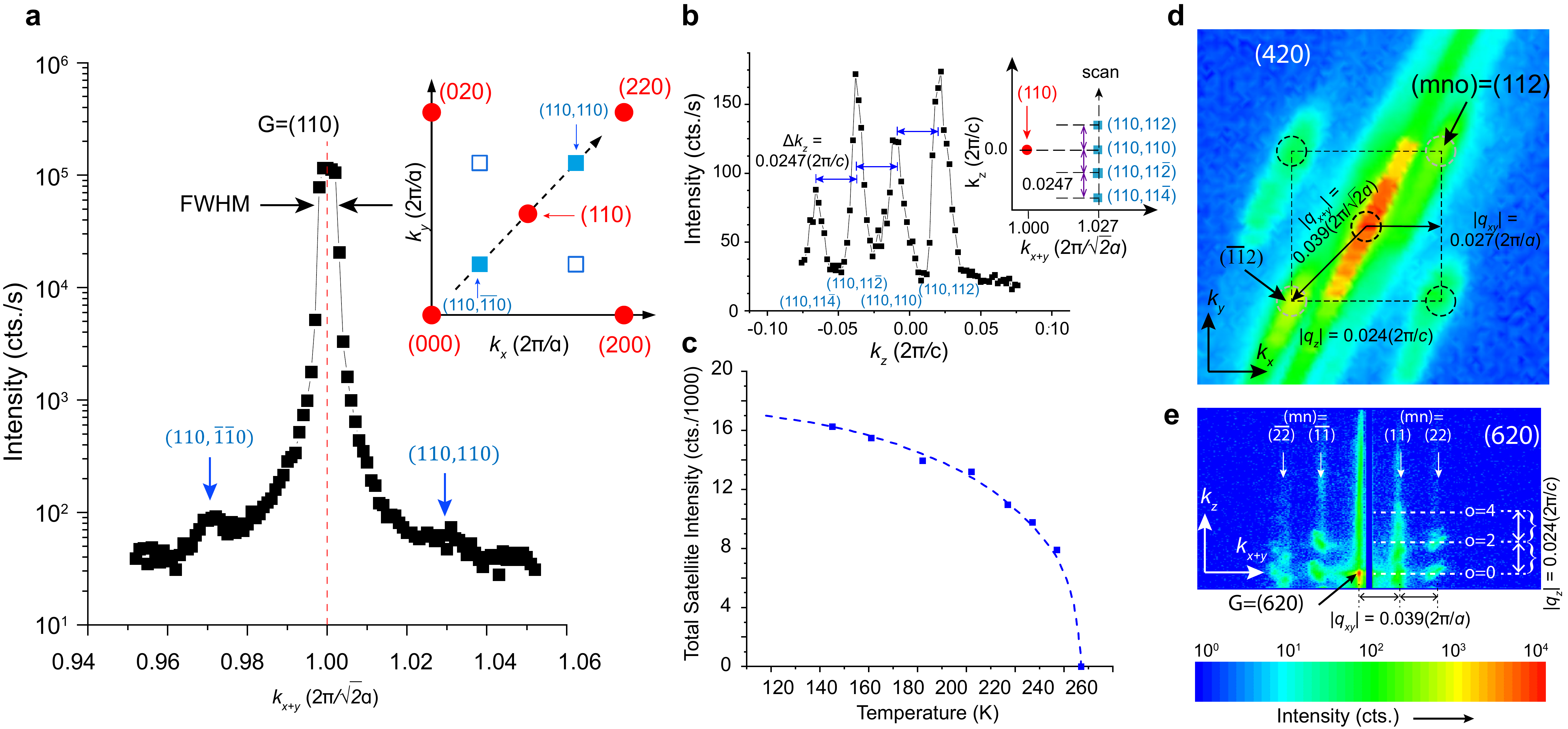}
\caption{(Color online) X-ray diffraction (XRD) data for the CDW phase of (TaSe$_4$)$_2$I.  (a) XRD line scan along the $k_{x+y}$ direction through the ${\bf G}=(110)$ main reflection.  (b) XRD line scan along the $k_{z}$ direction through the ${\bf Q}=(hkl,mno)=(110,110)$ satellite reflection.  (d,e) 2D reciprocal-space maps plotted on a logarithmic intensity scale and recorded near the (d) ${\bf G}$=$(420)$ and (e) $(620)$ main reflections collected in the $k_{x,y}$- and $k_{x+y,z}$-planes, respectively.  We attribute the elongated profile of the reflections in (d,e) to the logarithmic intensity scale and to the mosaicity of our sample, which we have measured through transverse angular scans to be $0.3^{\circ}$ to $0.5^{\circ}$ -- well within the range of typical high-quality samples ($\sim 0.01^{\circ}$ to $\sim 1.0^{\circ}$, see SI~H.1).  In (b), satellite reflections related by $C_{2,x+y}$ exhibit the same intensities, and in (d), pairs of satellite reflections related by $C_{2z}$ (respectively labeled with white and dark dashed circles) exhibit the same intensities within uncertainty, but satellite reflections related by $C_{4z}$ (\emph{e.g.} one satellite in a dashed white circle and one satellite in a dashed dark circle) exhibit intensities that differ by an order of magnitude.  This implies that our sample contains two macroscopic domains with different, $C_{4z}$-related CDW orderings, where the CDW order within each domain respects the symmetries of point group $D_{2}$ ($222$) in a setting with $C_{2z}$ and $C_{2,x\pm y}$ symmetry.  (c) Total measured satellite intensity in the vicinity of the ${\bf G}=(110)$ main reflection as a function of temperature.  We observe that all satellites simultaneously disappear at $T_{C}\approx 248~K$, representing a signature of a transition away from a CDW phase.  The measured $T_{C}$ in (c) is slightly lower than, but still in close agreement with, the value of $T_{C}= 260$~K previously reported in~\cite{Cava1986,260k,Fu1984qvec}. The data for (a,b,d,e) were collected within a temperature range of roughly $88$~K to $100$~K, which is well below the sample critical temperature $T_{C}\approx 248$~K determined in (c).  Taken together, the XRD data imply that the CDW order within each of the domains in our sample would either respect the symmetries of SG 22 ($F222$) or SG 16 ($P222$) if the underlying lattice were ignored or if the modulation vectors were tuned to a lattice-commensurate limit.  Further details of our XRD experiments and CDW symmetry analysis are provided in SI~H.1.}
\label{fig:XRD}
\newpage
\end{figure*}

\begin{figure*}[t]
\centering
\includegraphics[width=0.84\textwidth]{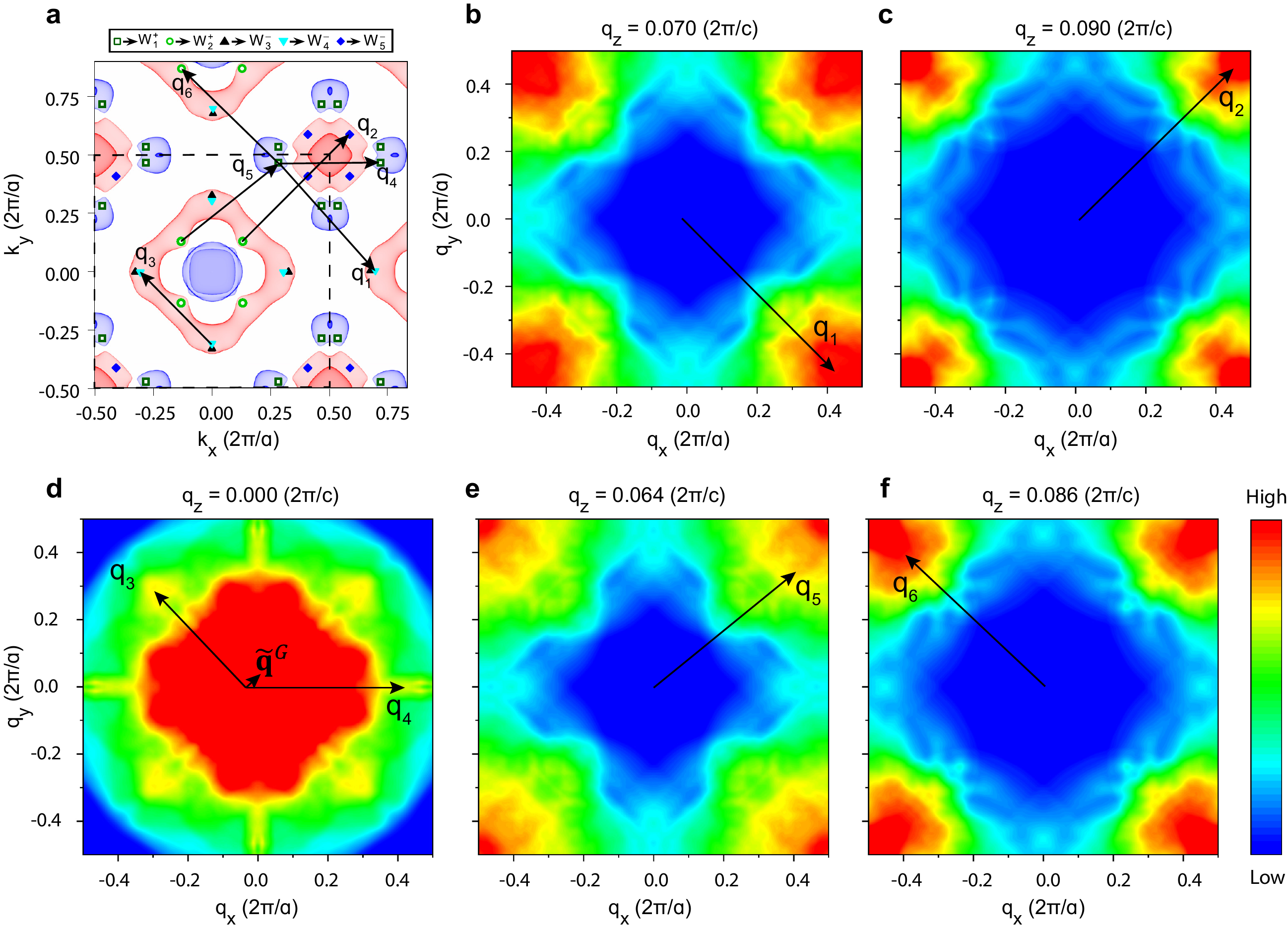}
\caption{(Color online) The $(001)$-projection of the bulk Fermi surface of (TaSe$_4$)$_2$I and the electronic susceptibility calculated from first-principles.  (a) The top view [$(001)$-projection] of the Fermi surface and the projected distribution of the FSWPs; electron (hole) pockets are plotted in blue (red) and the FSWPs are labeled following the convention established in Fig.~\ref{fig:wps}.  The black dashed square in (a) indicates the boundary of the 3D conventional (second) BZ (see Figs.~\ref{fig:Structure} and~\ref{fig:wps}).  (b-f) The electronic susceptibility $\chi_\bq$ calculated using the Fermi surface in (a) at (b) ${\bf q}_z=0.070 (\frac{2\pi}{c})$, (c) ${\bf q}_z=0.090 (\frac{2\pi}{c})$, (d) ${\bf q}_z=0.000 (\frac{2\pi}{c})$, (e) ${\bf q}_z=0.064 (\frac{2\pi}{c})$, and (f) ${\bf q}_z=0.086 (\frac{2\pi}{c})$, respectively (see SI~G for further calculation details).  Up to symmetry-equivalent scattering vectors, the strong peaks shown in (b-f) represent all of the discernible peaks in $\chi_{\bq}$ in the first 3D scattering BZ indexed by $q_{x,y,z}$ that coincide with FSWP nesting vectors (see Table~\ref{chipts}).  We have additionally labeled the vector $\tilde{\bq }^{G}$ within the large spot in (d) at ${\bf q}={\bf 0}$.  $\tilde{\bf q}^{G}$ coincides with the ${\bf q}=(110)$ CDW modulation vector observed in our XRD experiments, which we emphasize to be much shorter than the FSWP nesting vectors in (TaSe$_4$)$_2$I (Fig.~\ref{fig:XRD} and SI~G and~H.1).  This suggests that FSWP nesting is not itself the origin of the CDW in (TaSe$_4$)$_2$I.  Nevertheless, we find that the high-density peaks ${\bf q}_{i}$ in $\chi_\bq$ -- as well as the nesting vectors between all of the FSWPs in (TaSe$_4$)$_2$I with opposite chiral charges (SI~A) -- still match integer multiples of the experimentally-observed CDW modulation vectors.  This indicates that the CDW in (TaSe$_4$)$_2$I backfolds and gaps the FSWPs, consistent with the appearance of a CDW gap in our ARPES experiments (SI~H.2).}
\label{fig:BFS}
\newpage
\end{figure*}

\begin{table*}[t]
\caption{Symmetry-equivalent sets of peaks in the electronic susceptibility [$\chi_\bq$, plotted in Fig.~\ref{fig:BFS}(b-f) and SI~G] that match FSWP ``nesting'' vectors (see SI~A).  We respectively list the index $\bq_{i}$ of one vector within each symmetry-equivalent set of peaks in $\chi_{\bq}$, the coordinates of $\bq_{i}$ in the 3D conventional scattering BZ, the FSWPs nested by $\bq_{i}$ (Table~S1),  the closest integer multiple of the experimentally-observed CDW modulation vectors ${\bf q}=(m,n,o)=m{\boldsymbol \eta}_{1} + n{\boldsymbol \eta}_{2} + o{\boldsymbol \delta}= [m\eta(\frac{2\pi}{a}),n\eta(\frac{2\pi}{a}),o\delta(\frac{2\pi}{c})]$ where $\eta=0.027\pm 0.001$ and $\delta=0.012\pm 0.001$ (further details provided in Fig.~\ref{fig:XRD} and SI~H.1), and the value of $\text{Re}(\chi_{\bq_{i}})$ in relative units.  Among the vectors listed in this table, $\bq_{1}$, and $\bq_{2}$ notably nest FSWPs with opposite chiral charges, and $\bq_{2}$ is the strongest peak in $\chi_\bq$ away from the large central spot near $\bq={\bf 0}$ (see SI~G).  We have additionally listed the vector $\tilde{\bf q}^{G}$, which coincides with the high-intensity ${\bf q}=(110)$ satellite reflection observed in our XRD experiments (see SI~H.1).}
\begin{ruledtabular}
\begin{tabular}{ccccc}
    $\bq_{i}$ & Coordinates & Coupled   & $(m,n,o)$ & $\text{Re}(\chi_{\bq_{i}})$\\
              &($q_{x}\frac{2\pi}{a},q_{y}\frac{2\pi}{a},q_{z}\frac{2\pi}{c}$) &Weyl Points&\\
  \hline
${\bf q}_1$ &(0.41039,-0.46969,0.07045)&$W_1^+\rightarrow W_4^-$ &(15,-17,6)&151945 \\ 
${\bf q}_2$ &(0.45748,0.45748,0.09085) &$W_2^+\rightarrow W_5^-$ &(17,17,8) &235822 \\ 
${\bf q}_3$ &(-0.30640,0.30643,0.00000)&$W_4^-\rightarrow W_4^-$ &(-11,11,0)&112025 \\
${\bf q}_4$ &(0.43364,0.00000,0.00000) &$W_1^+\rightarrow W_1^+$ &(16,0,0)  &104570 \\
${\bf q}_5$ &(0.41496,0.33791,0.06479) &$W_2^+\rightarrow W_1^+$ &(16,13,5) &119227 \\
${\bf q}_6$ &(-0.41496,0.39853,0.08547)&$W_1^+\rightarrow W_2^+$ &(-15,15,7)&155170 \\
${\bf \tilde{q}}^G$ & (0.02700,0.02700,0.00000)& N.A.                            & (1,1,0)    &691542 \\
\end{tabular}
\end{ruledtabular}
\label{chipts}
\end{table*}

On both the TaSe$_4$-chain [Fig.~\ref{fig:FS}(a-c)] and I-atom [Fig.~\ref{fig:FS}(d-f)] terminations of the $(110)$-surface of (TaSe$_4$)$_2$I, eight topological Fermi-arc surface states are present within each surface BZ.  Like the bulk Fermi surface [Fig.~\ref{fig:wps}(b)], the surface Fermi arcs are largely localized within a narrow $k_{z}$ range near $k_{z}=\pm \pi/c$.  To diagnose the topology of the surface Fermi arcs, we calculate the surface-state energy dispersion on closed loops traversing the $(110)$-surface BZ [horizontal and vertical cuts in Fig.~\ref{fig:FS}(a,d) at $k_z=0.54(2\pi/c)$ and $k_{\parallel}=0.50(2\pi/\sqrt{2}a)$, respectively].  On both the TaSe$_4$-chain and I-atom terminations, the horizontal cut [Fig.~\ref{fig:FS}(b,e), respectively] exhibits a $C=-4$ topological spectrum, and the vertical cut [Fig.~\ref{fig:FS}(c,f), respectively] displays a trivial spectrum.  This can be understood by recognizing that the horizontal cut in Fig.~\ref{fig:FS}(a,d) along $k_z=0.54(2\pi/c)$ is equivalent to a loop around $\beta$.  Conversely, the vertical cut in Fig.~\ref{fig:FS}(a,d), which lies along the projection of a $\mathcal{T}$-invariant bulk plane, is required to exhibit a net-zero Chern number.  Consequently, on both possible terminations, no topological surface states cross $E_{F}$ along the vertical line at $k_{\parallel}=0.50(2\pi/\sqrt{2}a)$ [Fig.~\ref{fig:FS}(c,f)].  Interestingly, on the I-atom termination [Fig.~\ref{fig:FS}(d)], the four surface Fermi arcs at $k_{\parallel}<0$ exhibit a different connectivity than in (a), and the four surface Fermi arcs at $k_{\parallel}>0$ all intersect at a single (TRIM) point.  Because the bulk projections and Fermi level in Fig.~\ref{fig:FS}(a,d) are the same, then we attribute the difference in Fermi-arc connectivity between (a,d) to surface Lifshitz transitions driven by the additional layer of $(110)$-surface iodine atoms that is present in (d).  In SI~E and SI~F, we respectively analyze the quasiparticle interference patterns and temperature dependence of the $(110)$-surface Fermi arcs.

Finally, although (TaSe$_4$)$_2$I does not favor cleavage in the $(001)$-direction~\cite{PhysRevLett.110.236401}, the calculated $(001)$-surface Fermi arcs [Fig.~\ref{fig:FS}(g,h)] still provide useful topological information.  On the $(001)$-surface, the projections of the bulk Fermi pockets lie close to $k_{x,y}=0,\pi/a$, and are connected by eight, zone-spanning topological Fermi arcs [Fig.~\ref{fig:FS}(g)].  Calculating the $(001)$-surface states on a loop separating the projected Fermi pockets [Fig.~\ref{fig:FS}(h)], we find that the projected Fermi pockets exhibit the largest Chern numbers predicted to date ($|C|=8$).  The $(001)$-surface states of chiral (TaSe$_4$)$_2$I crystals are in this sense reminiscent of the experimentally confirmed large Fermi arcs of chiral crystals in the RhSi family~\cite{chang2017large,CoSi,NewFermions,KramersWeyl,AlPtObserve,CoSiObserveJapan,CoSiObserveHasan,CoSiObserveChina,PdGaObserve}, which also span the entire surface BZ and connect projected Fermi pockets with large Chern numbers.

{\bf Weyl-Point Coupling.}
Having theoretically established that the high-temperature phase of (TaSe$_4$)$_2$I is a Weyl semimetal, we will now demonstrate a relationship between the bulk FSWPs and the modulation vectors of the CDW phase.  First, we have performed experimental investigations of (TaSe$_4$)$_2$I samples using XRD and ARPES probes to measure the CDW modulation vectors and gap, respectively (SI~H).  Next, to characterize the electronic contribution to the CDW phase, we have calculated the electronic susceptibility from first principles (SI~G).  Lastly, for comparison, we have calculated the ``nesting'' vectors between the FSWPs.

To begin, we first performed XRD experiments on single-crystal (TaSe$_4$)$_2$I samples to infer the CDW modulation vectors and amplitude from satellite reflections.  Specifically, when a crystal with the lattice constants $a,b,c$ is periodically modulated, as occurs in a CDW phase, then satellite Bragg reflections begin to appear in XRD probes at the momentum-space locations ${\bf Q}={\bf G} + {\bf q}$, where ${\bf G} = h{\bf a}^{*} + k{\bf b}^{*} + l{\bf c}^{*}$ are the larger reciprocal lattice vectors of the smaller unit cell of the unmodulated (high-temperature) structure, and ${\bf q} = m{\bm \eta}_{1} + n{\bm \eta}_{2} + o{\bm \delta}$ are the smaller modulation vectors of the (typically incommensurate) CDW-modulated structure.  Examining the results of our XRD probes of (TaSe$_4$)$_2$I, we observed the appearance of satellite reflections in the vicinities of the ${\bf G} = (110)$, $(420)$, $(620)$, and $(554)$ main reflections after samples were cooled below $T_{C} \approx$ 248~K (see Fig.~\ref{fig:XRD} and SI~H.1), representing clear evidence of a CDW transition.  The value of $T_{C}\approx 248$~K observed in our sample is slightly lower than, but still in close agreement with, the value of $T_{C}= 260$~K previously measured in (TaSe$_4$)$_2$I~\cite{Cava1986,260k}.  In the XRD data collected below $T_{C}$ [Fig.~\ref{fig:XRD}(a-e)], we observe a tetragonal arrangement of satellite reflections whose modulation vectors (but not intensities) follow ${\bf q}=[m\eta(\frac{2\pi}{a}),n\eta(\frac{2\pi}{a}),o\delta(\frac{2\pi}{c})]$, where $m + n + o \in 2\mathbb{Z}$, $\eta=0.027\pm 0.001$, and $\delta=0.012\pm 0.001$ (further details provided in SI~H.1).

However, through a careful analysis of the satellite reflection intensities in SI~H.1, we determine that our sample contains two macroscopic domains in position space in which the CDW exhibits a lower point group symmetry [$D_{2}$ (222) in a setting with $C_{2z}$ and $C_{2,x\pm y}$ symmetries] than the high-temperature crystal structure in SG 97 ($I422$) [$D_{4}$ (422)].  This can be seen from the XRD data shown in Fig.~\ref{fig:XRD}(d), in which pairs of satellite reflections related by $C_{2z}$ exhibit the same intensities within experimental uncertainty (see SI~H.1), but satellite reflections related by $C_{4z}$ exhibit intensities that differ by an order of magnitude.  Isolating the satellite reflections within the domain of larger spatial volume -- which we term the majority domain -- we observe a pattern of satellite reflection vectors and intensities that would respect the symmetries of either SG 22 ($F222$) or SG 16 ($P222$) if the underlying lattice were ignored or if the modulation vectors were lengthened to a lattice-commensurate limit.  Notably, the spacing of the satellite reflections in the $q_{x,y}$-plane [Fig.~\ref{fig:XRD}(a,d,e)] indicates that the CDW order is weakly 3D, consisting of both a Peierls-like modulation along the $c$ axis, as well as weak modulation in the $xy$-plane.  In SI~H.1, we use the intensities of the satellite reflections to obtain an estimate for the strength of the in-plane modulation which we find to be small, but nonzero.  We attribute the relative weakness of the in-plane CDW modulation to the weak van der Waals interactions between the TaSe$_4$ chains~\cite{PhysRevLett.110.236401,Ta2Se8IPrepare}.  We emphasize that we were only able to obtain the CDW modulation vectors and estimate the magnitude of the in-plane CDW modulation because of the quality of our crystal sample and because of the high $k$- ($q$-) space resolution and dynamic intensity range of our experiments, which we further detail in SI~H.1.

Next, to characterize the electronic contribution of the Fermi surface of the high-temperature phase of (TaSe$_4$)$_2$I to the low-temperature CDW phase~\cite{Lorenzo1998,Fu1984qvec,Lee1985qvec}, we have calculated both the Fermi-surface nesting vectors between the FSWPs (see SI~A), as well as the electronic susceptibility~\cite{ElectronicSusceptibility}.  In Fig.~\ref{fig:BFS}(b-f), we plot the real part of the bare electronic susceptibility in the constant-matrix approximation $\chi_\bq$ (see SI~G for further details).  To understand the origin of the peaks in $\chi_{\bq}$, we compare the ${\bf q}$ vectors of the strong peaks to Fermi surface nesting vectors.  We find that most -- but not all -- of the peaks in $\chi_\bq$ match FSWP nesting vectors, and that the strongest peaks in $\chi_\bq$ away from $\bq=0$ [$\bq_{2}$ in Fig.~\ref{fig:BFS}(c) and Table~\ref{chipts}] coincide with nesting vectors between FSWPs with opposite chiral charges.  In SI~G, we detail the remaining peaks in $\chi_\bq$, which are weaker than ${\bf q}_{2}$, but comparable in magnitude to ${\bf q}_{1,3-6}$ in Fig.~\ref{fig:BFS}(b-f) and Table~\ref{chipts}.  Furthermore, as shown in Fig.~\ref{fig:XRD} and in SI~H.1, the CDW modulation vectors observed in our XRD experiments [\emph{e.g.} $\tilde{\bf q}^{G}$ in Fig.~\ref{fig:BFS}(d)] are much shorter than the FSWP nesting vectors, suggesting that FSWP nesting is not itself the origin of the CDW in (TaSe$_4$)$_2$I.  Hence, our XRD and electronic susceptibility analyses provide further support for the recognition in~\cite{ElectronicSusceptibility} that 3D CDWs rarely originate from electronic instabilities.  Nevertheless, as shown in SI~A, because all of the nesting vectors between FSWPs with opposite chiral charges can be expressed as integer-valued linear combinations of the much shorter, majority-domain CDW modulation basis vectors (SI~H.1), then we conclude that the CDW in (TaSe$_4$)$_2$I still backfolds and couples the FSWPs.

To further confirm that the CDW in (TaSe$_4$)$_2$I opens an insulating gap, which has been measured in several previous studies~\cite{PhysRevLett.110.236401,260k,Cava1986}, we performed ARPES probes of samples at $100$~K and $270$~K, which are respectively well below and above the CDW transition temperatures observed in our XRD experiments [$T_{C}\approx 248$~K, Fig.~\ref{fig:XRD}(c)] and in the aforementioned previous works ($T_{C}=260$~K).  In the low-temperature phase, we observe a gap of roughly 0.12~eV, which shrinks to less than $0.04$~eV when samples are warmed to $270$~K (see SI~H.2 for additional details).  We attribute this change in gap size to a transition from a low-temperature phase with a CDW-induced band gap at all crystal momenta into the high-temperature Weyl-semimetal phase predicted in this work.  Our ARPES experiments thus provide further evidence that the CDW couples all of the WPs with compensating chiral charges, because a gap cannot be opened by only coupling WPs with the same chiral charges~\cite{W5,ShouchengCDW,TaylorCDW}.

\vspace{0.05in}
{\bf Topology of the CDW Gap.}
Because there are a large number of FSWPs, then it is difficult -- and largely beyond the scope of this work -- to determine the precise topological nature of the CDW gap in (TaSe$_4$)$_2$I at a static value of the CDW phase angle $\phi$.  However, it is plausible, and bolstered by recent experimental findings performed concurrently with this work~\cite{AxionCDWExperiment}, that the CDW gap is topologically nontrivial.  Specifically, recent works have demonstrated that $\mathcal{T}$-symmetric Weyl-CDWs at fixed $\phi$ can be topologically equivalent to mean-field weak topological insulators whose weak-index vectors lie parallel to the CDW wavevector~\cite{WiederBradlynCDW,JiabinWiederCDW}.  More generally, because we have shown that the CDW order in (TaSe$_4$)$_2$I preserves twofold rotation symmetries (Fig.~\ref{fig:XRD} and SI~H.1), which along with $\mathcal{T}$ symmetry, can protect a variety of topological (crystalline) insulating phases~\cite{ChenRotation}, then it is also possible that the CDW gap at static $\phi$ is topologically nontrivial in a manner distinct from previously studied Weyl-CDWs.   This is further supported by analyzing the high-temperature electronic structure of (TaSe$_4$)$_2$I from the perspective of Topological Quantum Chemistry~\cite{QuantumChemistry} (see SI~I).  We leave for future works the precise question of whether the CDW gap in (TaSe$_4$)$_2$I access a lattice-incommensurate topological (crystalline) insulating phase with a single-particle description, or whether the CDW accesses a more exotic, correlated topological phase beyond mean-field theory.

{\bf Notes.}
During the preparation of this work, signatures of an axionic CDW phase were observed in (TaSe$_{4}$)$_2$I samples~\cite{AxionCDWExperiment}, in agreement with the predictions made in this work.  During the submission of this work, signatures of a high-temperature Weyl semimetal phase consistent with the predictions of this work were observed in ARPES probes of (TaSe$_4$)$_2$I samples~\cite{OtherARPES}.  After the submission of this work, theoretical studies of mean-field axionic band topology in Weyl-CDW systems were performed in~\cite{WiederBradlynCDW,JiabinWiederCDW}, and a first-principles study of the CDW instability in (TaSe$_4$)$_2$I was performed in~\cite{CDWTaSeIDFT}; the results of these studies are consistent with our theoretical analysis and experimental data.

{\bf Acknowledgments.}
We thank Barry Bradlyn, Katharina Franke, Yichen Hu, and Jeffrey C. Y. Teo for helpful discussions.  The first-principles calculations of the electronic structure, electronic susceptibility, and quasiparticle interference patterns of (TaSe$_4$)$_2$I were supported by DOE Grant No. DE-SC0016239.  B. J. W. and B. A. B. were additionally supported by NSF EAGER Grant No. DMR 1643312, NSF-MRSEC Grant Nos. DMR-2011750 and DMR-142051, Simons Investigator Grant No. 404513, ONR Grant Nos. N00014-14-1-0330 and N00014-20-1-2303, the BSF Israel US foundation Grant No. 2018226, the Packard Foundation, the Schmidt Fund for Innovative Research, and a Guggenheim Fellowship from the John Simon Guggenheim Memorial Foundation.  Z. W. was supported by the National Natural Science Foundation of China [Grant No. 11974395], the Strategic Priority Research Program of the Chinese Academy of Sciences (CAS) [Grant No. XDB33000000], and the Center for Materials Genome. H. L. M. acknowledges financial support from DFG through the priority program SPP1666 (Topological Insulators).  Technical support by F. Weiss is gratefully acknowledged.  H. L. M. thanks the staff of the ESRF for their hospitality during his stay in Grenoble, and additionally acknowledges helpful interactions with G. Castro, J. Rubio-Zuazo, K. Mohseni, and R. Felici during experiments performed at the ESRF.  W. S., Y. S., Y. Z., and C. F. were supported by ERC Advanced Grant No. 291472 `Idea Heusler', ERC Advanced Grant No. 742068--TOPMAT, and Deutsche Forschungsgemeinschaft DFG under SFB 1143. W. S. additionally acknowledges support from the Shanghai high repetition rate XFEL and extreme light facility (SHINE). Y. Q. acknowledges the support by the National Natural Science Foundation of China (Grant No. U1932217 and 11974246). A portion of the calculations were carried out at the HPC Platform of ShanghaiTech University Library and Information Services, and at the School of Physical Science and Technology.

{\bf Author contributions.}
This project was conceived by Z. W. and B. A. B.  The Weyl semimetal phase of (TaSe$_{4}$)$_2$I was discovered by W. S., Z. W., C. F., and B. A. B.  The first-principles calculations of the high-temperature electronic structure and electronic susceptibility of (TaSe$_{4}$)$_2$I were performed by W. S., Y. Z., Y. S., and Z. W.  The quasiparticle interference patterns of the surface Fermi arcs were computed by B. J. W., W. S., and Z. W.  The FPLO package and the Wannier function interface for first-principles calculations were written by K. K.  The theoretical analysis was performed by B. J. W., Z. W., and B. A. B.  The single-crystal bulk samples were synthesized by Y. Q.  The XRD experiments were performed by H. L. M., J. J., P. W., and S. P.  The ARPES experiments were performed by Y. L, L. S., L. Y., and Y. C.  The manuscript was written by B. J. W., W. S., H. L. M., Z. W., and B. A. B. with help from all authors.

{\bf Competing interests.} The authors declare that they have no competing interests.

{\bf Data availability.}  The source data for all of the figures in this work are available at~\url{https://dataverse.harvard.edu/dataset.xhtml?persistentId=doi:10.7910/DVN/FSRRE4}.  All other data supporting the findings of this study are available from the corresponding authors upon reasonable request.

{\bf Code availability.}  The source code for the calculations performed in this work is available from the corresponding authors upon reasonable request.


{\bf Methods.} We performed {\it ab-initio} calculations based on density functional theory (DFT) as implemented in the FPLO package~\cite{Koepernik1999}, and used the full-potential local-orbital basis within the generalized gradient approximation (GGA)~\cite{perdew1996}, fully incorporating the effects of spin-orbit coupling (SOC).  The projected atomic Wannier functions (PAWFs) were constructed using the Ta $d$, Se $p$, and I $p$ orbitals to reproduce the band structures obtained from {\it ab-initio} calculations. The surface states were obtained by calculating the surface Green's functions~\cite{Sancho1984,Sancho1985} of a semi-infinite tight-binding model constructed from the above PAWFs.  All calculations were performed employing the experimental lattice parameters~\cite{Ta2Se8IPrepare,gressier1984characterization,gressier1984electronic}.

We also performed XRD and ARPES experiments on (TaSe$_{4}$)$_2$I samples to study the CDW wavevector and electronic band structure, respectively.  The XRD experiments were performed at beamline 25B of the European Synchrotron Radiation Facility in Grenoble, France using a six-circle diffractometer and a wavelength of $\lambda$=0.71~\AA.  The whisker-shaped (TaSe$_{4}$)$_2$I sample -- which was $\sim$100 $\mu$m in diameter and grown using the method detailed in~\cite{AxionCDWExperiment} -- was first mounted on a copper sample holder oriented with its $c$-axis perpendicular to the incoming beam and cooled to a minimum temperature of 88~K using a flow of liquid nitrogen.  We then measured the intensities of the diffracted X-rays near several main Bragg reflections, employing a 2D pixel detector with pixel size 55 $\mu$m placed 1250 mm away from the sample to collect the data from both 1D line scans and 2D reciprocal-space maps.  ARPES measurements were performed at the high-resolution branch of beamline I05, Diamond Light Source (DLS) with a Scienta R4000 analyzer.  The photon-energy range for the DLS was 30-220~eV.  The angles of the emitted photoelectrons were measured with a resolution of 0.2$^\circ$, and their energies were measured at an overall resolution of $<15$~meV.  After samples were glued to the sample holder, they were then cleaved in situ to expose the $(110)$-surface, which is the favored cleavage plane of (TaSe$_{4}$)$_2$I~\cite{PhysRevLett.110.236401}.  Throughout our ARPES experiments, samples were kept at a pressure of $<1.5\times10^{-10}$ Torr, and measurements of the low- and high-temperature phases of (TaSe$_{4}$)$_2$I were performed at $100$~K and $270$~K, respectively.


\bibliography{TopMater_HLM}
\clearpage
\beginsupplement{}
\onecolumngrid

\begin{center}
{\bf Supplementary Information for ``A Charge-Density-Wave Topological Semimetal''}
\end{center}

\subsection{A. Distribution and Nesting Vectors of the Bulk Weyl Points in (TaSe$_4$)$_2$I}

\vspace{-0.11in}

In this section, we will detail the distribution of the bulk chiral fermions at the Fermi level [Fermi-suface Weyl points (FSWPs)] in (TaSe$_4$)$_2$I.  We will then demonstrate that all of the FSWPs with opposite chiral charges can be nested by the experimentally observed CDW modulation vectors, leading to a gapped CDW phase (see SI~H.1 and H.2, respectively).

To begin, in 3D (TaSe$_4$)$_2$I, the \emph{entire} Fermi surface is formed from topological bands connected to bulk chiral fermions (WPs).  Specifically, because (TaSe$_4$)$_2$I crystals in SG 97 ($I422$) are symmorphic, chiral, and have strong SOC, then all of their bulk degeneracies are necessarily point-like, and carry nontrivial chiral charges, as explicitly shown in~\cite{KramersWeyl}.  In the case of SG 97 with relevant SOC, bands are generically singly degenerate~\cite{WiederLayers}, and can only meet in conventional WPs away from high-symmetry lines, double-WPs along fourfold rotation axes~\cite{Xu2011,Huang02022016}, or Kramers-WPs at high-symmetry (TRIM) points~\cite{KramersWeyl}.  Therefore, independent of the details of the dispersion, every single band in (TaSe$_4$)$_2$I is connected to a bulk degeneracy with nontrivial chiral charge (though depending on the dispersion, the Fermi-arc surface states of chiral fermions below $E_{F}$ may be obscured by the projections of bulk Fermi pockets; for example, the surface Fermi
arcs of Kramers-WPs with weak Dresselhaus SOC are necessarily covered by surface-projecting Fermi pockets~\cite{KramersWeyl}).

In Figs.~1(c) and~2 of the main text, we show the distribution of the FSWPs in (TaSe$_4$)$_2$I obtained from first-principles calculations, as detailed in the Methods section.  Because (TaSe$_4$)$_2$I crystallizes in a body-centered tetragonal structure [space group (SG) 97 ($I422$)], then bulk BZ slices taken at different values of a perpendicular momentum exhibit a different shape [Fig.~1(c) of the main text].  Specifically, the larger octagon in Fig.~2(a) of the main text represents the boundary of the bulk $k_{z}=0$ plane [containing $\Gamma$, $\Sigma$, $Y$, and $X$ in Fig.~1(c) of the main text], and the smaller square represents the boundary of the $k_{z}=2\pi/c$ plane (containing $Z$, $\Sigma_{1}$, and $Y_{1}$).  In Fig.~2(b) of the main text, the larger octagon represents the boundary of the $k_{x} + k_{y} = 0$ plane (containing $\Gamma$, $X$, $P$, $Y_{1}$, and $Z$), and the smaller diamond represents the boundary of the $k_{x} + k_{y} = \frac{\pi}{a\sqrt{2}}$ plane (containing $Y$ and different, symmetry-related $X$ and $P$ points).

\begin{table}[!h]
 \caption{Fermi-Surface Weyl Points (WPs) in (TeSe$_4$)$_2$I.  The 48 WPs in (TeSe$_4$)$_2$I nearest the Fermi energy ($E_{F}$) -- which we designate the ``Fermi-surface Weyl Points'' (FSWPs) -- appear in five, symmetry-related sets.  For each set of FSWPs, we list the number of WPs within the set, the position of one WP within the set in reduced (conventional-cell) coordinates, the energy of all WPs within the set relative to $E_{F}$, and the Chern number (chiral charge) of each WP within the set (which is the same for all WPs within each set because (TeSe$_4$)$_2$I is a structurally chiral crystal~\cite{KramersWeyl}).}
  \label{tab:wp_position}
  \begin{tabular}{ccccc}
  \hline
  \hline
      WPs    & Multi-& Coordinates &$E-E_F$& Chern \\
             & plicity &($k_{x}\frac{2\pi}{a},k_{y}\frac{2\pi}{a},k_{z}\frac{2\pi}{c}$) & (meV)& Number\\
             \hline
  $W_1^+$    &16  &(0.28318, 0.46969, 0.51034) &-8.928 & $+1$\\
  $W_2^+$    & 8  &(0.13178, 0.13178, 0.57513) &6.195 & $+1$\\
  $W_3^-$    & 8  &(0.32795, 0.00000, 0.56275) &9.497 & $-1$\\
  $W_4^-$    & 8  &(0.30643, 0.00000, 0.56011) &12.666 & $-1$\\
  $W_5^-$    & 8  &(0.41074, 0.41074, 0.51572) &12.906 & $-1$\\
\hline
\hline
  \end{tabular}
\label{tbwps}
\end{table}

In (TaSe$_4$)$_2$I, we identify 24 pairs of WPs in the bulk within 15~meV of the Fermi energy ($E_{F}$), which we designate as the ``Fermi-surface'' WPs (FSWPs).  In Table~\ref{tbwps}, we list the coordinates and multiplicities of the 48 FSWPs.  The bulk FSWPs listed in Table~\ref{tbwps} divide into five sets of symmetry-related WPs.  Within each set, the FSWPs are related to each other by the symmetry elements of SG 97 ($I422$): $C_{4z}$, $C_{2x(2y)}$, $C_{2,x+y(2,x-y)}$, and $\mathcal{T}$.  Because all of these operations are proper rotations or time-reversal, which do not change the sign of the chiral charge of a WP, then all of the WPs within each set exhibit the same chiral charge.

To explicitly confirm and provide additional context for the contents of Table~\ref{tbwps}, we can use crystal symmetries to derive the momentum-space coordinates and multiplicities of the symmetry-related sets of FSWPs.  This can be accomplished by employing the~\href{http://www.cryst.ehu.es/cryst/get_kvec.html}{KVEC} tool on the Bilbao Crystallographic Server (BCS)~\cite{BCS1,BCS2}.  In the notation of~\href{http://www.cryst.ehu.es/cryst/get_kvec.html}{KVEC}, symmetry-related $k$ points are labeled as distinct reciprocal-space Wyckoff positions, which are also known more commonly as ``momentum stars''~\cite{BigBook}.

\begin{figure*}[b]
\centering
\includegraphics[width=0.59\textwidth]{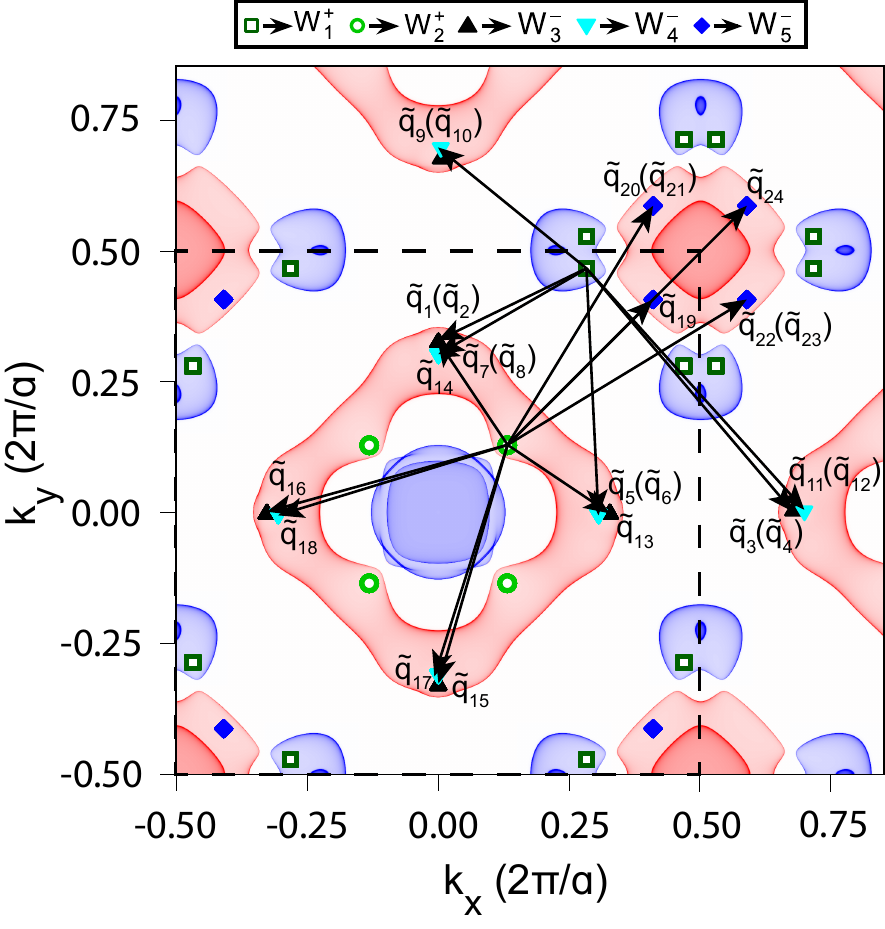}
\caption{(Color online) Independent nesting vectors $\tilde{\bf q}$ between FSWPs with opposite chiral charges ($W^{+}_{i}$ and $W_{j}^{-}$, respectively see Table~\ref{tbwps}) that match integer multiples of the experimentally-observed CDW modulation vectors.  In this figure, we show the top view [$(001)$-projection] of the Fermi surface of (TeSe$_4$)$_2$I and the projected distribution of the FSWPs; electron (hole) pockets are plotted in blue (red) and the FSWPs are labeled following the convention established in Fig.~2 of the main text.  The black dashed square indicates the boundary of the 3D conventional (\emph{i.e.} second) BZ (see Figs.~1 and~2 of the main text).  The nesting vectors $\tilde{\bf q}$ shown in this figure can be expressed as integer-valued linear combinations of the CDW modulation vectors, such that $\tilde{\bf q}=(m,n,o)=m{\boldsymbol \eta}_{1} + n{\boldsymbol \eta}_{2} + o{\boldsymbol \delta}= [m\eta(\frac{2\pi}{a}),n\eta(\frac{2\pi}{a}),o\delta(\frac{2\pi}{c})]$, where $\eta=0.027\pm 0.001$, $\delta=0.012\pm 0.001$, and $m+n+o \in 2\mathbb{Z}$ (see SI~H.1 for further details).  The specific coordinates of the linked FSWPs and nesting vectors $\tilde{\bf q}$ are provided in Table~\ref{wpnesting}.  The vectors shown in this figure, when taken along with their conjugates under the experimentally-observed symmetries of the CDW phase $C_{2z}$, $C_{2,x\pm y}$, and $\mathcal{T}$ (see SI~H.1), can nest and gap all of the FSWPs in (TeSe$_4$)$_2$I, consistent with the gapped CDW phase observed in our ARPES experiments (see SI~H.2).  We note that $\tilde{\bf q}_{12}$ and $\tilde{\bf q}_{24}$ respectively coincide with the electronic susceptibility peaks ${\bf q}_{1}$ and ${\bf q}_{2}$ in Fig.~5 and Table~I of the main text.}
\label{fig:nesting}
\end{figure*}

It is important to emphasize that, for each SG,~\href{http://www.cryst.ehu.es/cryst/get_kvec.html}{KVEC} lists the reciprocal-space Wyckoff positions generated by the unitary crystal symmetries of the type-I magnetic subgroup of the SG~\cite{BigBook}; specifically,~\href{http://www.cryst.ehu.es/cryst/get_kvec.html}{KVEC} \emph{does not} incorporate the action of time-reversal ($\mathcal{T}$) symmetry, which is present in real materials, such as (TaSe$_{4}$)$_2$I, and must also be included to correctly determine the locations of the symmetry-related FSWPs.  However, we can circumvent the absence of $\mathcal{T}$ symmetry in~\href{http://www.cryst.ehu.es/cryst/get_kvec.html}{KVEC} in the particular case of SG 97 by recognizing that inversion symmetry ($\mathcal{I}$) and $\mathcal{T}$ symmetry have the same action on crystal momenta:
\begin{equation}
\mathcal{T}\vec{k} = \mathcal{I}\vec{k} = -\vec{k},
\label{eq:Wyckoff1}
\end{equation}
even though they act differently on Bloch wavefunctions~\cite{bernevigbook}.  SG~97 is generated by:
\begin{equation}
\{\mathcal{T}|000\},\ \{C_{4z}|000\},\ \{C_{2x}|000\},
\label{eq:Wyckoff2}
\end{equation}
as well as body-centered tetragonal lattice translations~\cite{BigBook}.  We can add $\mathcal{I}$ symmetry to SG 97 to generate its index-2 supergroup SG 139 ($I4/mmm$):
\begin{equation}
I4/mmm \equiv E(I422)\cup \{\mathcal{I}|000\}(I422),
\label{eq:Wyckoff3}
\end{equation}
where $E$ is the identity element.  Crucially, Eqs.~(\ref{eq:Wyckoff1}) through~(\ref{eq:Wyckoff3}) imply that the reciprocal-space Wyckoff positions of the type-I magnetic subgroup of SG 139 have the same multiplicities and momentum-space coordinates [though not the same little groups or small irreducible representations] as those in $\mathcal{T}$-symmetric SG 97.  We are therefore able to use the output of~\href{http://www.cryst.ehu.es/cryst/get_kvec.html}{KVEC} for SG 139 to check our analysis of the symmetry-related FSWPs in (TaSe$_{4}$)$_2$I.  We note that, because~\href{http://www.cryst.ehu.es/cryst/get_kvec.html}{KVEC} lists the reciprocal-space Wyckoff positions with respect to the conventional (second) BZ, which is twice as large as the primitive BZ of body-centered SGs 97 and 139, the reciprocal-space Wyckoff positions and multiplicities referenced in this work contain half as many sites as those listed in~\href{http://www.cryst.ehu.es/cryst/get_kvec.html}{KVEC}.  For example, when the effects of $\mathcal{T}$ symmetry are incorporated, a (Kramers-~\cite{KramersWeyl}) WP lying at $\Gamma$ in SG 97 would be listed as occupying the $2a$ reciprocal-space Wyckoff position in~\href{http://www.cryst.ehu.es/cryst/get_kvec.html}{KVEC}, but would be labeled in this work as occupying the $1a$ position, because there is only one $\Gamma$ point in the first BZ of SG 97.

\begin{table}[b]
  \caption{The independent nesting vectors $\tilde{\bf q}$ between FSWPs with opposite chiral charges ($W^{+}_{i}$ and $W_{j}^{-}$, respectively see Table~\ref{tbwps}) depicted in Fig.~\ref{fig:nesting}.  Each nesting vector $\tilde{\bf q}$ in this table is related to additional nesting vectors under the symmetries of the high-temperature phase of (TeSe$_4$)$_2$I in SG 97 ($I422$) [see the text following Eq.~(\ref{eq:Wyckoff3})].  The nesting vectors $\tilde{\bf q}$ listed in this table match integer multiples of the experimentally-observed CDW modulation vectors, such that $\tilde{\bf q}=(m,n,o)=m{\boldsymbol \eta}_{1} + n{\boldsymbol \eta}_{2} + o{\boldsymbol \delta}= [m\eta(\frac{2\pi}{a}),n\eta(\frac{2\pi}{a}),o\delta(\frac{2\pi}{c})]$, where $\eta=0.027\pm 0.001$, $\delta=0.012\pm 0.001$, and $m+n+o \in 2\mathbb{Z}$ (see SI~H.1 for further details).  Combining the vectors listed in this table with their conjugates under the experimentally-observed symmetries of the CDW phase $C_{2z}$, $C_{2,x\pm y}$, and $\mathcal{T}$ (see SI~H.1), the resulting nesting vectors couple all of the FSWPs in (TeSe$_4$)$_2$I with opposite chiral charges.  We note that $\tilde{\bf q}_{12}$ and $\tilde{\bf q}_{24}$ respectively coincide with the electronic susceptibility peaks ${\bf q}_{1}$ and ${\bf q}_{2}$ in Fig.~5 and Table~I of the main text.}
  \begin{tabular}{ccc|cc|cc}
  \hline
  \hline
   $\bf \tilde{q}_{i}$  &($q_{x}\frac{2\pi}{a},q_{y}\frac{2\pi}{a},q_{z}\frac{2\pi}{c}$) &$(m,n,o)$  & $W_i^+$ & ($k_{x}\frac{2\pi}{a},k_{y}\frac{2\pi}{a},k_{z}\frac{2\pi}{c}$) & $W_j^-$ & ($k_{x}\frac{2\pi}{a},k_{y}\frac{2\pi}{a},k_{z}\frac{2\pi}{c}$) \\
  \hline
${\bf \tilde{q}}_{1}$  & (-0.28318,-0.14174,0.07309) & (-11,-5,6)  & $W_1^+$ & (0.28318,0.46969,0.48966) & $W_3^-$ & (0.00000,0.32795,0.56275) \\
${\bf \tilde{q}}_{2}$  & (-0.28318,-0.14174,-0.05241)& (-11,-5,-4) & $W_1^+$ & (0.28318,0.46969,0.48966) & $W_3^-$ & (0.00000,0.32795,0.43725) \\
${\bf \tilde{q}}_{3}$  & (0.38887,-0.46969,0.07309)  & (15,-17,6)  & $W_1^+$ & (0.28318,0.46969,0.48966) & $W_3^-$ & (0.67205,0.00000,0.56275) \\
${\bf \tilde{q}}_{4}$  & (0.38887,-0.46969,-0.05241) & (15,-17,-4) & $W_1^+$ & (0.28318,0.46969,0.48966) & $W_3^-$ & (0.67205,0.00000,0.43725) \\
${\bf \tilde{q}}_{5}$  & (0.02325,-0.46969,-0.04977) & (1,-17,-4)  & $W_1^+$ & (0.28318,0.46969,0.48966) & $W_4^-$ & (0.30643,0.00000,0.43989) \\
${\bf \tilde{q}}_{6}$  & (0.02325,-0.46969,0.07045)  & (1,-17,6)   & $W_1^+$ & (0.28318,0.46969,0.48966) & $W_4^-$ & (0.30643,0.00000,0.56011) \\
${\bf \tilde{q}}_{7}$  & (-0.28318,-0.16326,-0.04977)& (-10,-6,-4) & $W_1^+$ & (0.28318,0.46969,0.48966) & $W_4^-$ & (0.00000,0.30643,0.43989) \\
${\bf \tilde{q}}_{8}$  & (-0.28318,-0.16326,0.07045) & (-10,-6,6)  & $W_1^+$ & (0.28318,0.46969,0.48966) & $W_4^-$ & (0.00000,0.30643,0.56011) \\
${\bf \tilde{q}}_{9}$  & (-0.28318,0.22388,-0.04977) & (-10,8,-4)  & $W_1^+$ & (0.28318,0.46969,0.48966) & $W_4^-$ & (0.00000,0.69357,0.43989) \\
${\bf \tilde{q}}_{10}$ & (-0.28318,0.22388,0.07045)  & (-10,8,6)   & $W_1^+$ & (0.28318,0.46969,0.48966) & $W_4^-$ & (0.00000,0.69357,0.56011) \\
${\bf \tilde{q}}_{11}$ & (0.41039,-0.46969,-0.04977) & (15,-17,-4) & $W_1^+$ & (0.28318,0.46969,0.48966) & $W_4^-$ & (0.69357,0.00000,0.43989) \\
${\bf \tilde{q}}_{12}$ & (0.41039,-0.46969,0.07045)  & (15,-17,6)  & $W_1^+$ & (0.28318,0.46969,0.48966) & $W_4^-$ & (0.69357,0.00000,0.56011) \\
\hline
${\bf \tilde{q}}_{13}$ & (0.19617,-0.13178,0.13788)  & (7,-5,12)   & $W_2^+$ & (0.13178,0.13178,0.42487) & $W_3^-$ & (0.32795,0.00000,0.56275) \\
${\bf \tilde{q}}_{14}$ & (-0.13178,0.19617,0.13788)  & (-5,7,12)   & $W_2^+$ & (0.13178,0.13178,0.42487) & $W_3^-$ & (0.00000,0.32795,0.56275) \\
${\bf \tilde{q}}_{15}$ & (-0.13178,-0.45973,0.13788) & (-5,-17,12) & $W_2^+$ & (0.13178,0.13178,0.42487) & $W_3^-$ & (0.00000,-0.32795,0.56275)\\
${\bf \tilde{q}}_{16}$ & (-0.45973,-0.13178,0.13788) & (-17,-5,12) & $W_2^+$ & (0.13178,0.13178,0.42487) & $W_3^-$ & (-0.32795,0.00000,0.56275)\\
${\bf \tilde{q}}_{17}$ & (-0.13178,-0.43821,0.13524) & (-5,-16,11) & $W_2^+$ & (0.13178,0.13178,0.42487) & $W_4^-$ & (0.00000,-0.30643,0.56011)\\
${\bf \tilde{q}}_{18}$ & (-0.43821,-0.13178,0.13524) & (-16,-5,11) & $W_2^+$ & (0.13178,0.13178,0.42487) & $W_4^-$ & (-0.30643,0.00000,0.56011)\\
${\bf \tilde{q}}_{19}$ & (0.27896,0.27896,0.09085)   & (10,10,8)   & $W_2^+$ & (0.13178,0.13178,0.42487) & $W_5^-$ & (0.41074,0.41074,0.51572) \\
${\bf \tilde{q}}_{20}$ & (0.27896,0.45748,0.05941)   & (10,17,5)   & $W_2^+$ & (0.13178,0.13178,0.42487) & $W_5^-$ & (0.41074,0.58926,0.48428) \\
${\bf \tilde{q}}_{21}$ & (0.27896,0.45748,0.09085)   & (10,17,7)   & $W_2^+$ & (0.13178,0.13178,0.42487) & $W_5^-$ & (0.41074,0.58926,0.51572) \\
${\bf \tilde{q}}_{22}$ & (0.45748,0.27896,0.05941)   & (17,10,5)   & $W_2^+$ & (0.13178,0.13178,0.42487) & $W_5^-$ & (0.58926,0.41074,0.48428) \\
${\bf \tilde{q}}_{23}$ & (0.45748,0.27896,0.09085)   & (17,10,7)   & $W_2^+$ & (0.13178,0.13178,0.42487) & $W_5^-$ & (0.58926,0.41074,0.51572) \\
${\bf \tilde{q}}_{24}$ & (0.45748,0.45748,0.09085)   & (17,17,8)   & $W_2^+$ & (0.13178,0.13178,0.42487) & $W_5^-$ & (0.58926,0.58926,0.51572) \\
\hline
\hline
  \end{tabular}
\label{wpnesting}
\end{table}

Having established the action of~\href{http://www.cryst.ehu.es/cryst/get_kvec.html}{KVEC}, we now specifically analyze the 48 FSWPs in (TaSe$_{4}$)$_2$I.  First, in the most general reciprocal-space Wyckoff position (\emph{i.e.} the highest-multiplicity momentum star) in SG 97 ($16o$), crystal symmetries [specifically $C_{4z}$, $C_{2x(2y)}$, and $C_{2,x+y(x-y)}$] generate eight of the sites, and $\mathcal{T}$ symmetry generates the other eight sites.  In (TeSe$_4$)$_2$I, 16 symmetry-related FSWPs with the same Chern number ($C=+1$) lie below the Fermi energy and occupy $16o$; we denote them as $W^{+}_{1}$ in Table~\ref{tbwps}.  When placed at a higher-symmetry location in the BZ, a WP becomes related by symmetry to fewer WPs than one placed at a generic $k$ point.  In (TaSe$_4$)$_2$I, we additionally identify four groupings of 8 symmetry-related WPs at less general $k$ points than $16o$, all of which lie above $E_{F}$ (see Table~\ref{tbwps}).  A WP at ($k_x,0,k_z$) is left invariant under the combined symmetry $C_{2y}\times\mathcal{T}$~\cite{Soluyanov:2015WSM2}, whereas, under the other symmetries of SG 97, it is related to 7 other WPs, resulting in 8 total WPs at ($\pm k_x,0,\pm k_z$) and ($0,\pm k_x,\pm k_z$).  We observe two sets of WPs with these reciprocal-space coordinates (the $8m$ position), which we denote as $W_3^-$ and $W_4^-$, respectively.  Similarly, a WP at ($k_x,k_x,k_z$) is left invariant under $C_{2,x-y}\times\mathcal{T}$, whereas, under the other symmetries of SG 97, it is related to 7 other WPs, resulting in 8 total WPs at ($\pm k_x,\pm k_x,\pm k_z$).  We also observe two sets of WPs with these coordinates ($8n$), which we denote as $W_2^+$ and $W_5^-$, respectively.  In SI~I, we show how the WPs in (TaSe$_{4}$)$_2$I derive from the electronic structure of isolated TaSe$_4$ chains, which we find to be filling-enforced 1D semimetals.

As shown in Table~\ref{tbwps} and discussed in~\cite{KramersWeyl}, because crystals in SG 97 are structurally chiral, then WPs in (TaSe$_4$)$_2$I with opposite Chern numbers generically lie at different energies; of the 48 FSWPs in (TaSe$_4$)$_2$I in Table~\ref{tbwps}, only the 16 $C=+1$ WPs $W_{1}^{+}$ lie below $E_{F}$.  Specifically, in nonmagnetic ($\mathcal{T}$-symmetric) topological semimetals~\cite{novoselov2005two,QSHGraphene,WiederLayers,armitage2017weyl,Young2012,Liu:2014bf,Wang:2013is,Wang:2012ds,AlexeyTriple,HasanTriple,WCTriple,DDP,NewFermions,HingeSM,TMDHOTI,TaylorToy}, an energy offset between all chiral fermions with oppositely-signed chiral charges can only be realized in the absence of bulk improper rotation symmetries, \emph{i.e.}, in structurally chiral crystals~\cite{KramersWeyl} like (TaSe$_4$)$_2$I.  While examples of filling-enforced semimetals with energetically separated, oppositely charged chiral fermions have been identified in previous works~\cite{chang2017large,CoSi,2018Flicker}, (TaSe$_4$)$_2$I represents an extremely rare example of a \emph{non-minimally-connected} Weyl semimetal in which \emph{all} of the symmetry-related occupied chiral fermions nearest the Fermi energy exhibit the same Chern number.  The net $C=+16$ chiral charge of the occupied WPs in (TaSe$_4$)$_2$I is \emph{by far} the largest value yet predicted.  (TaSe$_4$)$_2$I samples should therefore exhibit a strongly amplified response in quantized bulk topological chiral probes, such as the chiral magnetic effect (CME)~\cite{PhysRevD.78.074033, PhysRevB.89.035142, PhysRevB.92.161110, PhysRevB.91.115203, PhysRevB.88.125105, PhysRevLett.111.027201} and the quantized circular photogalvanic effect (CPGE)~\cite{de2017quantized,2018Flicker}.  In particular, the CPGE has been predicted in the structurally chiral Ag$_2$Se, TlTeO$_6$, and RhSi families~\cite{KramersWeyl,KramersWeylExperiment,chang2017large,2018Flicker}, and was recently observed in RhSi~\cite{RhSiCPGE}.  Additionally, signatures of the long-sought ``axion insulator''~\cite{WiederAxion} as well as other, correlated topological phases have been proposed in chiral semimetals for which pairs of chiral fermions have become coupled by (typically incommensurate) charge- or spin-density waves (CDWs and SDWs, respectively)~\cite{RahulSDW,TaylorCDW,TitusCDW,ShouchengCDW,VladCDW}.  Therefore, (TaSe$_4$)$_2$I represents a highly promising platform for the observation of enhanced or quantized response effects in gapless topological phases.

Lastly, in our XRD experiments -- which we will explicitly detail in SI~H.1 -- we determined that the CDW phase of (TeSe$_4$)$_2$I respects $C_{2z}$, $C_{2,x\pm y}$, and $\mathcal{T}$ symmetries, and that the CDW modulation vectors ${\bf q}$ follow the pattern ${\bf q}=(m,n,o)=m{\boldsymbol \eta}_{1} + n{\boldsymbol \eta}_{2} + o{\boldsymbol \delta}= [m\eta(\frac{2\pi}{a}),n\eta(\frac{2\pi}{a}),o\delta(\frac{2\pi}{c})]$, where $\eta=0.027\pm 0.001$, $\delta=0.012\pm 0.001$, and $m+n+o \in 2\mathbb{Z}$.  Using the FSWP positions listed in Table~\ref{tbwps}, we have calculated all of the nesting vectors between the FSWPs with opposite chiral charges that match integer multiples of the experimentally-observed CDW modulation basis vectors; the resulting nesting vectors are shown in Fig.~\ref{fig:nesting} and listed in Table~\ref{wpnesting}.  We find that \emph{all} of the FSWPs in (TeSe$_4$)$_2$I can be pairwise annihilated by the nesting vectors in Fig.~\ref{fig:nesting} in Table~\ref{wpnesting}, when the nesting vectors are taken in conjunction with their conjugates under the experimentally-observed symmetries of the CDW phase $C_{2z}$, $C_{2,x\pm y}$, and $\mathcal{T}$ (see SI~H.1).  This is consistent with the gapped CDW phase that we observed in our ARPES experiments (see SI~H.2) and observed in previous works~\cite{260k,Cava1986,CDWTransport1}, because a gapped Weyl-CDW phase can only be realized by coupling WPs with opposite chiral charges~\cite{W5,ShouchengCDW,TaylorCDW}.

\subsection{B. Band Dispersions of the Bulk Weyl Points}

\begin{figure*}[!h]
\centering
\includegraphics[width=\textwidth]{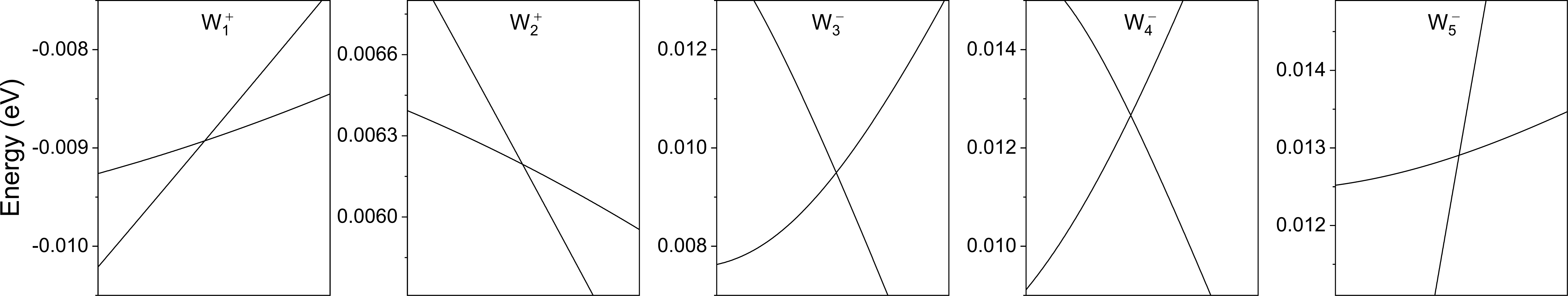}
\caption{(Color online) Typifying band dispersions of WPs within each of the five symmetry-related groupings of Fermi-surface WPs in (TaSe$_4$)$_2$I (Table~\ref{tbwps}).}
\label{fig:type}
\end{figure*}

To diagnose each bulk FSWP in (TaSe$_4$)$_2$I as type-I (untilted) or type-II (tilted)~\cite{Soluyanov:2015WSM2}, we perform detailed three-dimensional band structure calculations.  Specifically, by calculating the Fermi surface at different energy contours, we determine that $W_3^-$ and $W_4^-$ (upward- and downward-pointing triangles in Fig.~2 of the main text, respectively) are type-I WPs, whereas $W_1^+$, $W_2^+$, and $W_5^-$ (boxes, circles, and diamonds in Fig.~2 of the main text, respectively) are type-II  WPs.  In Fig.~\ref{fig:type}, we show typifying band dispersions of one FSWP within each of the five symmetry-related sets listed in Table~\ref{tbwps}.

\subsection*{C. Bulk Fermi Surface Projections in the Conventional-Cell $(110)$-Surface BZ}
\label{app:c}

In order to understand the distribution of the topological Fermi arcs in (TaSe$_{4}$)$_2$I, we determine the net chiral charges of the projections of the bulk Fermi surface onto the surface BZs.  Because (TaSe$_{4}$)$_2$I experimentally favors cleaving along the conventional-cell $(110)$-plane~\cite{PhysRevLett.110.236401}, we focus in this section on the conventional-cell $(110)$-surface projections of the bulk Fermi surface, where all momentum-space vectors are given with respect to the conventional-cell reciprocal lattice vectors $(k_x\frac{2\pi}{a}$, $k_y\frac{2\pi}{a}$, $k_z\frac{2\pi}{c})$.

\begin{figure*}[!h]
\centering
\includegraphics[width=\textwidth]{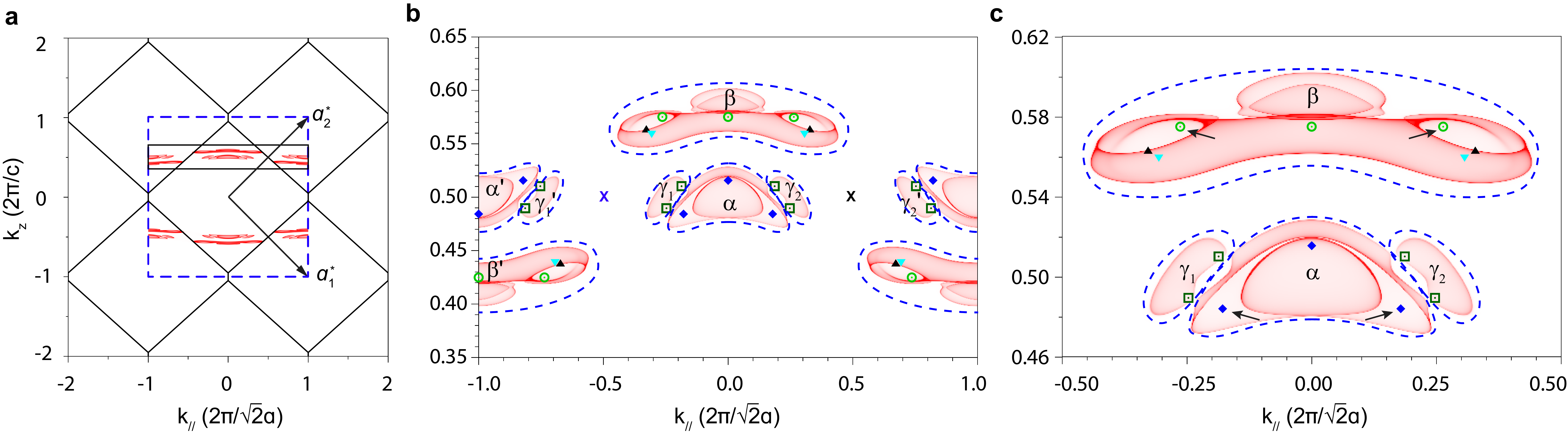}
\caption{(Color online) Bulk Fermi surface projected onto the conventional-cell $(110)$-surface BZ.  (a) The $(110)$-surface BZs (black squares), whose reciprocal lattice vectors are $a_{1,2}^{*}$, and the projections of the bulk Fermi surface; the blue dashed square in (a) indicates the conventional surface BZ indexed by $k_{\parallel}$ and $k_{z}$, which is twice as large as the primitive surface BZ (black squares).  (b) An enlarged and rescaled view of the projected Fermi pockets near $k_{z}=\pi/c$ in the solid black rectangle in (a); the projected bulk states in (b) are the same as those shown [along with the $(110)$-surface states] in Fig.~3(a) of the main text.  In (b), we denote the $(110)$-surface time-reversal-invariant momenta (TRIM points) at $-\frac{a_1^*}{2}$ and $\frac{a_2^*}{2}$ with $\times$ symbols.  The projections of the bulk Fermi pockets form four time-reversal pairs of islands, each of which is shown enclosed in a dashed blue line in (b) and (c), and carries a net topological chiral charge.  The four islands (and their time-reversal partners) are respectively denoted as $\alpha\ (\alpha ')$, $\beta\ (\beta ')$, $\gamma_1\ (\gamma_1')$, and $\gamma_2\ (\gamma_2')$.  (c) A closer view of the projected bulk Fermi surface in the vicinity of $k_{\parallel}=0$, $k_{z}=\pi/c$.  In both (b) and (c), we employ the labeling scheme of Fig.~2 of the main text, in which the projections of the bulk WPs $W_{1}^{+}$, $W_{2}^{+}$, $W_{3}^{-}$, $W_{4}^{-}$, and $W_{5}^{-}$ are respectively indicated by boxes, circles, upward-pointing triangles, downward-pointing triangles, and diamonds.  As previously in Fig.~2 of the main text, we use arrows to indicate FSWP projections with total chiral charges of $|C|=1$; the symbols without arrows, which correspond to the superposed projections of two bulk FSWPs with the same chiral charge, indicate projections with net chiral charges of $|C|=2$.}
\label{bulk}
\end{figure*}

To project the bulk Fermi surface onto the $(110)$-surface, we must carefully treat the bulk body-centered tetragonal position- and reciprocal-space lattice vectors of (TaSe$_4$)$_2$I.  In the units of the conventional-cell lattice translations $(a,a,c)$, the position-space lattice vectors of the $(110)$-surface are $a_1=(-a/2, a/2,-c/2)$ and $a_2=(-a/2,a/2, c/2)$.  Thus, the $(110)$-surface primitive reciprocal lattice vectors are $a^*_1=(-\pi/a, \pi/a,-2\pi/c)$ and $a^*_2=(-\pi/a,\pi/a, 2\pi/c)$.  To construct the $(110)$-surface BZ [Fig.~\ref{bulk}(a)], we first define $k_{\parallel}$ in the $(\bar 110)$ direction of the conventional cell [\emph{i.e.}, $k_{\parallel}=-k_{x} + k_{y}$], and then form the linear combinations:
\begin{equation}
k_{1,2} = k_{\parallel} \pm k_{z},
\end{equation}
which lie parallel to $a^{*}_{1,2}$.  For simplicity, however, in this work, even though the $(110)$-surface BZ is defined as the square indexed by the primitive-surface-cell momenta $k_{1,2}$, we will still use the more natural conventional-cell momenta $k_{\parallel}=-k_{x} + k_{y}$ and $k_{z}$ to index momentum-space coordinates along the $(110)$-surface.  $k_{\parallel}$ and $k_{z}$ are periodic with respect to the conventional $(110)$-surface BZ [dashed square in Fig.~\ref{bulk}(a)], which is twice as large as the primitive surface BZ [black squares in Fig.~\ref{bulk}(a)], because it contains both the first and second primitive-cell surface BZs.  Restricting to the dashed blue square in Fig.~\ref{bulk}(a), we project the bulk Fermi pockets of the 48 FSWPs listed in Table~\ref{tbwps}.  Because all of the projected pieces of the bulk Fermi surface lie in the vicinity of the $k_{z}=\pm \pi/c$ lines, which are related by linear combinations of the surface reciprocal lattice vectors $a^{*}_{1,2}$, then we are free to focus on the projected Fermi pockets near $k_{z}=\pi/c$ [solid black rectangle in Fig.~\ref{bulk}(a)], which is shown in a larger, rescaled view in Fig.~\ref{bulk}(b).  As discussed in Fig.~3 of the main text, the projected bulk Fermi pockets split into four pairs of islands related by the action of $\mathcal{T}$ symmetry.  In Fig.~\ref{bulk}, we enclose the islands of projected bulk states with dashed blue lines, and denote the surface TRIM points $-a^*_1/2$ and $a^*_2/2$ with $\times$ symbols.  Labeling islands (and their time-reversal partners) with unprimed (primed) symbols, we respectively denote the four islands (and their time-reversal partners), as $\alpha\ (\alpha')$, $\beta\ (\beta')$, $\gamma_1\ (\gamma_1')$, and $\gamma_2\ (\gamma_2')$.

To obtain the net chiral charge of each projected island, we project the bulk FSWPs in Table~\ref{tbwps} onto the $(110)$-surface BZ in Fig.~\ref{bulk}(b) (employing the labeling scheme of the WPs in Fig.~2 of the main text).  In Fig.~\ref{bulk}(c), we show an even closer view of the projected Fermi surface in the vicinity of $k_{\parallel}=0$, $k_{z}=\pi/c$.  Following the labeling employed in Fig.~2 of the main text, the arrows in Fig.~\ref{bulk}(c) point to FSWP projections with net chiral charges $|C|=1$; the symbols in Fig.~\ref{bulk}(c) without arrows indicate two bulk FSWPs of the same chiral charge $C=\pm 1$ that lie at the same values of $k_{\parallel}$ and $k_{z}$, and thus project to the same point in the $(110)$-surface BZ.  Therefore, the symbols without arrows in Fig.~\ref{bulk}(c) indicate bulk projections with net chiral charges $|C|=2$.  As shown in Fig.~\ref{fig:type}, $W_{3}^{-}$ and $W_{4}^{-}$ (upward- and downward-pointing triangles, respectively) are type-I WPs, whereas $W_1^+$ (boxes), $W_2^+$ (circles), and $W_5^-$ (diamonds) are type-II WPs.  To calculate the topological charge of each island, we simply sum the projected Chern numbers of the WPs that project to that island.  From this counting, we determine the net chiral charges of $\alpha$, $\beta$, $\gamma_1$, and $\gamma_2$ to be $C=-4,\ -4,\ +4$, and $+4$, respectively.  Because the operation of $\mathcal{T}$ does not change the sign of the chiral charge of a WP~\cite{Wan2011}, then the time-reversal partners of each island carry the same respective chiral charges.

\subsection*{D. Spin Texture of the $(110)$-Surface Fermi Arcs }

\begin{figure*}[!h]
\centering
\includegraphics[width=0.92\textwidth]{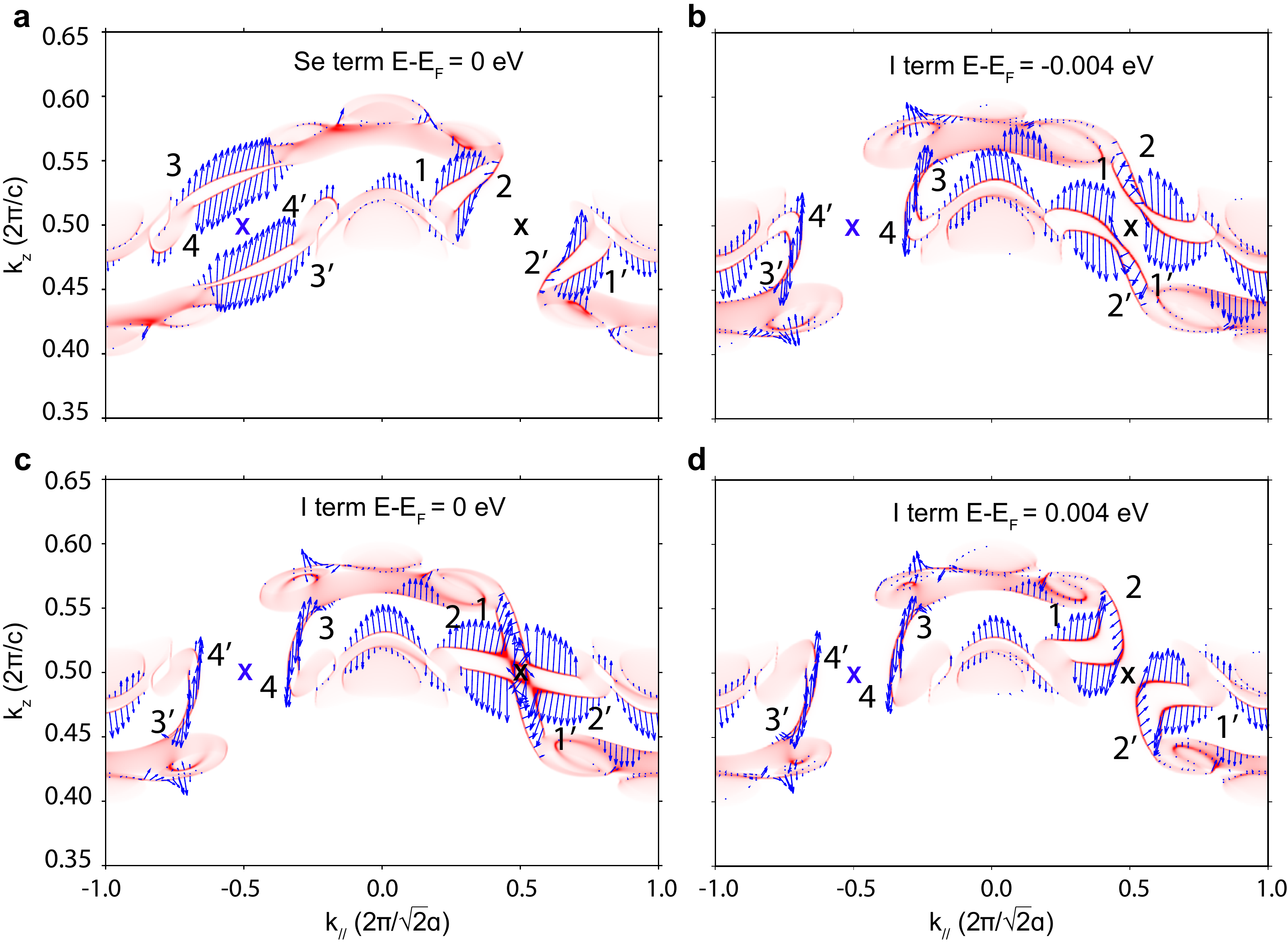}
\caption{(Color online) Spin texture of the conventional-cell $(110)$-surface Fermi arcs of (TaSe$_4$)$_2$I calculated from first principles.  In (a-d), the arrows indicate the magnitude of the in-plane spin expectation values $\left(\langle S^{\parallel} \rangle, \langle S^{z}\rangle\right)$; the out-of-plane component $\langle S^{\perp}\rangle $ of $\langle \vec{S}\rangle$ is vanishingly small.  In all panels, we mark the time-reversal-invariant momenta of the $(110)$-surface with $\times$ symbols.  In (a-d), we label four of the eight surface Fermi arcs within each primitive surface BZ (SI~C) $1-4$, and label their time-reversal partners $1'-4'$.  As expected, the spin textures of Fermi arcs related by $\mathcal{T}$ (\emph{e.g.} $3$ and $3'$) are reversed.  (a) Spin texture of the TaSe$_4$-chain termination at $E=E_{F}$.  (b-d) Spin textures of the I-atom termination at $E-E_{F}= -0.004,~0,$ and $0.004$ eV, respectively.  In (c), arcs $1$, $2$, $1'$, and $2'$ merge in a surface Lifshitz transition that represents the critical point between the Fermi-arc connectivity in (b) and the connectivity in (d).}
\label{fig:spintexture}
\end{figure*}

In Fig.~\ref{fig:spintexture}(a-d), we show the spin texture of the $(110)$-surface states of (TaSe$_4$)$_2$I calculated on a slab Wannier-based tight-binding Hamiltonian from first-principles.  Specifically, for each state within the $(110)$-surface Fermi arcs [Fig.~3(a,b) of the main text], we calculate the spin expectation value vector $\langle\vec{S}\rangle = \left(\langle S^{\parallel}\rangle,\langle S^{z} \rangle, \langle S^{\perp}\rangle\right)$, where $S^{\parallel,\perp} = -S^{x}\pm S^{y}$.  Because we find that $\langle S^{\perp} \rangle \ll \langle S^{\parallel}\rangle,\langle S^{z}\rangle$ at all surface $k$ points, then we only show the in-plane components $\left(\langle S^{\parallel} \rangle, \langle S^{z}\rangle\right)$ in Fig.~\ref{fig:spintexture}.  In Fig.~\ref{fig:spintexture}(a), we show the spin texture of the TaSe$_4$-chain-termination surface states at $E-E_{F}=0$, and in Fig.~\ref{fig:spintexture}(b-d), we respectively show the spin texture of the iodine-atom-termination surface states at $E-E_{F}=-0.004, 0$, and $0.004$ eV.  On both terminations, the surface Fermi arcs appear in pairs (\emph{e.g.} $3$ and $4$) with largely reversed in-plane spin textures.  Each Fermi arc (\emph{e.g.} $3$) is additionally related by $\mathcal{T}$ to a partner (\emph{e.g.} $3'$) with an opposite in-plane spin texture (surface TRIM points are indicated with $\times$ symbols in Fig.~\ref{fig:spintexture}).  At the Fermi energy ($E-E_{F}=0$), four of the Fermi arcs on the iodine-atom termination ($1$, $2$, $1'$, and $2'$) intersect in a junction that is representative of a surface Lifshitz critical point [Fig.~\ref{fig:spintexture}(c)] between the Fermi-arc connectivity in Fig.~\ref{fig:spintexture}(b) and the connectivity in Fig.~\ref{fig:spintexture}(d).  Specifically, the projected bulk states and their respective chiral charges are the same in Fig.~\ref{fig:spintexture}(a,c).  Because the bulk Fermi level is independent of the surface termination, then the Lifshitz critical point that appears in (c) and not in (a) is a consequence of the different surface atoms on the TaSe$_{4}$ and I terminations, as opposed to a shift of the Fermi energy.  A similar surface-dependent Lifshitz transition was recently realized in experiment in~\cite{AndreiLifshitz}.  Additionally, as shown in Fig.~3(a,b) of the main text, on the iodine-atom termination, Fermi arcs $3$ and $4$ and their time-reversal partners $3'$ and $4'$ exhibit a different connectivity in Fig.~\ref{fig:spintexture}(b-d) than they do on the TaSe$_4$-chain termination (a); this is also the result of a surface Lifshitz transition, because the chiral charges of the projected bulk states are the same in (a-d).

\subsection*{E. Quasiparticle Interference Patterns on the Conventional-Cell $(110)$-Surface}

In this section, we calculate the quasiparticle interference (QPI) patterns of the Fermi-arc surface states in Fig.~3(a,b) of the main text using the prescription developed in~\cite{Kourtis2016}.  First, because scanning tunneling spectroscopy  (STM) measurements are surface sensitive, and to emphasize topological contributions to the QPI, we set the surface Green's function to zero at each of the $k$ points in which it only contains contributions from the projected bulk states in Fig.~\ref{bulk}; the Green's function at the remaining $k$ points only contains contributions from surface-localized topological Fermi-arc states.  Next, we calculate and compare two different autocorrelators for each possible conventional-cell $(110)$-surface termination:
\begin{eqnarray}
\text{JDOS}(q,E) &=& \int\ \frac{d^{3}k}{(2\pi)^3}\ \sum_\sigma A_{\sigma\sigma}(k,E)\sum_{\sigma'} A_{\sigma'\sigma'}(k+q,E) \nonumber \\
\text{SSP}(q,E) &=& \int\ \frac{d^{3}k}{(2\pi)^3}\ \sum_{\sigma\sigma'}\left[A_{\sigma\sigma'}(k,E)A_{\sigma'\sigma}(k+q,E)\right], \nonumber \\
\label{eq:QPISM}
\end{eqnarray}
where the $\text{JDOS}(q,E)$ and the $\text{SSP}(q,E)$ respectively represent the joint density of states and the spin-dependent scattering probability.  First, we calculate the imaginary part of the surface Green's function $A_{ij\sigma\sigma'}(k,E)$ at each $(110)$-surface $k$ point.  We then trace out the orbital components indexed by $i,j$, leaving $A_{\sigma\sigma'}(k,E)$, which we take to approximate the spin-resolved Fourier-transformed density of states~\cite{Kourtis2016}.  Finally, we calculate the two autocorrelators in Eq.~(\ref{eq:QPISM}), which differ by whether we first matrix multiply $A_{\sigma\sigma'}(k,E)$ and $A_{\sigma'\sigma}(k+q,E)$ and then trace over the result (SSP), or whether we first individually trace over $A_{\sigma\sigma}(k,E)$ and $A_{\sigma'\sigma'}(k+q,E)$ and then multiply the scalar results (JDOS).

\begin{figure*}[!h]
\centering
\includegraphics[width=\textwidth]{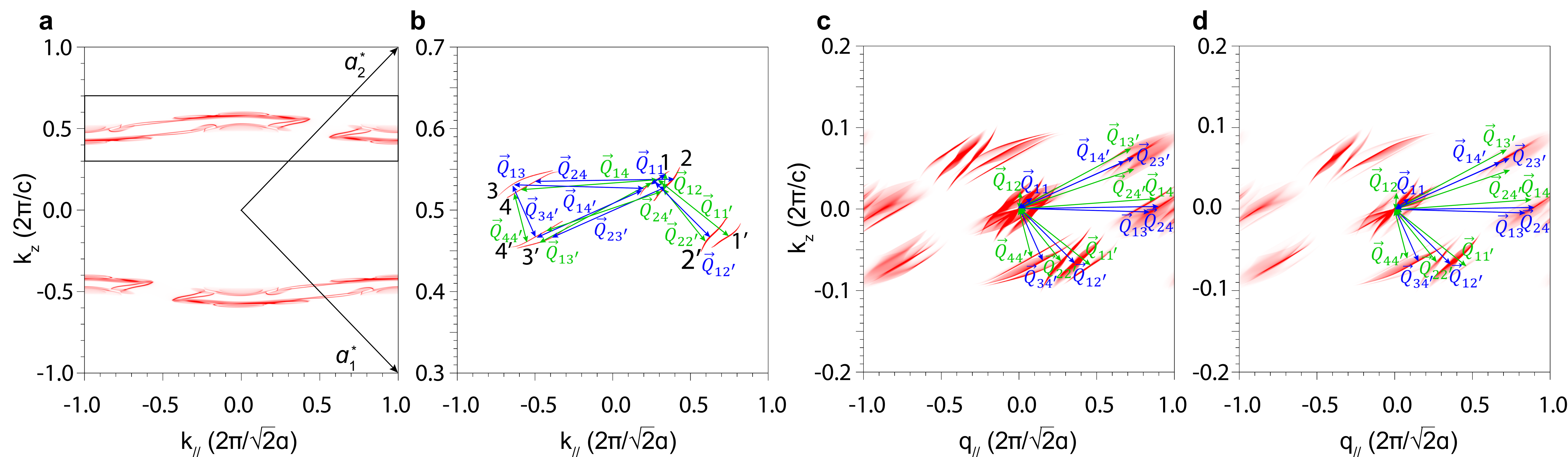}
\caption{(Color online) Quasiparticle interference (QPI) patterns of the surface Fermi arcs of the TaSe$_4$-chain termination of the conventional-cell $(110)$-surface of (TaSe$_4$)$_2$I at $E_{F}$.  (a) The $(110)$-surface states, as well as the projections of bulk states, in the full surface BZ, which is spanned by the surface-BZ reciprocal lattice vectors $a^{*}_{1}$ and $a^{*}_{2}$ (see SI~C for more details).  There are 16 topological Fermi arcs in (a); the 8 Fermi arcs in the black box at $k_{z}=\pi/c$ [\emph{i.e.}, in the region of the surface BZ shown in Fig.~3(a) of the main text] are related to the 8 Fermi arcs outside of the box at $k_{z}=-\pi/c$ by linear combinations of $a^{*}_{1,2}$.  (b) To emphasize the topological contributions to the QPI from the surface Fermi arcs, we filter out the projections of the bulk states and focus on the Fermi arcs in the top half of the conventional surface BZ [black box in (a)].  We then enumerate all of the spin-conserving (blue) and nonconserving (green) scattering vectors between the Fermi arcs for a scalar impurity based on the Fermi-arc spin polarization as calculated from first principles (see SI~D).  Because the Fermi pockets at $k_{z}=\pm\pi/c$ are related by surface reciprocal lattice vectors [$a^{*}_{1,2}$ in (a)], we only label in (b) the independent scattering processes within a single surface BZ; larger scattering vectors between surface states at $k_{z}=\pm\pi/c$ are also permitted, but are related to the vectors in (b) by linear combinations of $a^{*}_{1,2}$.  Next, to numerically characterize scattering between the surface Fermi arcs, we calculate and compare two approximations of the QPI: the joint density of states (JDOS) and the spin-dependent scattering probability~\cite{Kourtis2016}, the results of which are shown in (c,d) respectively, and are labeled using the scattering vectors identified in (b).  Because the Fermi arcs in (b) appear in pairs (\emph{e.g.} $1$ and $2$) with largely opposite spin polarization directions [see Fig.~\ref{fig:spintexture}(a) in SI~D], the JDOS exhibits characteristic pairs of scattering states that alternate between spin-conserving  (\emph{e.g.} $\vec{Q}_{12'}$) and nonconserving (\emph{e.g.} $\vec{Q}_{11'}$) scattering processes.  In the SSP (d), the spin-nonconserving scattering processes within each pair are visibly suppressed (though we still label them with green arrows to emphasize their relative absence in the SSP).  Near $\vec{q}=\vec{0}$, all of the pairs of surface Fermi arcs in (b) also contribute to a central set of three features in the JDOS (c) and one feature in the SSP (d), as highlighted in~\cite{Kourtis2016}.  The centralmost set of scattering states originates from scattering within the same Fermi arc (\emph{e.g.} $\vec{Q}_{11}$) and is spin-conserving; therefore it is visible in both the JDOS and the SSP.  Conversely, the two features to the left and the right of it (\emph{e.g.} $\vec{Q}_{12}$ and its time-reverse) are spin-nonconserving, and are largely absent in the SSP.  To infer the temperature dependence of the QPI pattern, we also calculate the surface Fermi-arc states at increasing effective temperatures, the results of which are shown in Fig.~\ref{fig:temp} in SI~F.}
\label{fig:seQPI}
\end{figure*}

\begin{figure*}[!h]
\centering
\includegraphics[width=0.88\textwidth]{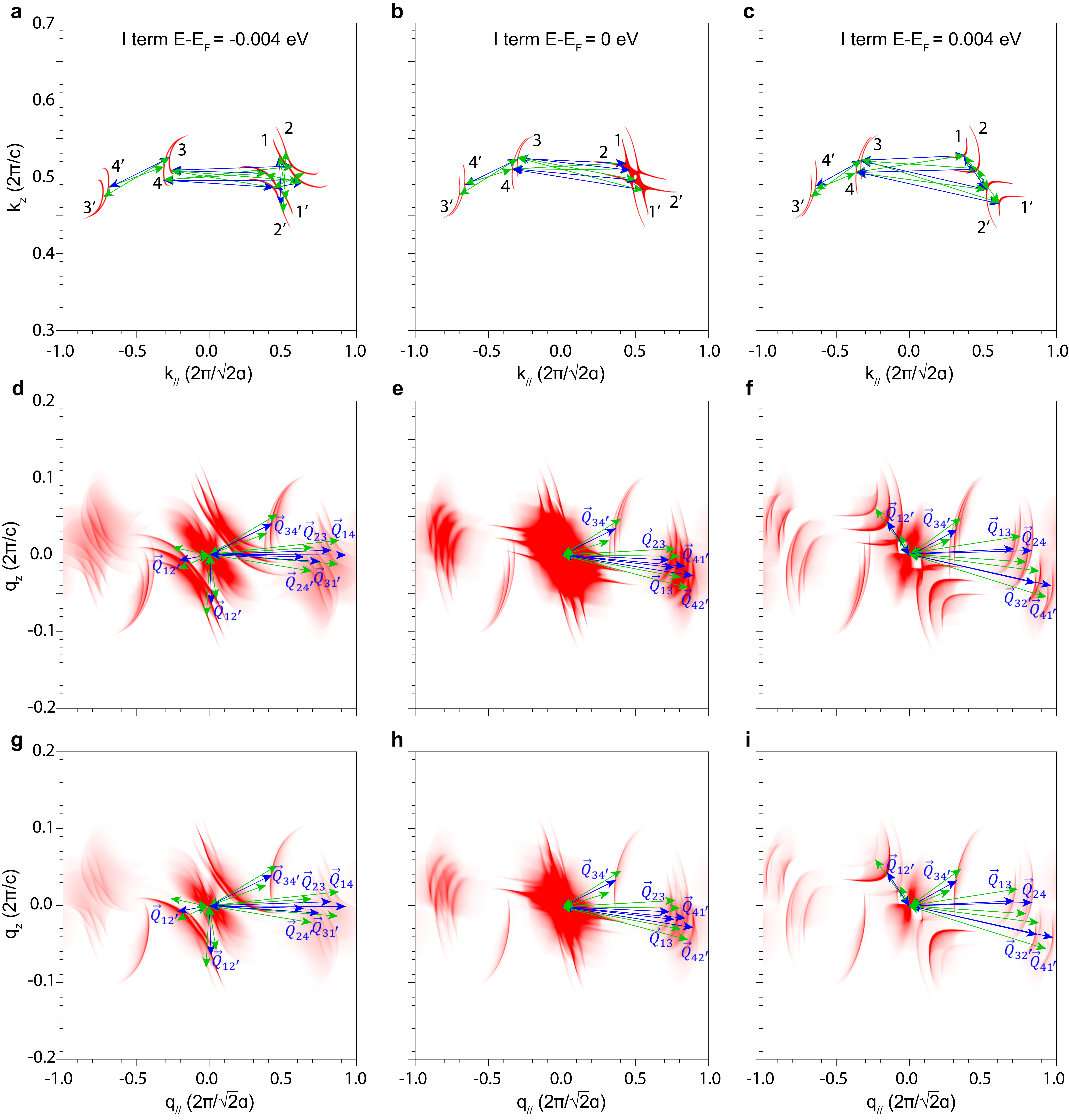}
\caption{(Color online) Quasiparticle interference (QPI) for the topological Fermi arcs on the iodine-atom termination of the conventional-cell $(110)$-surface of (TaSe$_4$)$_2$I at $E-E_{F}= -0.004,~0,~ 0.004$ eV.  In all panels, we label spin-conserving (nonconserving) scattering processes with blue (green) arrows.  (a,b,c) The $(110)$-surface Fermi arcs at $E-E_{F}= -0.004,~0,~ 0.004$ eV, respectively.  As in Fig.~\ref{fig:seQPI}(c,d), we have set the surface Green's function to zero at the $k$ points in which its only contributions come from bulk states.  (a,c) Above and below $E_{F}$, the surface Fermi arcs in appear in pairs (\emph{e.g.} $1$ and $2$) with largely opposite spin polarization directions [Fig.~\ref{fig:spintexture}(b)].   Therefore, the JDOS (d,f) exhibits characteristic pairs of scattering states that alternate between spin-conserving (\emph{e.g.} $\vec{Q}_{12'}$) and nonconserving (\emph{e.g.} $\vec{Q}_{11'}$) scattering processes.  Conversely, in the SSP (g,i), the spin-nonconserving scattering processes within each pair are visibly suppressed (though we again still label them with green arrows to emphasize their relative absence in the SSP).  Near $\vec{q}=\vec{0}$, all of the pairs of surface Fermi arcs in (a,c) also contribute to a central set of three features in the JDOS (d,f) and one feature in the SSP (g,i), as previously highlighted in~\cite{Kourtis2016} and in Fig.~\ref{fig:seQPI}.  Unlike in the TaSe$_4$-chain termination [Fig.~3(a) of the main text], at $E-E_{F}=0$, two of the surface Fermi arcs and their time-reversal partners ($1$, $2$, $1'$, and $2'$) intersect in (b) at a surface Lifshitz critical point (SI~D) between the Fermi-arc connectivity in (a) and the connectivity in (c).  Short-range scattering processes from surface states in the momentum-space vicinity of the Lifshitz point combine to produce a large signal near $\vec{q}=\vec{0}$ in the JDOS (e) and the SSP (h) at $E-E_{F}=0$.  In addition to all of the scattering processes depicted in this figure, like in Fig.~\ref{fig:seQPI}, scattering is also allowed at larger ${\bf q}$ vectors between the surface Fermi-arc states at $k_{z}=\pm \pi/c$ [Fig.~\ref{bulk}(a)].  However, because the eight surface Fermi arcs at $k_{z}=-\pi/c$ are related to the arcs at $k_{z}=\pi/c$ by surface reciprocal lattice vectors [$a^{*}_{1,2}$ in Fig.~\ref{bulk}(a)], then all of the long-range scattering processes between $k_{z}=\pm \pi/c$ can be expressed as the sum of the short-range processes shown here and a linear combination of $a^{*}_{1,2}$.}
\label{fig:iQPI}
\end{figure*}

In Fig.~\ref{fig:seQPI}, we show and analyze the QPI of the TaSe$_4$-chain termination of the $(110)$-surface states of (TaSe$_4$)$_2$I at $E-E_{F}=0$, calculated using the methodology detailed earlier in this section.  For the TaSe$_4$-chain termination at $E-E_{F}=0$, the surface Fermi surface consists of eight topological Fermi-arc states within each surface BZ [\emph{e.g.}, the eight Fermi arcs within the black rectangle in Fig.~\ref{fig:seQPI}(a), which is the same region of the surface BZ shown in Fig.~3(a) of the main text].  The eight Fermi arcs can be divided into four pairs (\emph{i.e.} two pairs and their time-reversal partners) with largely reversed in-plane spin textures (\emph{e.g.} Fermi arcs $3$ and $4$ have nearly opposite spin textures, as do their time-reversal partners, $3'$ and $4'$) [SI~D].  Because of the action of $\mathcal{T}$ symmetry, $\mathcal{T}$-related arcs (\emph{e.g.} $1$ and $1'$) also exhibit opposite spin polarization directions [see Fig.~\ref{fig:spintexture}(a) in SI~D for more details].  Calculating the QPI pattern of the surface Fermi arcs in Fig.~\ref{fig:seQPI}(b) for a scalar impurity [Fig.~\ref{fig:seQPI}(c)], we observe pairs of arc-like scattering states in the JDOS (\emph{e.g.} $\vec{Q}_{12'}$ and $\vec{Q}_{11'}$) formed from one set of spin-conserving (blue, \emph{e.g.} $\vec{Q}_{12'}$) and one set of spin-nonconserving (green, \emph{e.g.} $\vec{Q}_{11'}$) scattering processes.  In the SSP [Fig.~\ref{fig:seQPI}(d)], the spin nonconserving processes are visibly suppressed (though we still label them with green arrows to emphasize their relative absence).

We next analyze the QPI of the alternative, iodine-atom termination $(110)$-surface states.  In Fig.~\ref{fig:iQPI}(a-c), we show the JDOS and SSP [Eq.~(\ref{eq:QPISM})] of only the I-atom-termination $(110)$-surface topological Fermi arcs at $E-E_{F}=-0.004,~0,~0.004$ eV.  As with the TaSe$_4$-chain termination, the I-atom-termination surface Fermi arcs appear in pairs (\emph{e.g.} $1$ and $2$) with opposite spin textures [Fig.~\ref{fig:spintexture}(b-d)].  Therefore at all energies, scattering states appear in spin-conserving (blue, \emph{e.g.} $\vec{Q}_{12'}$) and nonconserving (green, \emph{e.g.} $\vec{Q}_{11'}$) pairs (Fig.~\ref{fig:iQPI}).  The spin-conserving scattering processes are present in both the JDOS and the SSP, whereas the spin-nonconserving processes are largely suppressed in the SSP (though we again still label them with green arrows to emphasize their relative absence).  Additionally, at energies above and below $E_{F}$ near $\vec{q}=\vec{0}$, all of the pairs of surface Fermi arcs contribute to a central set of three features in the JDOS (c) and one feature in the SSP (d), as highlighted in~\cite{Kourtis2016} and discussed in Fig.~\ref{fig:seQPI}.  Unlike with the TaSe$_4$-chain termination, at the Fermi energy, two of the I-atom surface Fermi arcs and their time-reversal partners ($1$, $2$, $1'$, and $2'$) merge in a surface Lifshitz transition [Fig.~\ref{fig:iQPI}(b) and SI~D] that represents the critical point separating the Fermi-arc connectivity in Fig.~\ref{fig:iQPI}(a) from the connectivity in Fig.~\ref{fig:iQPI}(c).  At $E=E_{F}$, short-range scattering processes from the four arcs at the Lifshitz point overwhelm both the JDOS and SSP at small $\vec{q}$ [Fig.~\ref{fig:iQPI}(e,h), respectively].


\clearpage
\subsection*{F. Conventional-Cell $(110)$-Surface States of the TaSe$_4$-Chain Termination at Different Temperatures}

\begin{figure*}[!h]
\centering
\includegraphics[width=0.8\textwidth]{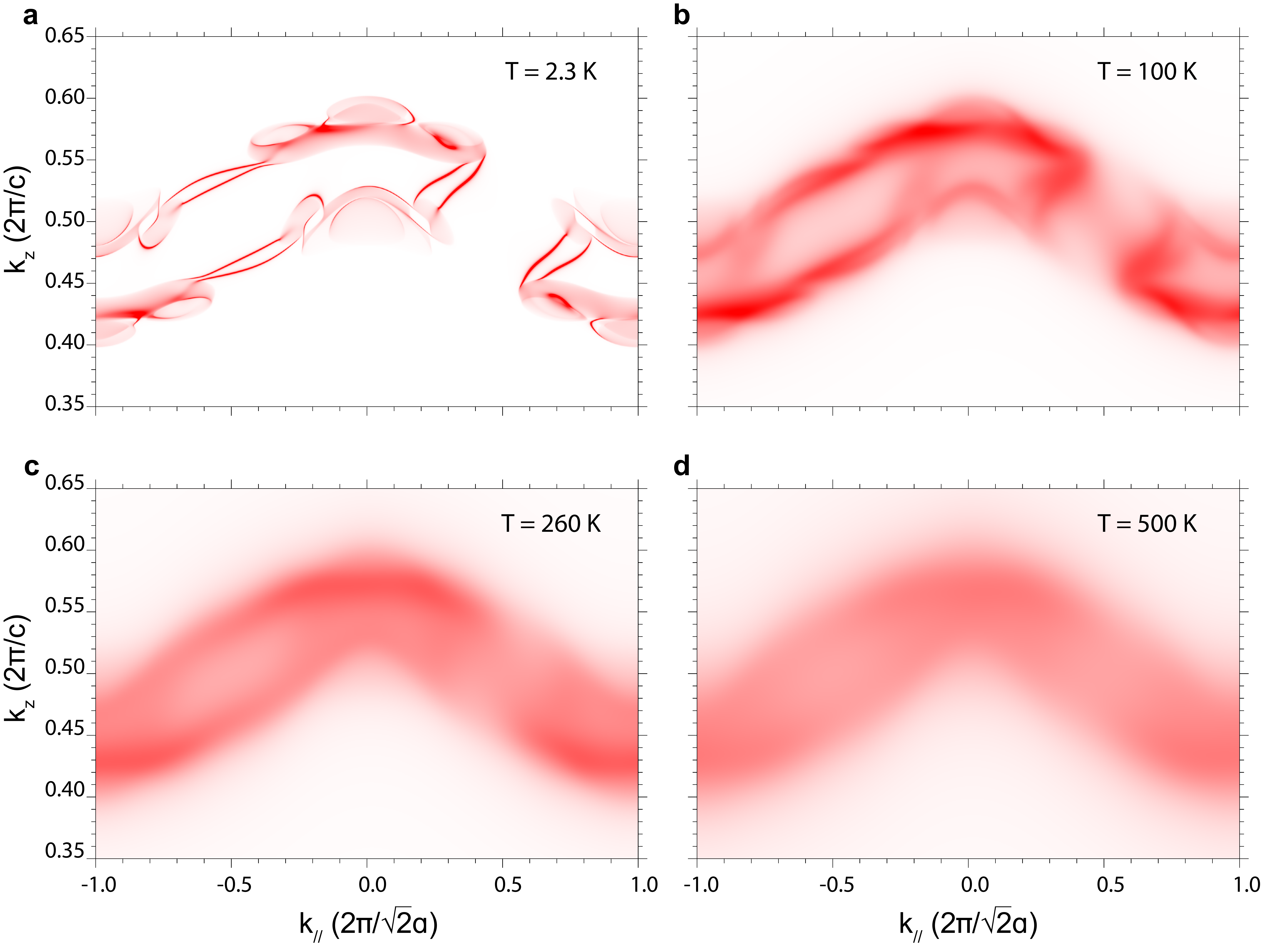}
\caption{(Color online) (a-d) The conventional-cell $(110)$-surface states of the TaSe$_4$-chain termination of (TaSe$_4$)$_2$I at $E=E_{F}$, calculated at the effective temperatures $T=2.3$~K, $100$~K, $260$~K, and $500$~K, respectively.  (a,b) While the surface Fermi arcs at $T = 2.3$~K and $T=100$~K are clear and distinct, they were calculated at effective temperatures below $T_{C}$ for the CDW phase, which is known to occur between $248$~K and $260$~K (Refs.~\cite{Cava1986,260k} and SI~H).  (c,d) However, even above $T_{C}$, the surface Fermi arcs are still visible as pieces of a continuous surface Fermi surface, though they quickly become difficult to resolve at room temperature and above.}
\label{fig:temp}
\end{figure*}

To show the temperature dependence of the topological surface Fermi arcs in (TaSe$_4$)$_2$I, we additionally calculated the conventional-cell $(110)$-surface states of the TaSe$_4$-chain termination at increasing effective temperatures by increasing the state-smearing parameter $\epsilon$ in the surface Green's function $A_{ij\sigma\sigma'}(k,E)=1/[E-H_{ij\sigma\sigma'}(k) + i\epsilon]$, where $\epsilon= k_{B}T$.  In Fig.~\ref{fig:temp}(a-d), we respectively plot the $(110)$-surface states at $\epsilon = 0.0001978$, $0.0086$, $0.02236$, and $0.043$, which respectively correspond to effective temperatures of $T = 2.3$~K, $100$~K, $260$~K, and $500$~K.  Above $T_{C}$ for the CDW phase, which lies between $248$~K and $260$~K (Refs.~\cite{Cava1986,260k} and SI~H), the surface Fermi arcs become increasingly difficult to resolve [Fig.~\ref{fig:temp}(c,d)].  However, even at high temperatures, the presence of surface Fermi arcs can still be inferred by observing that the surface Fermi surface is continuous, whereas the surface projections of the bulk Fermi surface form disconnected islands (SI~C).

\subsection*{G. Electronic Susceptibility}

In some cases, a charge-density wave (CDW) cannot be explained using the simple picture of Fermi-surface nesting vectors~\cite{ElectronicSusceptibility,CDWmisconceptions,CDWcomplication1,CDWcomplication2}.  Therefore, still focusing on the electronic contribution to the CDW instability (as opposed to the phonon contribution), we will perform the more accurate calculation of the electronic susceptibility.  Specifically, when the real part of the electronic susceptibility $\chi_\bq$ diverges, there is a strong electronic contribution towards a CDW transition~\cite{ElectronicSusceptibility}.  In the constant-matrix approximation:
\begin{eqnarray}
\chi_\bq=\sum_{\bk}[n_F(\epsilon_\bk)-n_F(\epsilon_{\bk+\bq})]/(\epsilon_{\bk}-\epsilon_{\bk+\bq}),
\label{eq:SMchi}
\end{eqnarray}
where $n_F(\epsilon)=1/[\text {exp}(\epsilon/k_{\text B}T)+1]$ is the Fermi-Dirac distribution function.  Using a Wannier-based tight-binding Hamiltonian, we compute $\chi_\bq$ in the full 3D BZ of (TaSe$_4$)$_2$I, which we plot in Fig.~5(b-d) of the main text.

\begin{table}[!h]
  \caption{Symmetry-equivalent sets of peaks in the electronic susceptibility ($\chi_\bq$) not shown in Fig.~5 and Table~I of the main text.  In this table, we respectively list the index ($\bar{\bf q}_{i}$) of one ${\bf q}$ vector within each symmetry-equivalent set of peaks in $\chi_{\bq}$, the coordinates of $\bar{\bq}_{i}$ in the 3D conventional scattering BZ, the closest integer multiple of the experimentally-observed CDW modulation vectors $[m\eta(\frac{2\pi}{a}),n\eta(\frac{2\pi}{a}),o\delta(\frac{2\pi}{c})]$ (SI~H.1), and the value of $\text{Re}(\chi_{\tilde{\bq_{i}}})$ in relative units.  In this work, we define a peak in $\chi_\bq$ to be an isolated spot with $\text{Re}(\chi_{\bq})>104000$ in the relative units of this table and Table~I of the main text.  This table includes the vector $\tilde{\bf q}_{2}^{G}$, which notably coincides with the high-intensity ${\bf q}=(112)$ satellite reflection observed in our XRD experiments (see SI~H.1).  Though the vectors listed in this table do not directly couple FSWPs, they match integer multiples of the much shorter, experimentally-observed CDW modulation vectors, which, as shown in SI~A, conversely do backfold and couple the bulk FSWPs.}
  \begin{tabular}{ccccc}
  \hline
  \hline
  $\bar{\bf q}_i$    & Coordinates   & $(mno)$    & $\text{Re}(\chi_{\bar{\bq}_{i}})$\\
                 &($q_{x}\frac{2\pi}{a},q_{y}\frac{2\pi}{a},q_{z}\frac{2\pi}{c}$) &\\
  \hline
  $\bar{\bf q}_1$    &  (0.0000,0.7760,0.0150)   & (0,29,1)  & 124739 \\
  $\bar{\bf q}_2$    &  (0.0000,0.2635,0.0240)   & (0,10,2)  & 164707 \\
  $\bar{\bf q}_3$    &  (0.1030,0.1590,0.0240)   & (4,6,2)   & 128628 \\
  $\bar{\bf q}_4$    &  (0.0606,-0.0606,0.0240)  & (2,-2,2)  & 167946 \\
  $\bar{\bf q}_5$    &  (-0.0606,0.0000,0.0240)  & (-2,0,2)  & 143021 \\
  $\tilde{\bf q}_2^G$  &  (0.0270,0.0270,0.0240)   & (1,1,2)   & 135097 \\
\hline
\hline
  \end{tabular}
\label{suppChipts}
\end{table}

\begin{figure*}[!h]
\centering
\includegraphics[width=0.8\textwidth]{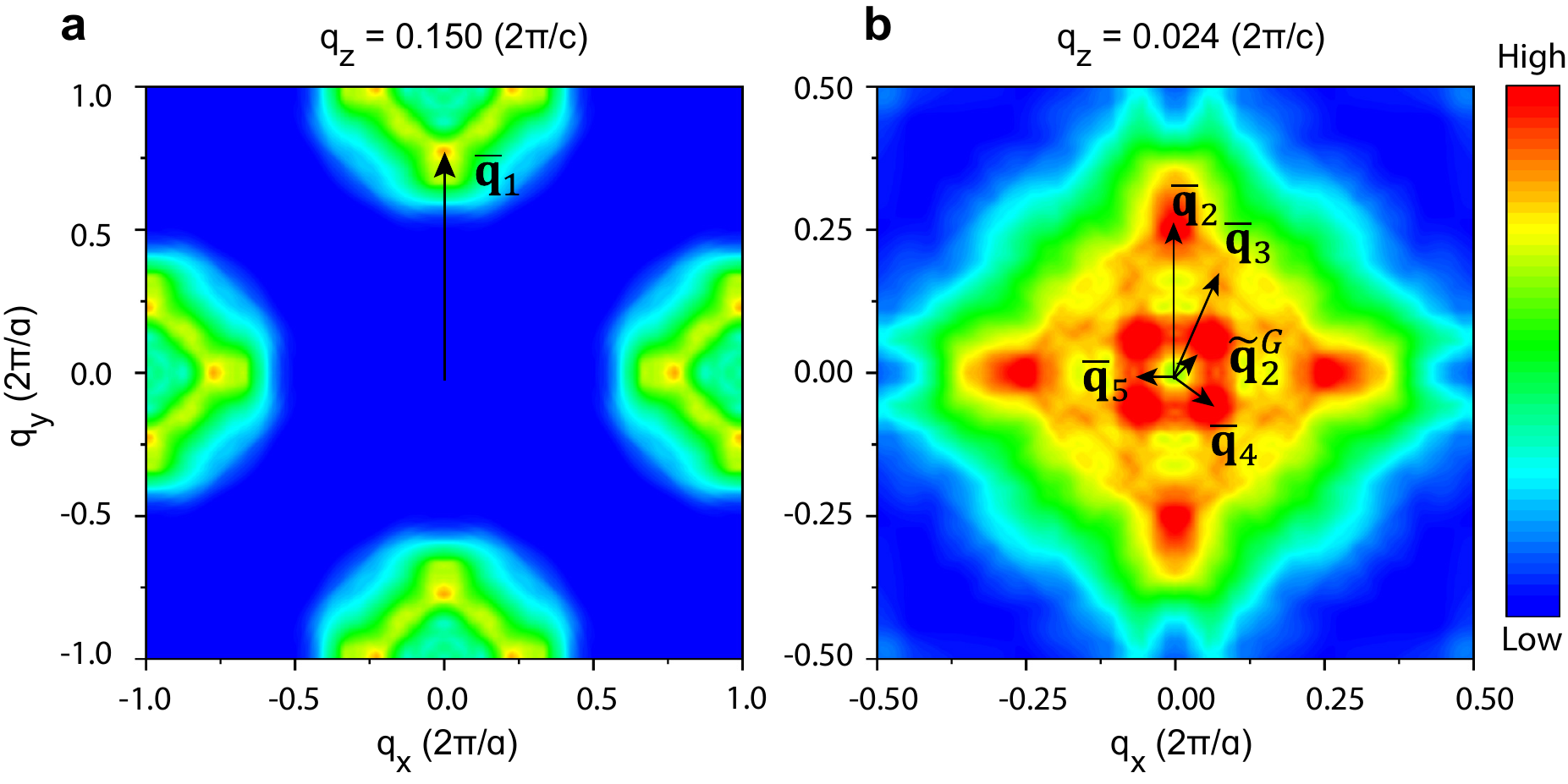}
\caption{(Color online) The electronic susceptibility peaks not shown in Fig.~5 and Table~I of the main text.  The vector $\tilde{\bf q}_{2}^{G}$ notably coincides with the high-intensity ${\bf q}=(112)$ satellite reflection observed in our XRD experiments (see SI~H.1).  Unlike the electronic susceptibility vectors in Fig.~5 and Table~I of the main text, the $\bar{\bf q}$ vectors labeled in this figure (listed in Table~\ref{suppChipts}) instead match nesting vectors between bulk Fermi pockets that are farther away in momentum (${\bf k}$) from the nodal centers of the FSWPs.  Nevertheless, the $\bar{\bf q}$ vectors shown in this figure also lie at integer multiples of the much smaller, experimentally observed CDW modulation vectors, which we have shown in SI~A to nest and couple the bulk FSWPs.  Therefore, the labeled vectors $\bar{\bf q}_{i}$ in this figure remain consistent with our conclusion that (TaSe$_4$)$_2$I is a Weyl semimetal whose WPs become linked and gapped by a CDW.}
\label{fig:suppChi}
\end{figure*}

We find that many -- but not all -- of the observed peaks in the electronic susceptibility can be attributed to nesting vectors between the nodal centers of the bulk FSWPs (\emph{i.e.} Bloch states that lie in a close range in ${\bf k}$-space to the FSWP band degeneracy, see SI~A).  In Table~I of the main text, we quantitatively detail the strong spots in $\chi_\bq$ that correspond to nesting vectors between bulk FSWPs, which we define to be the isolated spots in $\chi_\bq$ with $\text{Re}(\chi_{\bq})>104000$ in the relative units employed in this work.  In Fig.~\ref{fig:suppChi} and Table~I, we additionally detail the remaining peaks in $\chi_\bq$ that are not listed in Table~I of the main text.  We find that all of the strong peaks in $\chi_\bq$ (Table~I of the main text and Table~\ref{suppChipts}) lie at ${\bf q}$ vectors that closely match integer multiples of the experimentally observed CDW modulation vectors (SI~H.1), providing further support for our conclusion that (TaSe$_4$)$_2$I is a Weyl semimetal whose WPs become linked and gapped by a CDW.

However, as shown in Fig.~\ref{fig:suppChi} and Table~\ref{suppChipts}, the strong peaks in the electronic susceptibility that do not originate from FSWP nesting are comparable in magnitude to the peaks that do originate from FSWP coupling [see Table~I of the main text].  Furthermore, the CDW modulation vectors observed in our XRD experiments [\emph{e.g.} $\tilde{\bf q}_{2}^{G}$ in Fig.~\ref{fig:suppChi}(b), see SI~H.1] are much shorter than the ${\bf q}$ vectors of the electronic susceptibility peaks away the large spot at ${\bf q}=0$ [see Figs.~\ref{fig:suppChi} and~5 of the main text and Tables~\ref{suppChipts} and~I of the main text].  This suggests that Fermi-surface nesting is not itself the origin of the CDW in (TaSe$_4$)$_2$I, providing a further evidence for the recognition in~\cite{ElectronicSusceptibility} that 3D CDWs rarely originate from electronic instabilities.

\subsection*{H. Experimental Data}
\label{app-exp-cdw}

We also performed X-ray diffraction (XRD) and angle-resolved photoemission spectroscopy (ARPES) experiments to study the CDW modulation vectors and the electronic band structure of (TaSe$_{4}$)$_2$I, respectively.  In SI~H.1 and SI~H.2, we respectively provide details of our XRD and ARPES investigations.

\subsubsection*{H.1 X-Ray Diffraction Experiments}

In order to determine the direction and magnitude of the CDW modulation vectors, and to confirm the CDW transition temperature $T_{C}$ of (TaSe$_{4}$)$_2$I, we performed X-ray diffraction (XRD) experiments. In this section, we detail the methods and results of our XRD experiments, which are additionally summarized in Fig.~4 of the main text.  Our XRD experiments were performed  at beamline 25B of the European Synchrotron Radiation Facility in Grenoble, France using a six-circle diffractometer and a wavelength of $\lambda$=0.71~\AA.  The whisker-shaped (TaSe$_{4}$)$_2$I sample -- which was $\sim$100 $\mu$m in diameter and grown using the method detailed in~\cite{AxionCDWExperiment} -- was first mounted on a copper holder, oriented with its $c$-axis perpendicular to the incoming beam, and cooled using a flow of liquid nitrogen to reach a minimum temperature of 88~K.  Throughout this section, distances in reciprocal space are given with respect to the previously established conventional-cell lattice constants $a=b=9.59$~\AA\ and $c=12.64$~\AA\ of the high-temperature phase of (TaSe$_{4}$)$_2$I~\cite{Ta2Se8IPrepare,gressier1984characterization,gressier1984electronic}.

As the temperature of the sample was lowered, we searched for satellite reflections as evidence of the onset of a CDW phase. Specifically, when a crystal with the lattice constants $a,b,c$ is periodically modulated, as occurs in a CDW phase, then XRD probes begin to exhibit satellite Bragg reflections at the momentum-space locations ${\bf Q}={\bf G} + {\bf q}$, where ${\bf G} = h{\bf a}^{*} + k{\bf b}^{*} + l{\bf c}^{*}$ are the larger reciprocal lattice vectors of the smaller unit cell of the unmodulated (high-temperature) structure, and ${\bf q} = m{\bm \eta}_{1} + n{\bm \eta}_{2} + o{\bm \delta}$ are the smaller modulation vectors of the (typically incommensurate) CDW-modulated structure.  In ${\bf G}$ (${\bf q}$), the conventional reciprocal-space lattice (modulation) vectors along $k_{x}$ ($q_{x}$), $k_{y}$ ($q_{y}$), and $k_{z}$ ($q_{z}$) are respectively given by ${\bf a}^{*}$ (${\bm \eta_{1}}$), ${\bf b}^{*}$ (${\bm \eta_{2}}$), and ${\bf c}^{*}$ (${\bm \delta}$).  The main purpose of our XRD investigations was to experimentally obtain the values of ${\bm \eta}_{1,2}$ and ${\bm \delta}$; we additionally used the appearance of satellite reflections to obtain an estimate of $T_{C}$ in our (TaSe$_{4}$)$_2$I sample, as detailed below.

\begin{figure*}[!h]
\centering
\includegraphics[width=0.7\textwidth]{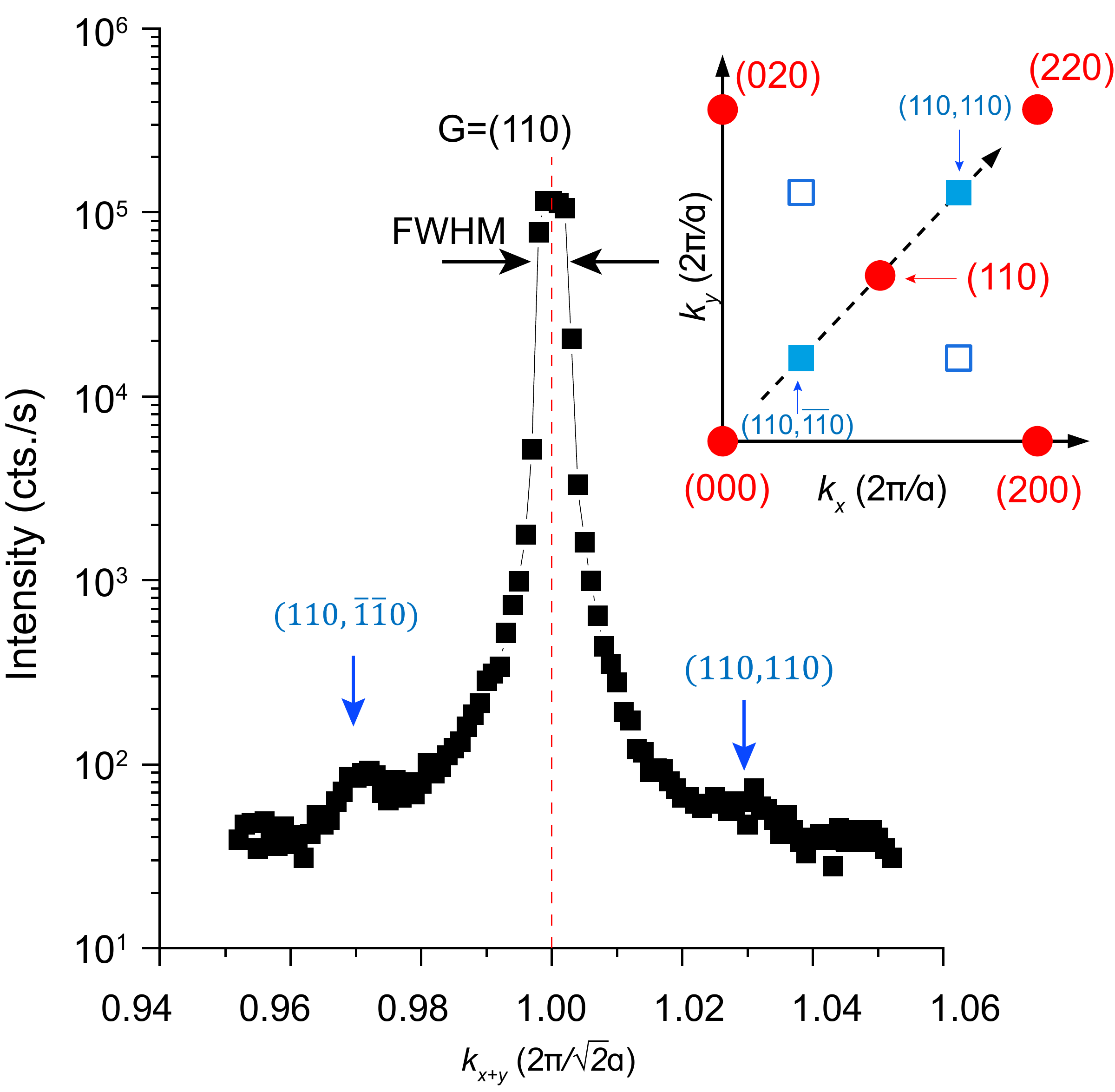}
\caption{(Color online) XRD line scan along the $k_{x+y}$ direction across the ${\bf G}=(hkl)=(110)$ main reflection of our (TaSe$_{4}$)$_2$I sample, plotted on a logarithmic intensity scale and collected within a temperature range of roughly $88$~K to $100$~K, which is well below the reported CDW transition temperature of $260$~K~\cite{Cava1986,260k}.  We use blue arrows to indicate the locations of the ${\bf q} = (mno)= (110)$ and $(\bar{1}\bar{1}0)$ satellite reflections, which appear at $k_{x+y} = \eta\sqrt{2}(\frac{2\pi}{a})$, where $\eta = 0.027\pm 0.001$.  The full-width half-maximum (FWHM) of the intensity peak of the ${\bf G}=(110)$ main reflection is labeled with black arrows; in the line scan shown in this figure, the FWHM lies close to the maximum, because intensity is plotted on a logarithmic scale.  We emphasize that the first-order satellites ${\bf q}=(110)$ and $(\bar{1}\bar{1}0)$ are extremely faint; they appear at an intensity that is about four orders of magnitude weaker than the main reflection at ${\bf G} = (110)$.  In previous investigations of the CDW phase in (TaSe$_{4}$)$_2$I, the ${\bf q}=(110)$ and $(\bar{1}\bar{1}0)$ satellite reflections were not reported at all~\cite{Fu1984qvec,Lee1985qvec}, most likely due to the limited resolution and laboratory X-ray beam intensities available at the time.  In the inset panel, we show a schematic depicting the relative locations in the $k_{x,y}$-plane of the ${\bf G}=(110)$ main reflection (central red circle) and the four nearest satellite reflections (open and filled blue squares) at $k_{z}=0$ ($l=o=0$) [distances in the inset panel are not drawn to scale].  As will be discussed in more detail in the text surrounding Eqs.~(\ref{eq:majorityLatVecCDW}) and~(\ref{eq:majorityLatVecCDW2}), we find that our (TaSe$_{4}$)$_2$I sample contains two macroscopic domains in position space with distinct, $C_{4z}$-symmetry-related CDW orders that individually break $C_{4z}$ symmetry.  Because the satellite reflection intensity peaks shown in this figure [${\bf q} = (110)$ and $(\bar{1}\bar{1}0)$, filled squares in the inset panel] exhibit the same intensity and are related by $C_{2z}$ symmetry, then we conclude that $C_{2z}$ symmetry is preserved in the CDW order within each domain [see the text surrounding Eqs.~(\ref{eq:majorityLatVecCDW}) and~(\ref{eq:majorityLatVecCDW2})].  We additionally depict two $C_{4z}$-symmetry related satellites [open squares in the inset panel] with different intensities than the satellites at ${\bf q} = (110)$ and $(\bar{1}\bar{1}0)$; the reflections depicted with open squares in the inset panel correspond to the first-order, $o=0$ satellites from a different domain than the ${\bf q} = (110)$ and $(\bar{1}\bar{1}0)$ reflections.}
\label{S1HLM}
\end{figure*}

\begin{figure*}[!h]
\centering
\includegraphics[width=0.7\textwidth]{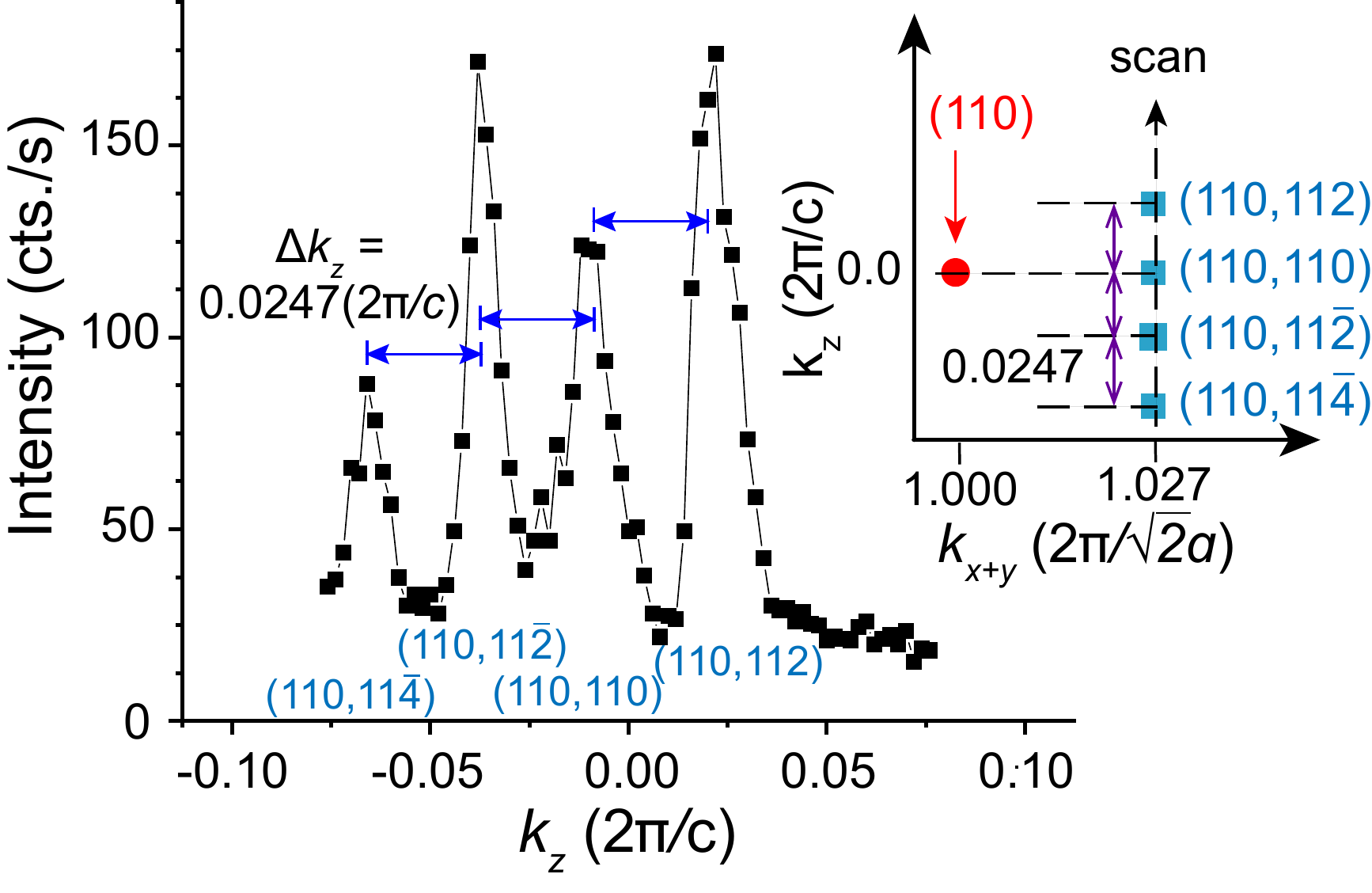}
\caption{(Color online) XRD line scan along the $k_{z}$ direction through the ${\bf Q}=(hkl,mno)=(110,110)$ satellite reflection in Fig.~\ref{S1HLM}, plotted on a linear intensity scale and collected within a temperature range of roughly $88$~K to $100$~K, which is well below the reported CDW transition temperature of $260$~K~\cite{Cava1986,260k}.  The satellite peaks along this scan exhibit a nearly regular spacing of $[0.0247\pm 0.0020](\frac{2\pi}{c})$ in the $q_{z}$ direction.  Anticipating that the CDW-modulated structure, like the unmodulated parent structure of (TaSe$_{4}$)$_2$I, may be body-centered~\cite{korekawa,Korekawa1967,Popescu2003}, we interpret the satellite peaks observed in this scan as lying at a separation of $2\delta$ [see the inset panel for a schematic depicting the relative locations (not drawn to scale)].  The ${\bf q}=(mno)=(112)$ and ${\bf q}=(11\bar{2})$ peaks exhibit the same intensity, implying that they belong to the same CDW domain, and that the CDW order within the domain of the ${\bf q}=(112)$ and ${\bf q}=(11\bar{2})$ satellite reflections has at least $C_{2,x+y}$ symmetry.  Along with the $C_{2z}$ symmetry of the CDW order within each sample domain implied by the data in Fig.~\ref{S1HLM}, this implies that the CDW order within each sample domain is symmetric under the action of point group $D_{2}$ ($222$) in a setting with $C_{2,x\pm y}$ and $C_{2z}$ symmetry [or the $C_{4z}$-symmetric supergroup $D_{4}$ ($422$), though we will shortly exclude this possibility using the data in Fig.~\ref{cdwdata}]~\cite{BCS1,BCS2,BigBook}.  Along with the line scan data presented in Fig.~\ref{S1HLM}, this implies that the satellite reflection modulation vectors ${\bf q}$ in our (TaSe$_{4}$)$_2$I sample (but not the satellite intensities), lie in a tetragonal arrangement specified by Eqs.~(\ref{eq:BCTLatVecCDW}) and~(\ref{eq:BCTLatVecCDW2}).}
\label{S2HLM}
\end{figure*}

We identified the locations of the satellite reflections by performing systematic XRD scans along several directions in $k$ space, employing a two-dimensional (2D) pixel detector to collect the data from both 1D line scans and 2D reciprocal-space intensity maps (RSMs) near the ${\bf G} = (110)$, $(420)$, $(620)$, and $(554)$ Bragg (main) reflections.  In our experiments, we placed the 2D pixel detector 1250 mm away from the (TaSe$_{4}$)$_2$I sample.  We chose to focus on the ${\bf G} = (110)$, $(420)$, $(620)$, and $(554)$ Bragg reflections because the $(110)$, $(420)$, and $(620)$ reflections were the simplest to measure in our $c$-axis directed, whisker-shaped sample, and because satellite reflections near the $(554)$ reflection were previously studied in~\cite{Lee1985qvec}.  As detailed in the text below, we have specifically measured the intensities and coordinates of several satellite reflections with the same values of $(mno)$ near main (Bragg) reflections with different values of $(hkl)$, which provide independent verification of the CDW modulation vectors and symmetry.  We emphasize that all of the line scans and RSMs used in this work to infer the direction and magnitude of the modulation vectors were performed within a temperature range of roughly $88$~K to $100$~K, which is well below the reported CDW transition temperature of $260$~K~\cite{Cava1986,260k}. In Fig.~\ref{S1HLM}, we show the results of an XRD intensity line scan along the $k_{x+y}$ direction at $k_{z}=0$ near the ${\bf G} = (110)$ main Bragg reflection. The intense peak at $k_{x+y}=\sqrt{2}(\frac{2\pi}{a})$ (${\bf Q} = {\bf G}$) reflects the average structure, which coincides with the unmodulated (high-temperature) crystal structure of (TaSe$_{4}$)$_2$I.  We observe two weak satellite peaks at ${\bf q}=[\pm\eta (\frac{2\pi}{a}), \pm\eta (\frac{2\pi}{a}), 0]$, where $\eta = 0.027\pm 0.001$.

Because, as we will show, across \emph{all} measured main reflections, we did not observe any satellite reflections closer to their main reflections than ${\bf q}=(mno)=(110)$ is to ${\bf G}=(110)$, then we label the two ${\bf q}$ vectors in Fig.~\ref{S1HLM} with the full indices ${\bf Q} = (hkl,mno) = (110,110)$ and $(110,\bar{1}\bar{1}0)$, respectively.  As we will shortly see, the data collected in our XRD line scans and RSMs imply that our (TaSe$_{4}$)$_2$I sample hosts two macroscopically large domains in position space with unequal size, where, within each domain, the CDW order breaks $C_{4z}$ symmetry, but preserves $C_{2,x+y}$, $C_{2,x-y}$, and $C_{2z}$.  However, we observe a pattern of satellite reflections that implies that, if the domains were of equal size, the CDW order would preserve $C_{4z}$ when averaged across the sample.  For these reasons, we will see that the modulation vectors ${\bf q}$ of the satellite reflections in our (TaSe$_{4}$)$_2$I sample, but not the satellite intensities (which differ depending on the volume of the domain from which they originate), are periodic with respect to the tetragonal reciprocal basis vectors:
\begin{equation}
{\bf q}^{G}_{1} = \left[0,\eta\left(\frac{2\pi}{a}\right),\delta\left(\frac{2\pi}{c}\right)\right],\ {\bf q}^{G}_{2} = \left[\eta\left(\frac{2\pi}{a}\right),0,\delta\left(\frac{2\pi}{c}\right)\right],\ {\bf q}^{G}_{3} = \left[\eta\left(\frac{2\pi}{a}\right),\eta\left(\frac{2\pi}{a}\right),0\right],
\label{eq:BCTLatVecCDW}
\end{equation}
such that satellite reflections only appear at values of ${\bf q}$ that satisfy:
\begin{equation}
{\bf q} = \left[m\eta\left(\frac{2\pi}{a}\right),n\eta\left(\frac{2\pi}{a}\right),o\delta\left(\frac{2\pi}{c}\right)\right],\text{ where }m+n+o\in 2\mathbb{Z},
\label{eq:BCTLatVecCDW2}
\end{equation}
implying that the domain-averaged CDW order is body-centered tetragonal (though we will shortly see that the CDW order within each domain does not have $C_{4z}$ symmetry).

This is consistent with the recognition, discussed by Korekawa \emph{et al.}~\cite{korekawa,Korekawa1967}, that CDW-modulated structures, when averaged across a sample, frequently preserve the point group symmetries and Bravais lattice centering (\emph{e.g.} $P$, $I$, $F$), but not the same lattice spacing, as the high-temperature crystal structure.  This effect has been observed in both natural crystals, such as feldspars, and in ultra-thin films~\cite{Popescu2003}.  In Fig.~\ref{S1HLM}, we more generally only observe satellite reflections for which $m+ n \in 2\mathbb{Z}$, which is consistent with Eqs.~(\ref{eq:BCTLatVecCDW}) and~(\ref{eq:BCTLatVecCDW2}) [we have not yet introduced satellite spots with nonzero $o$, though in a body-centered CDW phase (or set of CDW domains that average to body-centered tetragonal order), they will satisfy the more general relation $m+n+o \in 2\mathbb{Z}$, \emph{i.e.}, Eq.~(\ref{eq:BCTLatVecCDW2})].

The locations of the satellites in the ${\bf q}_{xy}$ plane in Fig.~\ref{S1HLM} also provide a new perspective on previous XRD probes of the CDW phase of (TaSe$_{4}$)$_2$I.  In earlier works~\cite{Fu1984qvec,Lee1985qvec}, satellite reflections were observed at ${\bf q}_{xy}=[\pm\eta (\frac{2\pi}{a}), \pm\eta (\frac{2\pi}{a})]$, where $\eta\approx0.05$ and  ${\bf q} = {\bf q}_{xy} + q_{z}{\bm \delta}$; specifically, the authors of those works \emph{did not} observe the faint $(mn)=(11)$ and $(\bar{1}\bar{1})$ satellite reflections that we observe in Fig.~\ref{S1HLM}.  We therefore now recognize that the satellite reflections with the smallest in-plane ${\bf q}$ vectors previously observed in~\cite{Fu1984qvec,Lee1985qvec} were in fact \emph{higher-order} satellites with ${\bf q}_{xy} = (mn) = (22)$ and $(\bar{2}\bar{2})$, and that the earlier experiments in those works most likely did not observe the $(mn)=(11)$ and $(\bar{1}\bar{1})$ satellite reflections due to the limited detector resolution and laboratory X-ray beam intensities available at the time.

\begin{figure*}[!t]
\centering
\includegraphics[width=0.9\textwidth]{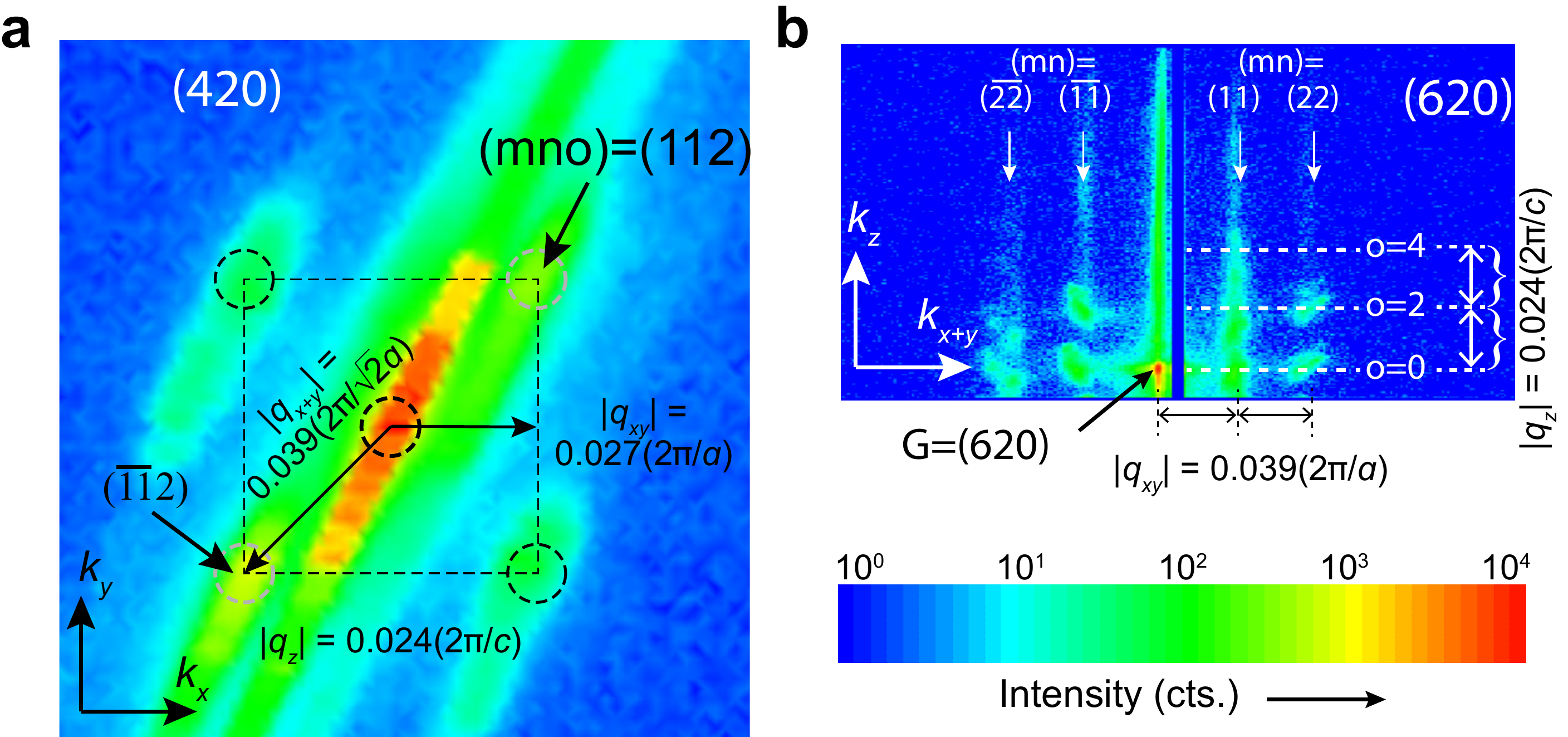}
\caption{(Color online) 2D reciprocal space maps (RSMs) recorded near the ${\bf G}$=$(420)$ (a) and $(620)$ (b) main reflections in our (TaSe$_{4}$)$_2$I sample, collected in the $k_{x,y}$- and $k_{x+y,z}$-planes, respectively.  In both (a,b), satellite intensities are plotted on a logarithmic color scale, as indicated by the color bar below (b).  As in Figs.~\ref{S1HLM} and~\ref{S2HLM}, data for the RSMs shown in Fig.~\ref{cdwdata} was collected within a temperature range of $88$~K to $100$~K, which is well below the reported CDW transition temperature of $260$~K~\cite{Cava1986,260k}.  In both (a,b), we label the indices ${\bf q} = (mno)$ of the visible satellite reflections closest to the main reflections ${\bf G}$, whose positions agree with the modulation vectors obtained from the line scans shown in Figs.~\ref{S1HLM} and~\ref{S2HLM} $\bq={\bf q}_{xy} + q_{z}{\bm \delta}=(mno)=[m\eta (\frac{2\pi}{a}), n\eta (\frac{2\pi}{a}), o\delta(\frac{2\pi}{c})]$, where $\eta = 0.027\pm 0.001$ and $\delta = 0.012 \pm 0.001$ [Eqs.~(\ref{eq:BCTLatVecCDW}) and~(\ref{eq:BCTLatVecCDW2})].  Data for the RSM in (a) was collected at $q_{z}=0.024(\frac{2\pi}{c})$, \emph{i.e.} $2\delta$ above the main ${\bf G}=(420)$ reflection, which is itself only residually visible.  In (a), the locations of the four satellite reflections nearest the main reflection (white and dark dashed circles) are $C_{4z}$-symmetric, but their intensities are only $C_{2z}$-symmetric, within uncertainty ($\sim3.8\times 10^{2}$ cts.).  This indicates that the CDW phase in our sample consists of two macroscopic domains of unequal volume, where the CDW order within each domain breaks the $C_{4z}$ symmetry of the high-temperature crystal structure [SG 97 ($I422$)].  Using dashed white circles, we highlight the brighter ${\bf q} = (112)$ and $(\bar{1}\bar{1}2)$ reflections in (a), which correspond to the domain of larger spatial volume in our sample (\emph{i.e.}, the majority domain).}
\label{cdwdata}
\end{figure*}

Next, we determine the satellite spacing along the $k_{z}$ direction by performing a line scan with varying $q_{z}$ at a fixed position in the $k_{x,y}$ plane ${\bf Q}_{xy} = (hk,mn)=(11,11)$, where ${\bf Q}_{xy} = {\bf G}_{xy} + {\bf q}_{xy}$ and ${\bf G} = {\bf G}_{xy} + g_{z}{\bf c}^{*}$. In this notation, when $q_{z}=0$, our scan passes through the previous satellite reflection ${\bf Q} = (hkl,mno) = (110,110)$ in Fig.~\ref{S1HLM}.  The results of our scan along $q_{z}$ near the ${\bf G}=(110)$ main reflection are plotted in Fig.~\ref{S2HLM} on a linear intensity scale.  We observe that the satellite reflections in Fig.~\ref{S2HLM} exhibit a nearly regular $q_{z}$ separation of $[0.0247\pm 0.0020](\frac{2\pi}{c})$, and that the satellite reflections with positive and negative values of $o$ exhibit the same intensity.   This indicates that all of the satellite reflections along the ${\bf Q}_{xy}=(11)$ line belong to the same CDW domain, and that the CDW order within that domain respects $C_{2,x+y}$ symmetry.  Along with the $C_{2z}$-symmetric intensities that we previously observed in Fig.~\ref{S1HLM}, this implies that the CDW order within the domain of the ${\bf q}=(110)$ reflection is symmetric under the action of point group $D_{2}$ ($222$) in a setting with $C_{2,x\pm y}$ and $C_{2z}$ symmetry [or the $C_{4z}$-symmetric supergroup $D_{4}$ ($422$), though we will shortly exclude this possibility]~\cite{BCS1,BCS2,BigBook}.  As discussed previously in the text surrounding Eq.~(\ref{eq:BCTLatVecCDW}), anticipating that the CDW phase in (TaSe$_{4}$)$_2$I exhibits a domain-averaged body-centered modulated structure, which implies that only satellite reflections satisfying $m + n + o\in 2\mathbb{Z}$ are visible in XRD probes~\cite{korekawa,Korekawa1967,Popescu2003}, then we interpret the satellite reflections in Fig.~\ref{S2HLM} as lying at a separation of $2\delta$.  This implies that the first visible satellite reflection in Fig.~\ref{S2HLM}, which lies at ${\bf q} = [0.027(\frac{2\pi}{a}),0.027(\frac{2\pi}{a}),0.0247(\frac{2\pi}{c})]$, is \emph{actually} the ${\bf q}=(112)$ reflection, \emph{i.e.} a reflection located $q_{z}=2\delta$ above the $k_{z}=0$ plane. Along with the in-plane satellite positions $\bq_{xy}=(mn)$ previously determined in Fig.~\ref{S1HLM}, this implies that, if the CDW order were body-centered tetragonal (which we will shortly see to indeed be the domain-averaged ordering), then the locations of the satellite reflections would be given by Eqs.~(\ref{eq:BCTLatVecCDW}) and~(\ref{eq:BCTLatVecCDW2}) with:
\begin{equation}
\eta = 0.027\pm 0.001,\ \delta = 0.012 \pm 0.001.
\label{eq:ExpDeltaParams}
\end{equation}

In order to complete our determination of the CDW symmetry and modulation vectors, and to confirm the values of $\eta$ and $\delta$ obtained from the line scans in Fig.~\ref{S1HLM} and~\ref{S2HLM} respectively [Eq.~(\ref{eq:ExpDeltaParams})], we additionally recorded detailed 2D RSMs near the ${\bf G} = (420)$ and $(620)$ main reflections.  In Fig.~\ref{cdwdata}(a) [Fig.~\ref{cdwdata}(b)], we show an RSM collected in the $k_{x,y}$- ($k_{x+y,z}$)- plane in the vicinity of the ${\bf G}=(420)$ [$(620)$] main reflection, plotted on a logarithmic intensity scale.  As in Figs.~\ref{S1HLM} and~\ref{S2HLM}, data for the RSMs shown in Fig.~\ref{cdwdata} was collected within a temperature range of $88$~K to $100$~K, which is well below the reported CDW transition temperature of $260$~K~\cite{Cava1986,260k}.  In particular, in the vicinity of the ${\bf G}$=$(420)$ main reflection [Fig.~\ref{cdwdata}(a)], we observe four satellite reflections whose spacing is consistent with Eqs.~(\ref{eq:BCTLatVecCDW}) and~(\ref{eq:BCTLatVecCDW2}) and the values of $\eta$ and $\delta$ obtained from the data in Fig.~\ref{S1HLM} and~\ref{S2HLM} [Eq.~(\ref{eq:ExpDeltaParams})].  However, while the ${\bf q}$ vectors of the satellite reflections in Fig.~\ref{cdwdata}(a) are $C_{4z}$ symmetric, the satellite reflection intensities are \emph{only} $C_{2z}$-symmetric within experimental uncertainty ($\sim3.8\times 10^{2}$ cts.).  This implies that our sample does not contain a single CDW ordering, but contains two different, coexisting CDW orderings of unequal volume, which are known as \emph{domains}.  Within each domain, $C_{4z}$ symmetry is broken, but $C_{2z}$ is preserved.  Additionally, as shown in Figs.~\ref{S1HLM} and~\ref{S2HLM}, $C_{2,x+y}$ symmetry is also preserved.  Along with $C_{2z}$, this implies a point group symmetry of $D_{2}$ ($222$), but crucially, in the setting in which the in-plane twofold rotation axes ($C_{2,x\pm y}$) are oriented along the $x\pm y$ directions, and \emph{not} along the $x$ and $y$ directions.  Using dashed white circles, we highlight the $C_{2z}$-symmetry-related satellites in Fig.~\ref{cdwdata}(a), which we label as ${\bf q} = (112)$ and $(\bar{1}\bar{1}2)$.  The ${\bf q} = (112)$ and $(\bar{1}\bar{1}2)$ satellite reflections are considerably brighter than the $C_{4z}$-symmetry-related satellites labeled with dashed dark circles in Fig.~\ref{cdwdata}(a); this implies that the ${\bf q} = (112)$ and $(\bar{1}\bar{1}2)$ satellites correspond to the CDW domain of larger spatial volume in our sample, which we term the \emph{majority} domain.

Having established the symmetries of the CDW ordering within each of the domains in our sample, and having separated the satellite reflections into the domains from which they originated, we will now determine the space groups (SGs)~\cite{BigBook} corresponding to the (inverse) Fourier-transformed satellite reflections.  Specifically, if we were to ignore the underlying lattice, or if we were to tune the lengths of the modulation vectors to realize a limit in which the CDW order within each domain was lattice-commensurate, then the CDW order in each domain would respect the symmetries of one of the 230 $\mathcal{T}$-symmetric SGs.  To determine of the SG of the CDW order in the majority domain, we begin by re-establishing that the line scans in Figs.~\ref{S1HLM} and~\ref{S2HLM} and the RSM in Fig.~\ref{cdwdata} imply that the majority domain has an SG with point group $D_{2}$ ($222$) and contains the specific symmetries $C_{2,x\pm y}$ and $C_{2z}$.  Because SGs with point group $D_{2}$ ($222$) are described in terms of coordinate axes parallel to the twofold axes, we next rotate the previous domain-averaged modulation vectors ${\bf q}^{G}_{1,2,3}$ [Eqs.~(\ref{eq:BCTLatVecCDW}) and~(\ref{eq:BCTLatVecCDW2})] into the majority-domain modulation vectors $\tilde{\bf q}^{G}_{1,2,3}$, where:
\begin{eqnarray}
\tilde{\bf q}^{G}_{1} = -\frac{\eta}{\sqrt{2}}\left(\frac{2\pi}{a}\right)\left[\frac{\hat{\bf x} + \hat{\bf y}}{\sqrt 2}\right] &+& \frac{\eta}{\sqrt{2}}\left(\frac{2\pi}{a}\right)\left[\frac{\hat{\bf x} - \hat{\bf y}}{\sqrt 2}\right] + \delta\left(\frac{2\pi}{c}\right)\hat{\bf z}, \nonumber \\
\tilde{\bf q}^{G}_{2} = \frac{\eta}{\sqrt{2}}\left(\frac{2\pi}{a}\right)\left[\frac{\hat{\bf x} + \hat{\bf y}}{\sqrt 2}\right] &-& \frac{\eta}{\sqrt{2}}\left(\frac{2\pi}{a}\right)\left[\frac{\hat{\bf x} - \hat{\bf y}}{\sqrt 2}\right] + \delta\left(\frac{2\pi}{c}\right)\hat{\bf z}, \nonumber \\
\tilde{\bf q}^{G}_{3} = \frac{\eta}{\sqrt{2}}\left(\frac{2\pi}{a}\right)\left[\frac{\hat{\bf x} + \hat{\bf y}}{\sqrt 2}\right] &+& \frac{\eta}{\sqrt{2}}\left(\frac{2\pi}{a}\right)\left[\frac{\hat{\bf x} - \hat{\bf y}}{\sqrt 2}\right] - \delta\left(\frac{2\pi}{c}\right)\hat{\bf z}.
\label{eq:majorityLatVecCDW}
\end{eqnarray}
Along with the point group symmetries $C_{2,x\pm y}$ and $C_{2z}$, the modulation vectors $\tilde{\bf q}^{G}_{1,2,3}$ in Eq.~(\ref{eq:majorityLatVecCDW}) imply that the majority-domain CDW order would respect the symmetries of SG 22 ($F222$)~\cite{BigBook,BCS1,BCS2} in the limit in which it were lattice-commensurate (or if one neglects the atoms in the underlying lattice).  After the submission of this work, a first-principles study of the CDW instability in (TaSe$_4$)$_2$I was performed in~\cite{CDWTaSeIDFT}; in agreement with the analysis in the text surrounding Eq.~(\ref{eq:majorityLatVecCDW}), the authors of~\cite{CDWTaSeIDFT} determined that the CDW order in (TaSe$_4$)$_2$I, when tuned to a lattice-commensurate limit, respects the symmetries of SG 22 ($F222$).

For completeness, we note that the domain-averaged modulation vectors ${\bf q}^{G}_{1,2,3}$ in Eqs.~(\ref{eq:BCTLatVecCDW}) and~(\ref{eq:BCTLatVecCDW2}) and the majority-domain modulation vectors $\tilde{\bf q}^{G}_{1,2,3}$ in Eq.~(\ref{eq:majorityLatVecCDW}) imply the existence of satellite reflections at ${\bf q}=(mno) = (101)$ and $(011)$, which do not lie within the momenta sampled in our line scans (Figs.~\ref{S1HLM} and~\ref{S2HLM}) and RSMs (Fig.~\ref{cdwdata}).  Therefore, it is possible that the majority-domain CDW order in (TaSe$_4$)$_2$I does not actually include reflections with ${\bf q}_{z} = o = (1)$.  If reflections with odd values of $o$ are not present, then the CDW order would not respect the symmetries of SG 22 ($F222$) [Eq.~(\ref{eq:majorityLatVecCDW}) and the surrounding text].  In this case, the majority-domain modulation vectors $\tilde{\bf q}^{G'}_{1,2,3}$ would instead be given by:
\begin{equation}
\tilde{\bf q}^{G'}_{1} = \eta\sqrt{2}\left(\frac{2\pi}{a}\right)\left[\frac{\hat{\bf x}+\hat{\bf y}}{\sqrt{2}}\right],\
\tilde{\bf q}^{G'}_{2} = \eta\sqrt{2}\left(\frac{2\pi}{a}\right)\left[\frac{\hat{\bf x}-\hat{\bf y}}{\sqrt{2}}\right],\
\tilde{\bf q}^{G'}_{3} = 2\delta\left(\frac{2\pi}{c}\right)\hat{\bf z}.
\label{eq:majorityLatVecCDW2}
\end{equation}
Along with the point group symmetries $C_{2,x\pm y}$ and $C_{2z}$, the modulation vectors $\tilde{\bf q}^{G'}_{1,2,3}$ in Eq.~(\ref{eq:majorityLatVecCDW2}) imply that the majority-domain CDW order would respect the symmetries of SG 16 ($P222$)~\cite{BigBook,BCS1,BCS2} in the limit in which it were lattice-commensurate (or if one neglects the atoms in the underlying lattice).  However, as shown in SI~A, the nesting vectors between all of the Weyl points at the Fermi energy with opposite chiral charges obtained in our first-principles calculations (Fig.~\ref{fig:nesting} and Table~\ref{wpnesting}) lie at integer multiples of $\tilde{\bf q}^{G}_{1,2,3}$ [Eq.~(\ref{eq:majorityLatVecCDW})], but not $\tilde{\bf q}^{G'}_{1,2,3}$ in Eq.~(\ref{eq:majorityLatVecCDW2}).  Because our ARPES experiments indicate that the CDW phase is gapped (see SI~H.2), which can only be achieved by coupling all of the Weyl points with opposite chiral charges~\cite{W5,ShouchengCDW,TaylorCDW}, then our first-principles calculations provide further evidence that the CDW in (TaSe$_4$)$_2$I is characterized by the modulation vectors $\tilde{\bf q}^{G}_{1,2,3}$ in Eq.~(\ref{eq:majorityLatVecCDW}).

It has been well established in previous works that, while the locations of satellites reflections are governed by the reciprocal modulation vectors [here Eqs.~(\ref{eq:BCTLatVecCDW}),~(\ref{eq:BCTLatVecCDW2}),~(\ref{eq:majorityLatVecCDW}), and~(\ref{eq:majorityLatVecCDW2})], the satellite intensities can vary in a complicated manner as a function of the indices of ${\bf Q} = (hkl,mno)$~\cite{korekawa,Overhauser,Lee1985qvec,Popescu2003}.  The (satellite) reflection intensities scattered from modulated structures are frequently modeled using a harmonic modulation ansatz with a transverse (sinusoidal or Bessel-function) modulation wave.  To quantitatively determine the modulation in the majority domain in our (TaSe$_4$)$_2$I sample, we specifically consider a harmonic modulation ansatz in which the intensity of the $n^{\text{th}}$-order satellite reflection is proportional to $[J_{n}({\bf Q}\cdot {\bf U})]^{2}$, where $J_{n}(x)$ is the $n^{\text{th}}$-order Bessel function, ${\bf Q}$ is the satellite reflection vector, and ${\bf U}$ is a position-space vector characterizing the magnitude and direction of the CDW modulation~\cite{korekawa,Overhauser,Lee1985qvec,Popescu2003}.

The incommensurate Peierls-like modulation along the $c$-axis in (TaSe$_4$)$_2$I has been studied in detail in several previous works~\cite{Fu1984qvec,Fujishita1985,Cava1986,260k} whose results are consistent with the majority-domain modulation vectors obtained in this section [Eqs.~(\ref{eq:majorityLatVecCDW}) and~(\ref{eq:majorityLatVecCDW2})].  However, it is important to note that the CDW order in (TaSe$_4$)$_2$I is \emph{not} purely one-dimensional -- indeed, the spacing of the satellite reflections in the $q_{x,y}$-plane (Figs.~\ref{S1HLM},~\ref{S2HLM}, and~\ref{cdwdata}) indicates that the CDW consists of both a Peierls-like modulation along the $c$ axis as well as weak modulation in the $xy$-plane.  The magnitude of the in-plane, transverse displacement $\Delta r$ between adjacent TaSe$_4$ chains can be estimated through the formula:
\begin{equation}
 \Delta r = |{\bf U}_{xy}|\left(\frac{D}{M}\right),
\label{eq:displacementMagnitude}
\end{equation}
where $|{\bf U}_{xy}|$ is the displacement amplitude in the $xy$-plane, $M$ is the modulation period, and $D = a/\sqrt{2}=0.67\ \text{nm}$ is the distance between adjacent TaSe$_4$ chains [see~\cite{Ta2Se8IPrepare} and Fig.~1(a) of the main text].  First, to determine the in-plane displacement amplitude $|{\bf U}_{xy}|$, we analyze the intensity ratio $R$ between the ${\bf G}=(110,000)$ main reflection ($n=0$) and the ${\bf Q}=(110,\bar{1}\bar{1}0)$ satellite reflection ($n=1$).  Using the line scan in Fig.~\ref{S1HLM}, we determine that $R\approx 6.25\times 10^{-4}$.  Within the harmonic modulation ansatz, this intensity ratio is given by:
\begin{equation}
R=[J_{1}({\bf G}\cdot {\bf U})]^{2}/[J_{0}({\bf G}\cdot {\bf U})]^{2},
\label{eq:intensityRatio}
\end{equation}
where we have employed the approximation that $J_{n}({\bf Q}\cdot {\bf U})\approx J_{n}({\bf G}\cdot {\bf U})$, because $|{\bf Q}-{\bf G}|\ll |{\bf G}|$ for leading-order satellite reflections.  Using $R\approx 6.25\times 10^{-4}$ and $|{\bf G}| = |(110,000)| \approx 0.89~\AA^{-1}$, we obtain an estimate of $|{\bf U}_{xy}| \sim 0.057~\AA$, which is roughly $1\%$ of the in-plane distance ($D=a/\sqrt{2}=0.67\ \text{nm}$) between adjacent TaSe$_4$ chains [see~\cite{Ta2Se8IPrepare} and Fig.~1(a) of the main text].  Next, we recognize that the modulation period $M = 1/|{\bf Q}-{\bf G}| = 1/{\eta\sqrt{2}} \sim 25\ \text{nm}$, where $\eta$ is defined in the text surrounding Eqs.~(\ref{eq:BCTLatVecCDW}) and~(\ref{eq:BCTLatVecCDW2}).  Having established the values of $|{\bf U}_{xy}|$, $D$, and $M$, we use Eq.~(\ref{eq:displacementMagnitude}) to determine that $\Delta r \sim 0.0015\ \AA$.  In combination with the modulation period $M \sim 25\ \text{nm}$, which is large on the atomic scale, $\Delta r \sim 0.0015\ \AA$ indicates that the in-plane (transverse) CDW modulation is weak (though nonzero).  We attribute the small value of $\Delta r$ in the CDW phase of (TaSe$_4$)$_2$I to the weak van der Waals interactions between the TaSe$_4$ chains~\cite{PhysRevLett.110.236401,Ta2Se8IPrepare}.

\begin{figure*}[!t]
\centering
\includegraphics[width=0.65\textwidth]{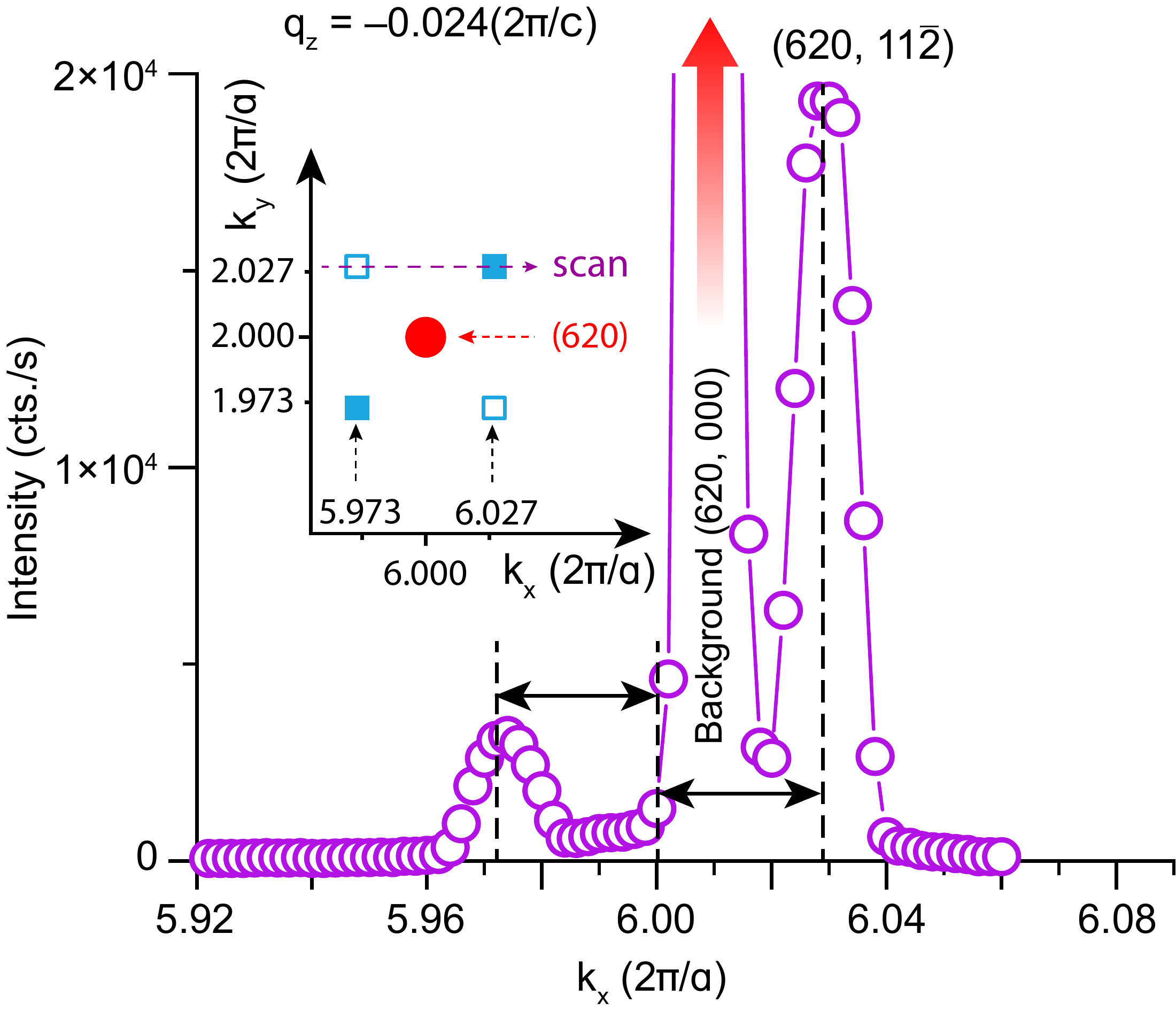}
\caption{(Color online) Longitudinal XRD line scans near the ${\bf Q}=(hkl,mno)=(620,11\bar{2})$ satellite reflection.  In the inset panel, we show a schematic depiction of the locations of the main ${\bf G}=620$ reflection (red circle) and the surrounding satellite reflections (blue squares) in the $k_{x,y}$- ($q_{x,y}$-) plane.  The direction of the line scan in this figure is indicated with a pink arrow in the inset panel.  The satellite positions and intensities observed in the line scan are consistent with the data in Figs.~\ref{S1HLM},~\ref{S2HLM}, and~\ref{cdwdata}, providing further support for the domain-averaged and majority-domain modulation vectors established in Eqs.~(\ref{eq:BCTLatVecCDW}),~(\ref{eq:BCTLatVecCDW2}),~(\ref{eq:majorityLatVecCDW}), and~(\ref{eq:majorityLatVecCDW2}) and the surrounding text.  Specifically, though the positions of the two satellite reflections in the line scan are related by $C_{4z}$ symmetry, the satellites display dramatically different intensities; this provides additional evidence for the existence of two macroscopic CDW order domains in our sample that individually break $C_{4z}$ symmetry.}
\label{figXRDNEW1}
\end{figure*}

\begin{figure*}[!t]
\centering
\includegraphics[width=\textwidth]{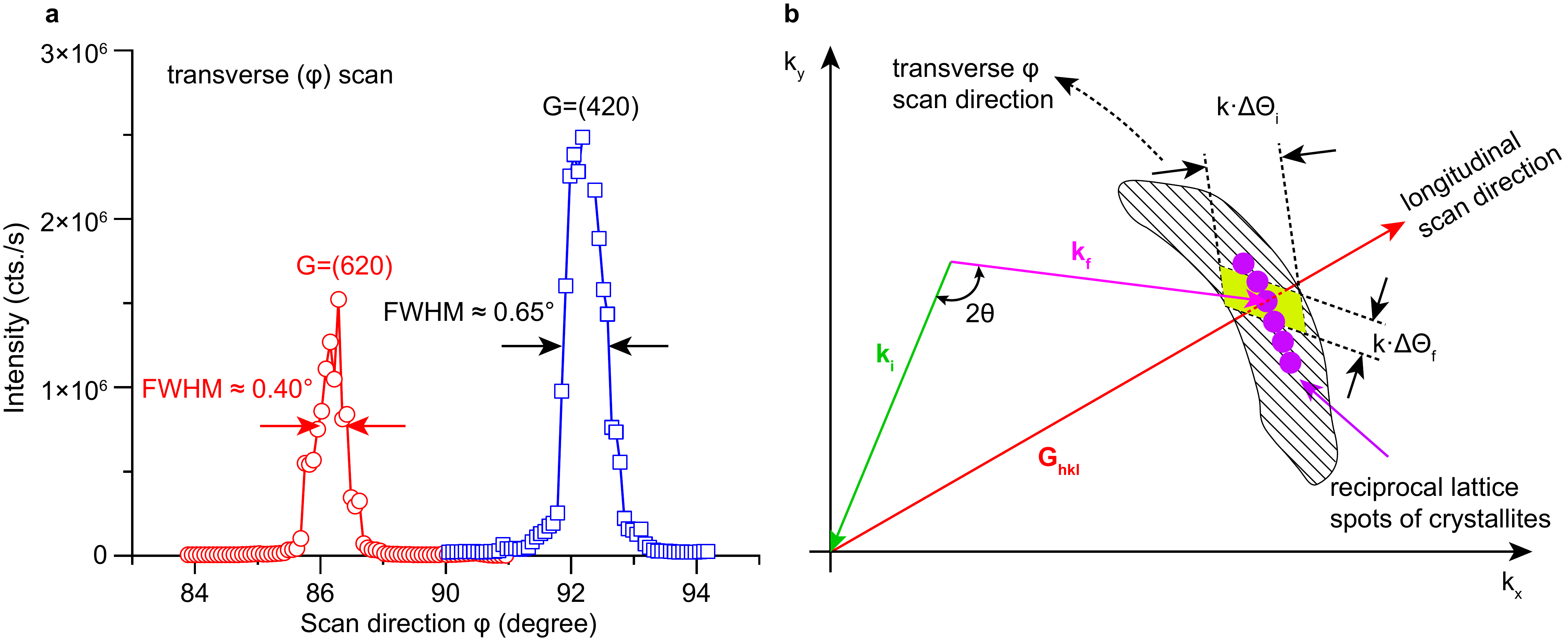}
\caption{(Color online) Transverse angular ($\varphi$) scans.  (a) $\varphi$ scans across the ${\bf G} = (620)$ (red) and the ${\bf G} =(420)$ (blue) main reflections.  In (a), we indicate the full-width half-maxima (FWHM) of the mosaic spread around the ${\bf G} = (620)$ and $(420)$ main reflections.  (b) A schematic depiction of the scattering geometry in $k$ ($q$) space projected onto the $k_{x,y}$-plane, and of the resulting experimental resolution (not shown to scale).  In (b), we respectively label the incoming (green) and scattered (red) X-ray beams ${\bf k}_{i}$ and ${\bf k}_{f}$, and label the resulting vector between the beams with ${\bf G}_{hkl}$ to indicate that it corresponds to the (main) Bragg reflection ${\bf G}=(hkl)$.  In the shaded area in (b), we depict the region of $k$ space covered in a transverse angular scan by an angle of $\varphi$ through ${\bf G}$.  The dark purple circles within the shaded area represent values of ${\bf G}_{hkl}$ (reciprocal lattice reflections) originating from individual crystallites within the sample~\cite{Nielsen2001}.  The resolution volume projected onto the $k_{x,y}$-plane (yellow trapezoid) is determined by the angular divergence of the incoming and the scattered beams ($\Delta\Theta_{i}$ and $\Delta\Theta_{f}$, respectively).}
\label{figXRDNEW2}
\end{figure*}

To demonstrate the reliability of our estimate of the in-plane CDW modulation $\Delta r \sim 0.0015\ \AA$, we will now apply the harmonic modulation ansatz in Eq.~(\ref{eq:intensityRatio}) to satellite reflections in the vicinity of the higher-order ${\bf G} = (hkl) = (620)$ main reflection.  In Fig.~\ref{figXRDNEW1}, we show the results of an XRD line scan performed near the ${\bf G} = (620)$ main reflection in the $k_{x}$- ($q_{x}$-) direction that passes through the ${\bf Q} = (620,11\bar{2})$ satellite reflection, as a well as a different, $C_{4z}$-symmetry-related satellite reflection.  The two satellite reflections in the line scan in Fig.~\ref{figXRDNEW1} differ in intensity, because they belong to different CDW domains within the sample [see the text surrounding Eqs.~(\ref{eq:majorityLatVecCDW}) and~(\ref{eq:majorityLatVecCDW2})].  Continuing to use the approximation ${\bf G}\approx {\bf Q}$ discussed in the text surrounding Eq.~(\ref{eq:intensityRatio}), we use the first-order ${\bf Q}=(620,11\bar{2})$ satellite reflection in Fig.~\ref{figXRDNEW1} to obtain a Bessel-function ratio of $R = [J_{1}({\bf G}\cdot {\bf U})]^{2}/[J_{0}({\bf G}\cdot {\bf U})]^{2} = 0.013$, which is in good agreement with the observed ratio of $\sim 0.014$ between the intensities of the ${\bf Q}=(hkl,mno)=(620,11\bar{2})$ satellite reflection [$\sim 2.0\times10^{4}$ cts/s, see Fig.~\ref{figXRDNEW1}] and the ${\bf G} = (620,000)$ main reflection [$\sim 1.4\times 10^{6}$ cts/s, see Fig.~\ref{figXRDNEW2}(a)].  This calculation provides further support for the harmonic modulation ansatz that we used in Eq.~(\ref{eq:intensityRatio}), in turn providing further support for the value of $\Delta r \sim 0.0015\ \AA$ obtained through Eq.~(\ref{eq:displacementMagnitude}).

We emphasize that we were only able to obtain the CDW modulation vectors in Eqs.~(\ref{eq:majorityLatVecCDW}) and~(\ref{eq:majorityLatVecCDW2}) and estimate the magnitude of the in-plane CDW modulation [Eq.~(\ref{eq:displacementMagnitude}) and the surrounding text] because of the quality of our crystal sample, $k$- ($q$-) space resolution, and dynamic intensity range.  Below, we will provide specific evidence demonstrating the quality of our samples and experimental configuration.  To begin, in a sample of a crystalline material with mosaicity, several internal regions of perfect crystalline order -- known as \emph{crystallites} -- coexist in a mosaic in which, generically, some or all of the symmetry axes within each crystallite are misaligned.  The quality of a sample (single crystal) is frequently quantified by the average size of the crystallites within the sample, and by the angular (mosaic) spread of the misaligned crystallite symmetry axes.  Using XRD data collected within the vicinity of the same main reflection ${\bf G}= (hkl)$, the average crystallite size can be obtained from the spread of a longitudinal line scan along the direction of ${\bf G}$, and the mosaic spread can be obtained from a transverse (angular) scan perpendicular to ${\bf G}$.

To determine the average crystallite size, we first recognize that the line scan across the ${\bf G} = (110)$ main reflection in Fig.~\ref{S1HLM} previously used to obtain the CDW modulation vectors [Eqs.~(\ref{eq:majorityLatVecCDW}) and~(\ref{eq:majorityLatVecCDW2})] was performed along the $q_{x+y}$-direction.  The full-width half-maximum (FWHM) of the intensity distribution of the ${\bf G}=(110)$ main reflection in Fig.~\ref{S1HLM} is approximately given by $\sim 0.005 (2\pi/a)$.  This implies an average crystallite size of 200 unit cells, or an average crystallite in-plane diameter of $\sim 0.2\ \mu \text{m}$, which is in the typical range of high-quality single crystals.  Indeed, it is in fact undesirable to have samples with larger crystallite sizes ($\geq 1\ \mu \text{m}$) in quantitative XRD intensity investigations, because larger crystallite sizes facilitate the onset of multiple scattering, leading to a breakdown of the single-scattering approximation~\cite{Warren1990}.

Next, to obtain the mosaic spread, we have performed additional transverse scans through the ${\bf G} = (420)$ and $(620)$ main reflections, the results of which are shown in Fig.~\ref{figXRDNEW2}(a).  The transverse ($\varphi$) angular scans in Fig.~\ref{figXRDNEW2}(a) indicate an angular mosaic spread of between $0.3^{\circ}$ to $0.5^{\circ}$.  This lies well within the range of common mosaic crystals and epitaxial films, for which the FWHM of the mosaic spread can vary between a few arc minutes up to several degrees.  For the purposes of the XRD experiments in this work -- which were performed to obtain the CDW modulation vectors and symmetry -- the mosaicity of our (TaSe$_4$)$_2$I sample did not impede our determination that the majority-domain CDW order respects the symmetries of point group $D_{2}$ ($222$) and is specified by the modulation vectors in Eqs.~(\ref{eq:majorityLatVecCDW}) and~(\ref{eq:majorityLatVecCDW2}).  Specifically, we were able to obtain Eqs.~(\ref{eq:majorityLatVecCDW}) and~(\ref{eq:majorityLatVecCDW2}) because of the large crystallite size within our sample, and because of the high $k$-space and intensity resolution of our experiments.

To demonstrate how the $k$-space resolution of our experiments is related to the experimental geometry and sample dimensions, we have provided a schematic in Fig.~\ref{figXRDNEW2}(b).  In Fig.~\ref{figXRDNEW2}(b), we use a yellow trapezoid to indicate the experimental $k$-space resolution, which is determined by the divergence of the primary (incident, $\Delta\Theta_{i}$) and diffracted (scattered, $\Delta\Theta_{f}$) beams.  In our experiments, the divergences of the incident and diffracted beams were roughly given by $\Delta\Theta_{i}\approx 100\ \mu\text{rad}$ and $\Delta\Theta_{f}\approx 80\ \mu\text{rad}$ respectively, which we have estimated primarily from the diameter of our needle-shaped sample ($\approx 100\ \mu \text{m}$) and the distance between the sample and the detector ($1250\ \text{mm}$).  This implies a resolution of approximately $\Delta Q = 10^{-3}(2\pi/a)$ in both the in- and out-of-plane directions.  Additionally, in Fig.~\ref{figXRDNEW2}(a), we observe that the intensity of the main reflections is extremely strong (reaching values as large as $10^{6}$ cts. per second); we were able to record such large intensities by employing absorbers.  Furthermore because of the high quality of our crystals (\emph{i.e.}, the large crystallite size and reasonable mosaic spread), we were able to collect intensity data across a range spanning over six orders of magnitude.  Along with the high $k$-space resolution achieved in our experiments (Fig.~\ref{figXRDNEW2}), this has enabled us to obtain consistent and precise measurements of the CDW symmetry and modulation vectors [Eqs.~(\ref{eq:majorityLatVecCDW}) and~(\ref{eq:majorityLatVecCDW2}) and the surrounding text], and to determine the subtle in-plane displacement amplitude of the CDW [Eq.~(\ref{eq:displacementMagnitude}) and the surrounding text].

We additionally emphasize that our determination of the CDW modulation vectors in Eqs.~(\ref{eq:majorityLatVecCDW}) and~(\ref{eq:majorityLatVecCDW2}) did not depend on the role of twinning within the sample.  Specifically, in crystals with structural chirality, samples typically contain both left- and right-handed crystallites, representing an effect known as \emph{twinning}.  Additionally, recent experiments and theoretical studies have demonstrated that the electronic properties of structurally chiral crystals, such as the helicity of the surface Fermi arcs, depend on the enantiomer with the largest spatial volume present throughout a twinned sample~\cite{KramersWeyl,AlPtObserve,RhSnChiralityReversal}.  Though our (TaSe$_{4}$)$_2$I sample did not display any optical signatures of twinning when viewed under a microscope, because the sample was not grown in a manner designed to isolate a single enantiomer~\cite{AxionCDWExperiment}, then the sample is likely twinned to some degree.  However, for the purposes of the XRD experiments performed for this work, the possible presence of twinning in the high-temperature phase of our (TaSe$_{4}$)$_2$I sample did not affect our determination of the CDW modulation vectors.  To avoid drawing conclusions dependent on sample twinning, we specifically extracted satellite reflection intensities  corresponding to a single sample domain of constant chirality, through which we obtained modulation vectors consistent with a structurally chiral, $\mathcal{T}$-symmetric CDW phase with point group $D_{2}$ ($222$) [Eqs.~(\ref{eq:majorityLatVecCDW}) and~(\ref{eq:majorityLatVecCDW2}) and the surrounding text].  Notably, $D_{2}$ ($222$) is the highest-possible point group symmetry for a chiral CDW that breaks the $C_{4z}$ symmetry of the parent high-temperature phase in SG 97 ($I422$)~\cite{BigBook}, which has been confirmed to be the correct high-temperature SG for a single enantiomer of (TaSe$_{4}$)$_2$I in several previous works~\cite{Ta2Se8IPrepare,Fu1984qvec,Fujishita1985,Cava1986,260k}.  Because the observed CDW order is chiral, then it is clear that we have successfully extracted modulation vectors that are compatible with a single enantiomer of the CDW phase of (TaSe$_{4}$)$_2$I, -- otherwise, we would have observed a pattern of intensities with higher symmetry than Eqs.~(\ref{eq:majorityLatVecCDW}) and~(\ref{eq:majorityLatVecCDW2}), and would have instead reported modulation vectors corresponding to an \emph{achiral} SG in the limit that the CDW were tuned to be lattice-commensurate [in contrast to the chiral SGs 22 ($F222$) and 16 ($P222$) reported in the text surrounding Eqs.~(\ref{eq:majorityLatVecCDW}) and~(\ref{eq:majorityLatVecCDW2})].  Furthermore, as shown in SI~A, integer multiples of the majority-domain modulation vectors in Eqs.~(\ref{eq:majorityLatVecCDW}) nest all of the bulk Weyl points with opposite chiral charges.  Therefore, regardless of possible twinning within our sample, we have unambiguously determined that a single enantiomer of the chiral, majority-domain CDW order observed in our (TaSe$_{4}$)$_2$I sample can fully account for the Weyl-point coupling and resulting CDW-induced gap in (TaSe$_{4}$)$_2$I (see SI~H.2).

\begin{figure*}[!t]
\centering
\includegraphics[width=0.6\textwidth]{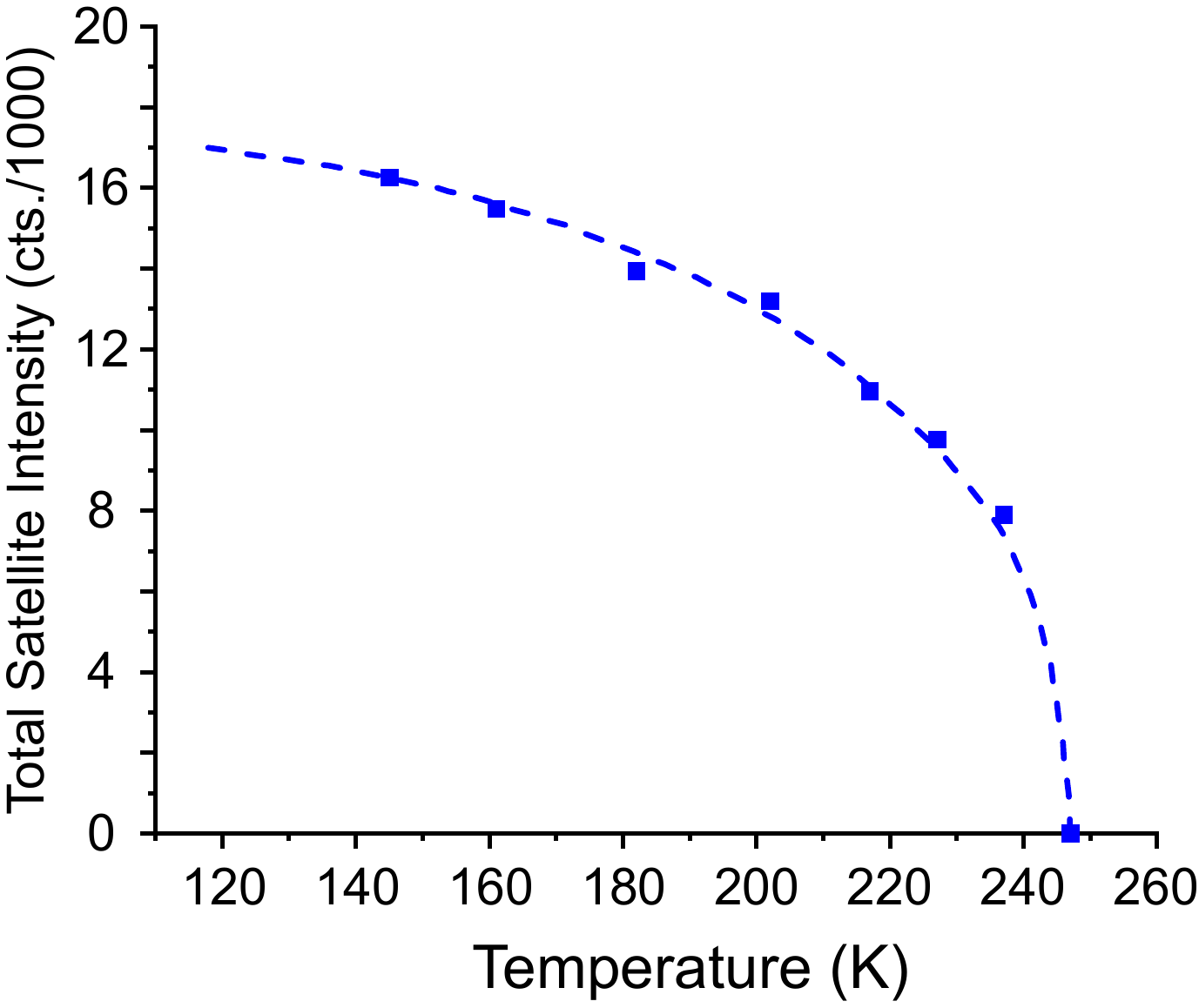}
\caption{(Color online) Total measured satellite intensity in our (TaSe$_{4}$)$_2$I sample in the vicinity of the ${\bf G}=(110)$ main reflection as a function of temperature.  The blue squares represent collected data and the dashed line is a smooth, close-fit interpolation to guide the eye.  We observe that all satellites simultaneously disappear at $T_{C}\approx 248~K$, representing a signature of a transition away from a CDW phase.}
\label{S3HLM}
\end{figure*}

Finally, to obtain an estimate for the critical temperature $T_{C}$ of the transition in (TaSe$_{4}$)$_2$I from the low-temperature CDW phase to the high-temperature Weyl-semimetal phase, we have also measured the temperature dependence of the satellite reflection intensities.  In Fig.~\ref{S3HLM}, we plot the summed intensities of all satellite reflections in the vicinity of the ${\bf G}=(110)$ main reflection as a function of temperature.  We observe that all satellite reflections simultaneously disappear at a transition temperature of $T_{C}\approx 248$~K, which is representative of a transition away from a CDW phase.  This value of $T_{C}$ is slightly lower than, but still in close agreement with, the value of $T_{C}=260$~K measured in previous works~\cite{Fu1984qvec,Fujishita1985,Cava1986,260k}.

\vspace{0.2in}
\subsubsection*{H.2 ARPES Experiments}

ARPES measurements were performed at the high-resolution branch of beamline I05 of the Diamond Light Source (DLS) with a Scienta R4000 analyzer. The photon energy range for the DLS was 30-220~eV.  During measurements, (TaSe$_{4}$)$_2$I samples were kept at a pressure of $<1.5\times10^{-10}$ Torr.  The angles of the emitted photoelectrons were measured with a resolution of 0.2$^\circ$, and their energies were measured at an overall resolution of $<15$~meV.  After samples were glued to the sample holder, they were then cleaved in situ to expose the $(110)$-surface, which is the favored cleavage plane of (TaSe$_{4}$)$_2$I~\cite{PhysRevLett.110.236401}.

\begin{figure*}[t]
\centering
\includegraphics[width=\textwidth]{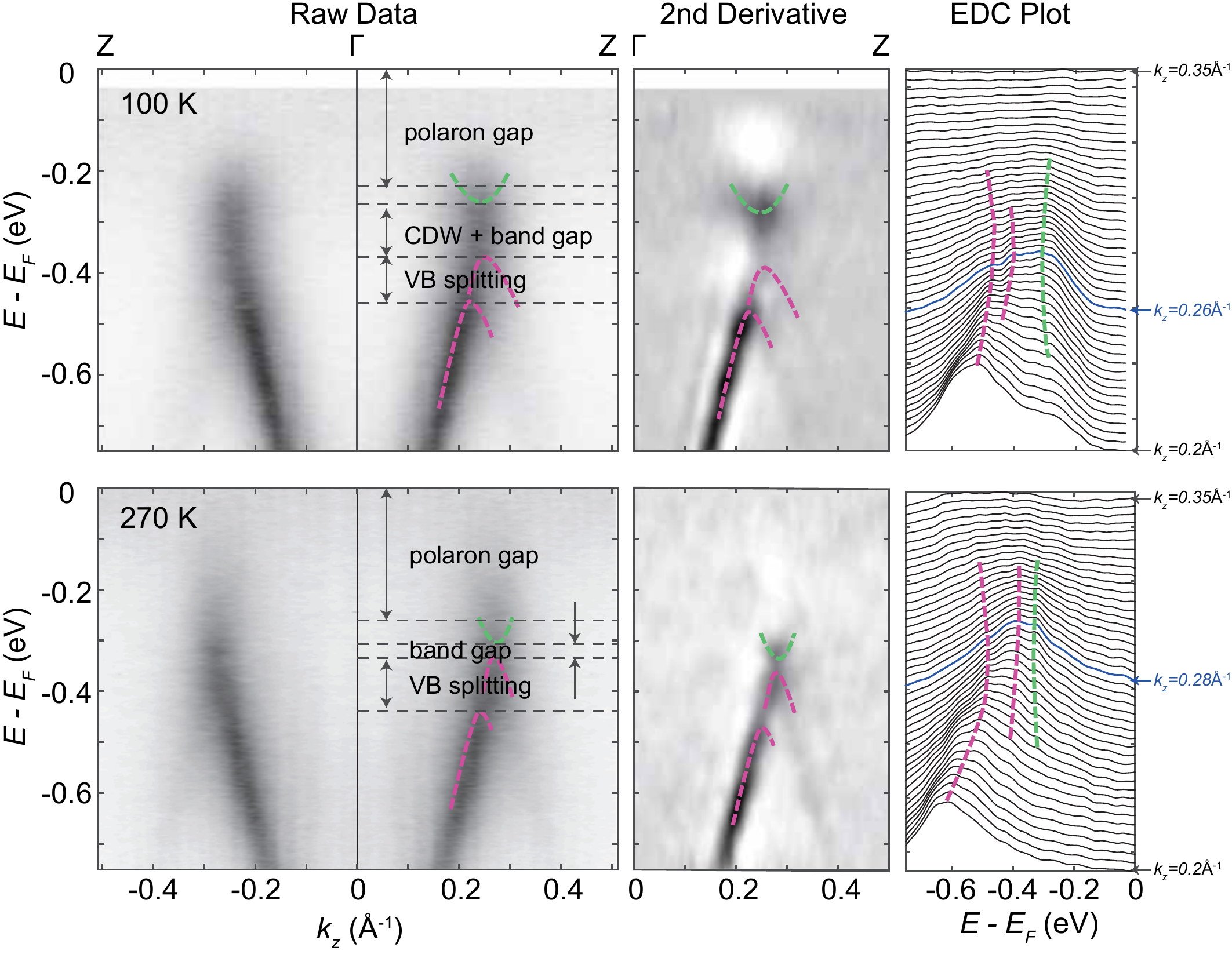}
\caption{(Color online) Band dispersion of (TaSe$_{4}$)$_2$I  as measured in angle-resolved photoemission (ARPES) experiments.  In the upper (lower) panels, we respectively plot the raw ARPES data, the second derivative of the data along $\Gamma Z$, and the stacking energy distribution curves (EDCs) at $100$~K ($270$~K).  The valence and conduction bands are respectively indicated by dashed magenta and green lines.  In the raw data, we mark the polaron gap, the CDW-induced band gap (in the low-temperature phase), and the valence band splitting.  The blue curves in the EDC plots (rightmost panels) mark the value of $k_{z}$ at which the gap between the top of the valence band and the bottom of the conduction band is at a minimum.}
\label{arpesdata}
\end{figure*}

In Fig.~\ref{arpesdata}, we plot the high-resolution band dispersion along $\Gamma Z$ measured using an incident photon energy of 72~eV.  Measurements of the low-temperature and high-temperature phases of (TaSe$_{4}$)$_2$I were performed at 100~K and 270~K, respectively, where previous literature has determined that a CDW transition occurs in (TaSe$_{4}$)$_2$I at $T_{C}= 260$~K~\cite{Cava1986,260k} [though our XRD data in SI~H.1 (Fig.~\ref{S3HLM}) shows the emergence of a CDW at the slightly lower temperature of $T_{C}\approx 248$~K].  We were unable to perform measurements below $100$~K due a significant charging effect, which we attribute to a rapid increase in resistance.  Specifically, in insulating materials, such as our (TaSe$_{4}$)$_2$I sample in its CDW phase, the emission of photoelectrons depletes the total sample charge.  This results in a finite surface voltage that reduces the kinetic energies of subsequent outgoing photoelectrons, which we observe in our ARPES measurements.  Conversely, in the high-temperature data [Fig.~\ref{arpesdata}(b)], the spectrum does not exhibit a charging effect, but is instead visibly affected by thermal broadening.  To set the Fermi energy, we placed our (TaSe$_{4}$)$_2$I sample in contact with polycrystalline Au, whose photoemission spectrum is well known.  In agreement with previous works~\cite{PhysRevLett.110.236401,perfetti2001spectroscopic}, both the high- and low-temperature ARPES spectra of our (TaSe$_{4}$)$_2$I sample showed signatures of strong polaronic effects that renormalize the electronic band structure to higher binding energies by $E_p\approx0.22$~eV.

\vspace{0.2in}
In Fig.~\ref{arpesdata}, we show the band dispersion of our (TaSe$_{4}$)$_2$I sample measured along $\Gamma Z$ through ARPES.  In both the low- and high-temperature phases (upper and lower panels, respectively, in Fig.~\ref{arpesdata}), the valence bands show two components with a separation of about  $0.1$~eV.  We observe that the energy gap between the top of the valence band and the bottom of the conduction band is roughly 0.12~eV in the low-temperature phase [Fig.~\ref{arpesdata} (upper panels)], but is significantly smaller ($< 0.04$~eV) in the high-temperature phase [Fig.~\ref{arpesdata} (lower panels)].  We attribute the change in gap size between the upper and lower panels in Fig.~\ref{arpesdata} to a transition from a low-temperature phase with a CDW-induced band gap into the high-temperature Weyl-semimetal phase predicted in this work.  At both of the sample temperatures used in our ARPES probes ($100$~K and 270~K), the energy bands display strong dispersion in the $k_{z}$ direction (see Fig.~\ref{arpesdata}).  Taken together with transport measurements performed in earlier works~\cite{260k,Cava1986,CDWTransport1}, which imply a single-particle gap of $\sim 0.138$~eV in the CDW phase, the ARPES data in the upper panels of Fig.~\ref{arpesdata} indicate that the CDW phase of (TaSe$_{4}$)$_2$I is fully gapped, with the smallest gap lying within a close vicinity of $\Gamma Z$.

\subsection*{I. Understanding the Electronic Structure of (TaSe$_4$)$_2$I from the Perspective of Filling-Enforced Gaplessness}

As shown in Fig.~1(d) of the main text and in Fig~\ref{fig:seQPI}(a), in the high-temperature Weyl semimetal phase of (TaSe$_4$)$_{2}$I in the body-centered tetragonal SG 97 ($I422$), the electronic structure is quasi-1D, and the entire Fermi surface lies in the vicinity of the $k_{z}=\pm \pi/c$ planes.  However, in SG 97, generic points in the $k_{z}=\pm \pi/c$ planes are not fixed by a symmetry (\emph{e.g.} $C_{2z}\times\mathcal{T}$), as they would be in a primitive tetragonal structure with a periodicity of $c$ in the $z$ direction [though the high-symmetry and TRIM (\emph{i.e.} non-generic) points $P$ and $N$ conversely \emph{do} lie in the $k_{z}=\pm\pi/c$ planes in SG 97, see Fig.~1(c) of the main text and SI~A]~\cite{BigBook,BCS1,BCS2}.  Therefore, it is natural to ask, in a real material such as (TaSe$_4$)$_{2}$I, whose electronic structure is not fine-tuned like that of a toy model, why the entire Fermi surface remains concentrated near $k_{z}=\pm \pi/c$ in the absence of a pinning symmetry.  In this section, we will show how the shape and localization of the Fermi surface of (TaSe$_4$)$_{2}$I derive from the electronic structure of decoupled TaSe$_4$ chains, which \emph{do} exhibit nonsymmorphic-symmetry- and electronic-filling-enforced band crossings near $k_{z}=\pi/c$.

We begin by considering a single, isolated TaSe$_4$ chain [total chemical formula (TaSe$_4$)$_4$ per chain unit cell] that is infinite in the $z$ ($c$-axis) direction.  Within an isolated TaSe$_4$ chain, the Se rectangles and Ta atoms follow a regular eightfold screw arrangement along the $c$-axis in which consecutive Se rectangles (and Ta atoms) are related by $45^{\circ}$ turns and $c/4$ translation in the $z$ direction [Fig.~1(b) of the main text].  Because the rectangle of Se atoms within each layer lies far from the limit of a perfect square, then the nearly-preserved eightfold screw symmetry generates a sixteen-fold pattern of Se atoms when each chain is viewed along the $c$ axis, as shown in Fig.~1(a) of the main text [if the Se rectangles were deformed to squares, then the Se atoms would only form an eightfold pattern].

Compared to the more familiar space group (SG) symmetry elements~\cite{BigBook}, the symmetry generators for an isolated TaSe$_4$ chain are relatively unusual.  In addition to $\mathcal{T}$ symmetry, twofold rotations about the $x$ axis, and $z$-direction lattice translation, a TaSe$_4$ chain is generated by an $s_{8_{2}}$ screw (the aforementioned combination of $45^{\circ}$ rotation about the $z$ axis and $c/4$ translation in the $z$ direction):
\begin{equation}
\mathcal{T} = \bigg\{\mathcal{T}\bigg|0\bigg\},\ C_{2x} = \bigg\{C_{2x}\bigg|0\bigg\},\ T_{z} = \bigg\{E\bigg|c\bigg\},\ s_{8_{2}} = \bigg\{C_{8z}\bigg|\frac{c}{4}\bigg\},
\label{eq:funny8Rod}
\end{equation}
where $E$ is the identity element.  An isolated TaSe$_4$ chain is therefore invariant under the \emph{non-crystallographic} chiral rod group $(p8_{2}22)_{RG}$~\cite{eightfoldRodGroups,PointGroupTables}, given in the notation of~\cite{HingeSM}.  Specifically, because the eightfold screw symmetry $s_{8_{2}}$ is not an element of any of the 230 3D SGs~\cite{BigBook}, then $(p8_{2}22)_{RG}$ does \emph{not} have a space supergroup, unlike the 75 \emph{crystallographic} rod groups~\cite{ITCA,subperiodicTables}.  We will later show in this section that the process of forming a 3D crystal by coupling an array of TaSe$_{4}$ chains in the $xy$-plane, along with the introduction of iodine atoms, breaks the non-crystallographic $s_{8_{2}}$ symmetry while preserving its crystallographic square:
\begin{equation}
s_{4_{2}} = (s_{8_{2}})^{2}=\bigg\{C_{4z}\bigg|\frac{c}{2}\bigg\},
\label{eq:42screw}
\end{equation}
which we will see to be a symmetry of the conventional cell of (TaSe$_{4}$)$_{2}$I when translations in the $xy$-plane are taken into account.

As discussed in~\cite{WPVZ,WiederLayers}, in nonsymmorphic symmetry groups, the glide and screw symmetries place constraints on the electronic fillings at which a symmetry-preserving [single-particle (band) or short-range-entangled interacting] gap can ever be present.  Specifically, in a $\mathcal{T}$-symmetric, nonsymmorphic 1D crystal with a $c$- ($z$-) directed screw axis of the form:
\begin{equation}
s_{n_{B}} = \bigg\{C_{n}\bigg|c\left(\frac{B}{n}\right)\bigg\},
\end{equation}
where $n,B\in\mathbb{Z}^{+}$ and $n>B$, insulating gaps are only permitted (absent non-minimally connected bands~\cite{QuantumChemistry,Bandrep1,Bandrep2,Bandrep3} or long-range-entangled topological order~\cite{WPVZ}) at the electronic fillings:
\begin{equation}
\nu \in 2\left(\frac{n}{B}\right)\mathbb{Z}.
\label{eq:WPVZmain}
\end{equation}
This is because the $(B/n)c$ translation in $s_{n_{B}}$ divides each unit cell into $n/B$ symmetry-related segments in which states in position-space (atomic orbitals in the single-particle limit) are twofold degenerate due to Kramers' theorem ($\mathcal{T}$ symmetry).  Consequently, in the terminology of~\cite{WPVZ,WiederLayers}, $2(n/B)\mathbb{Z}$ is the ``minimal-insulating filling.''  Absent interactions, Eq.~(\ref{eq:WPVZmain}) implies that spinful bands in a 1D screw-symmetric crystal appear with a minimal connectivity of $2(n/B)$~\cite{WPVZ,WiederLayers,WPVZfollowUp}, and that they exhibit filling-enforced (nodal) degeneracies~\cite{WiederLayers,DDP} (\emph{i.e.}, filling-enforced gaplessness) at other (frequently even) values of the system filling $\nu$.

\begin{figure*}[t]
\centering
\includegraphics[width=0.97\textwidth]{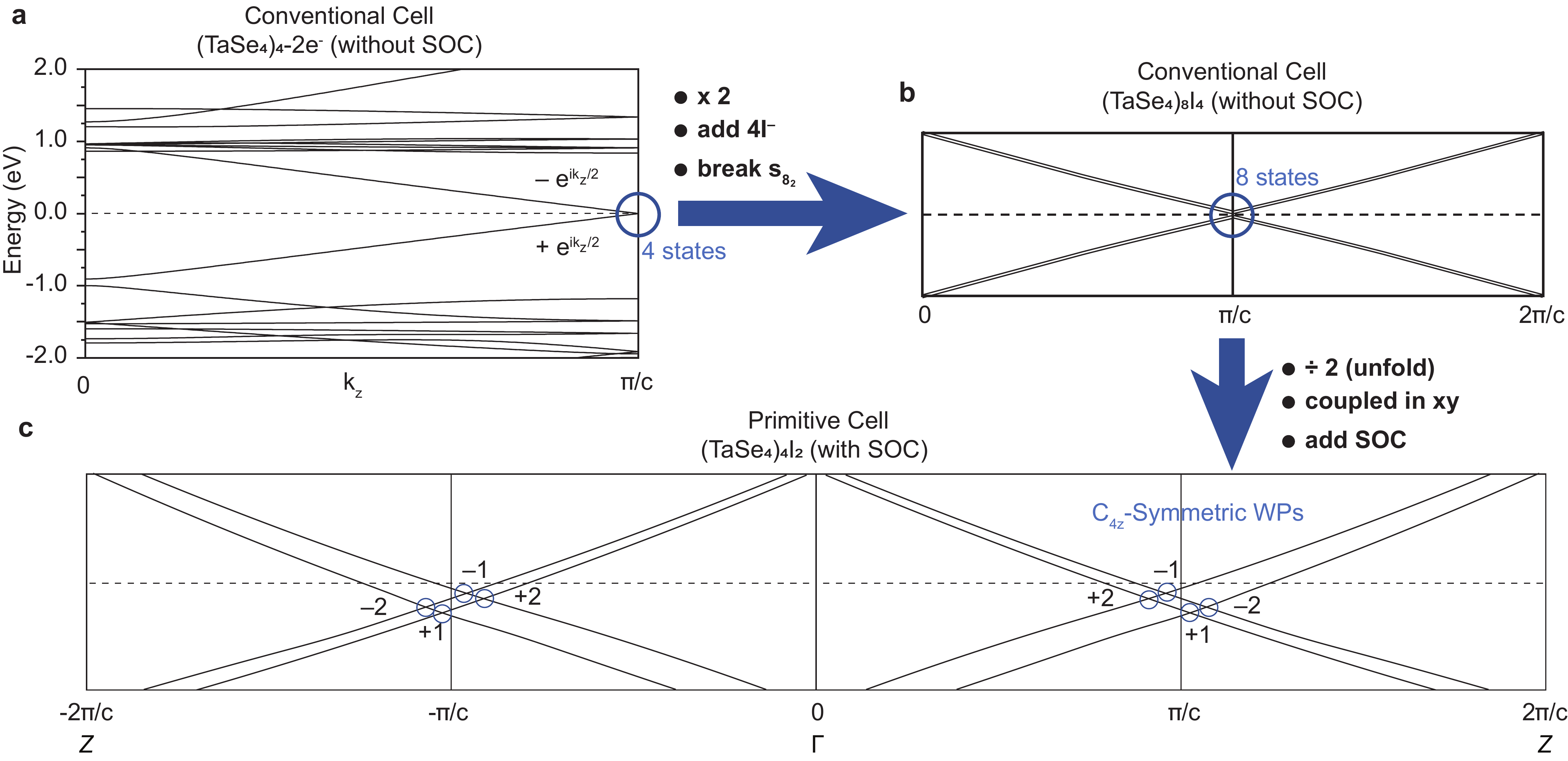}
\caption{(Color online) (a) The electronic structure of an isolated TaSe$_4$ chain in the non-crystallographic chiral rod group $(p8_{2}22)_{RG}$ [Refs.~\cite{eightfoldRodGroups,PointGroupTables,ITCA,subperiodicTables} and Eq.~(\ref{eq:funny8Rod})] calculated from first principles in the absence of SOC, with the Fermi level set two electrons lower to that of (TaSe$_4$)$_2$I, as discussed in the text surrounding Eq.~(\ref{eq:chainDoped}).  A $4_{2}$-screw [Eq.~(\ref{eq:42screw})], $\mathcal{T}$-symmetry, and filling-enforced~\cite{WPVZ,WiederLayers,WPVZfollowUp} fourfold nodal degeneracy (twofold per spin) is visible at $k_{z}=\pi/c$ (spinless $s_{4_{2}}$ eigenvalues shown in dark blue), which is a TRIM point in the isolated chain.  Away from $k_{z}=0,\pi/c$, the bands in (a) are spin-degenerate, because SOC is absent at this stage of the calculation.  (b) Schematic band structure of two superposed TaSe$_4$ chains and four iodine atoms [total chemical formula (TaSe$_4$)$_8$I$_4$ per chain unit cell] in crystallographic rod group $(p4_{2}22)_{RG}$.  An \emph{eightfold} spinless degeneracy is present at $k_{z}=\pi/c$; unlike the fourfold nodal degeneracy in (a), the eightfold degeneracy in (b) is \emph{not} filling-enforced [text surrounding Eq.~(\ref{eq:fullIodineChainTimesTwo})].  Away from $k_{z}=0,\pi/c$, the bands in (b) are fourfold degenerate (indicated with narrowly split black lines), because they represent two superposed, decoupled copies of the bands in (a).  (c) Schematic band structure of the bands closest to the Fermi energy in a 3D, body-centered crystal of (TaSe$_4$)$_2$I (Fig.~1 of the main text) in SG 97 ($I422$).  The introduction of body-centered 3D lattice translations to rod group $(p4_{2}22)_{RG}$ [Eq.~(\ref{eq:BCtranslation})] has unfolded the eight states at $k_{z}=\pi/c$ in (b) into fourfold nodal planes at $k_{z}=\pm\pi/c$ in (c), which then subsequently split in the presence of interchain coupling and SOC into the Fermi-surface WPs in Table~\ref{tbwps}, as well as eight $C_{4z}$-enforced WPs below the Fermi energy [Fig.~1(d) of the main text].  Because generic points in the $k_{z}=\pm\pi/c$ planes are not fixed by a symmetry (\emph{e.g.} $C_{2z}\times\mathcal{T}$) in SG 97 [text preceding Eq.~(\ref{eq:funny8Rod})], then the Fermi pockets in (TaSe$_4$)$_2$I are not pinned by symmetry to lie near $k_{z}=\pm\pi/c$.  Nevertheless, because the interchain coupling in (TaSe$_4$)$_2$I is perturbatively weak~\cite{Ta2Se8IPrepare,gressier1984characterization,gressier1984electronic}, then the Fermi pockets in (c) remain close to their original, filling-enforced location(s) ($k_{z}=\pi/c$) in (a).}
\label{fig:feSem}
\end{figure*}

Using Eq.~(\ref{eq:WPVZmain}), we determine that an isolated TaSe$_4$ chain in rod group $(p8_{2}22)_{RG}$, for which $n=8$ and $B=2$ [Eq.~(\ref{eq:funny8Rod})], exhibits a minimal-insulating filling of:
\begin{equation}
\nu_{chain}\in 8\mathbb{Z}.
\label{eq:chainMinimal}
\end{equation}
We next compare $\nu_{chain}$ to the number of electrons in a single TaSe$_4$ chain at both charge neutrality and at its oxidation state when paired with iodine atoms in 3D (TaSe$_4$)$_2$I crystals.  To begin, the electronic configurations of Ta and Se are respectively given by~\cite{mcQuarriePchem}:
\begin{equation}
\text{Ta} \equiv [\text{Xe}]4f^{14}5d^{3}6s^{2},\ \text{Se}\equiv [\text{Ar}]3d^{10}4s^{2}4p^{4},
\end{equation}
such that Ta and Se respectively carry the core and valence electron numbers:
\begin{equation}
N_{Ta}^{c} = 68,\ N_{Ta}^{v} = 5,\ N_{Se}^{c} = 28,\ N_{Se}^{v} = 6.
\label{eq:electronicCounting}
\end{equation}
A single TaSe$_4$ chain [total chemical formula (TaSe$_4$)$_4$ per chain unit cell] consists of four repeated units of Ta atoms and rectangles with Se atoms on each of the four corners [Fig.~1(a,b) of the main text].  Therefore, a single TaSe$_4$ chain carries the total core and valence electron numbers:
\begin{equation}
N^{c}_{T} = 4N_{Ta}^{c} + 16N_{Se}^{c} = 720,\ N^{v}_{T} = 4N_{Ta}^{v} + 16N_{Se}^{v} = 116.
\label{eq:chainTotalEs}
\end{equation}
First, because $N^{c}_{T}\text{ mod }8 = 0$, then a gap is permitted between bands induced from the core orbitals and those from the valence atomic orbitals.  Therefore, for now, we will focus our analysis on the valence atomic orbitals and electrons.  In a 3D crystal of (TaSe$_4$)$_2$I, there are $2$ iodine atoms per each TaSe$_4$ chain.  Because iodine atoms prefer to realize the ionic oxidation state I$^{-}$ by removing an electron from another atom, then, anticipating constructing a 3D crystal of TaSe$_4$ chains and iodine atoms, we remove two valence electrons from the isolated TaSe$_4$ chain that we are presently analyzing:
\begin{equation}
\tilde{N}^{v}_{T} = N^{v}_{T} - 2 = 114,
\label{eq:chainDoped}
\end{equation}
realizing an overall chemical formula of [(TaSe$_{4}$)$_{4}]^{2+}$.  Because $\tilde{N}^{v}_{T}\text{ mod }8 \neq 0$, then we predict that an isolated [(TaSe$_{4}$)$_{4}]^{2+}$ chain is a filling-enforced semimetal.  Specifically, if all of the bands in an isolated [(TaSe$_{4}$)$_{4}]^{2+}$ chain. are minimally connected, then $\tilde{N}^{v}_{T}\text{ mod }8 = 2$ implies the presence of a quarter-filled set of eight connected, spinful bands (four connected bands per spin in the limit of vanishing SOC) at the Fermi level.  In Fig.~\ref{fig:feSem}(a), we show the band structure of a single, isolated TaSe$_4$ chain calculated from first principles with the Fermi level set to that of [(TaSe$_{4}$)$_{4}]^{2+}$.  The band structure in Fig.~\ref{fig:feSem}(a) was specifically obtained by removing all of the atoms in the conventional cell of a 3D crystal of (TaSe$_4$)$_2$I except for a single TaSe$_4$ chain, increasing the cell volume to fully decouple the remaining TaSe$_4$ chains, manually restoring the weakly broken $s_{8_{2}}$ screw symmetry of the decoupled chains [Fig.~1(a,b) of the main text and Eq.~(\ref{eq:funny8Rod})], and calculating the electronic structure along the $k_{z}$ direction with the Fermi level lowered by two electrons to that of [(TaSe$_{4}$)$_{4}]^{2+}$.  To highlight the role of filling-enforced gaplessness, the band structure in Fig.~\ref{fig:feSem}(a) was calculated without spin-orbit coupling (SOC); therefore, away from $k_{z}=0,\pi/c$, the bands in (a) are spin-degenerate.  At $k_{z}=\pi/c$ in Fig.~\ref{fig:feSem}(a), we observe a half-filled, fourfold nodal degeneracy (two degenerate states per spin), which is locally protected by the combination of spinless $\mathcal{T}$ and $(s_{8_{2}})^{2}=s_{4_{2}}$ screw symmetry [Eq.~(\ref{eq:42screw})].  Specifically, as shown in Fig.~\ref{fig:feSem}(a), the spin-degenerate bands forming the nodal degeneracy at $E_{F}$ at $k_{z}=\pi/c$ carry the spinless $s_{4_{2}}$ screw eigenvalues $\pm \exp{(i k_{z}/2)}=\pm i$.  This confirms that [(TaSe$_{4}$)$_{4}]^{2+}$ is a filling-enforced semimetal with a quarter-filled eightfold band connectivity at $E_{F}$ (four connected bands per spin without SOC), as discussed in the text following Eq.~(\ref{eq:chainDoped}).

We pause to note that, as shown in~\cite{WiederLayers,Steve2D,HourglassInsulator,Cohomological,DiracInsulator}, if we were to introduce SOC, which is generically present and non-negligible in a real [(TaSe$_{4}$)$_{4}]^{2+}$ chain, then the nodal degeneracy at $k_{z}=\pi/c$ in Fig.~\ref{fig:feSem}(a) would split into a pair of twofold linear crossings at time-reversed values of $k_{z}$, realizing an ``hourglass''-like band structure.  However, whether or not SOC is present, a symmetry-preserving, noninteracting (short-range-entangled) gap still cannot be opened in $(p8_{2}22)_{RG}$ at $\nu=\tilde{N}^{v}_{T}$ [Eqs.~(\ref{eq:chainMinimal}) and~(\ref{eq:chainDoped})].  Therefore, [(TaSe$_{4}$)$_{4}]^{2+}$ would remain a filling-enforced (hourglass) semimetal if SOC were introduced.  However, for simplicity, we will continue to artificially neglect SOC until the final stage of the discussion in this section, and will for now consider all of the bands in the electronic structure of a single [(TaSe$_{4}$)$_{4}]^{2+}$ chain to be spin-degenerate.

While continuing to focus on a single [(TaSe$_{4}$)$_{4}]^{2+}$ chain, we next restore charge neutrality by introducing two iodine ($I^{-}$) ions.  Specifically, we add two I$^{-}$ ions to each 1D unit cell of the previous isolated [(TaSe$_{4}$)$_{4}]^{2+}$ chain -- one at $z=0$ and another at $z=c/2$.  The resulting chain carries a total chemical formula of (TaSe$_4$)$_4$I$_2$ per unit cell.  For simplicity, for now, we will not discuss the locations of the iodine atoms in the $xy$-plane, and will instead only consider their effects on the rod group symmetries and the electronic filling.  Because there are only two iodine ions, then the ions do not respect the non-crystalllographic $s_{8_{2}}$ screw symmetry [Eq.~(\ref{eq:funny8Rod})], though they do respect its square $s_{4_{2}}$ [Eq.~(\ref{eq:42screw})].  Therefore, the addition of two iodine atoms has lowered the symmetry of the chain to that of the \emph{crystallographic} rod group $(p4_{2}22)_{RG}$~\cite{ITCA,subperiodicTables}, which is generated by:
\begin{equation}
\mathcal{T} = \bigg\{\mathcal{T}\bigg|0\bigg\},\ C_{2x} = \bigg\{C_{2x}\bigg|0\bigg\},\ T_{z}=\bigg\{E\bigg|c\bigg\},\ s_{4_{2}} = \bigg\{C_{4z}\bigg|\frac{c}{2}\bigg\}.
\label{eq:normal4Rod}
\end{equation}
We can determine the minimal insulating filling of $(p4_{2}22)_{RG}$ by recognizing that:
\begin{equation}
P4_{2}22 \equiv E(p4_{2}22)_{RG}\cup T_{x}(p4_{2}22)_{RG},
\label{eq:spaceSuperGroup}
\end{equation}
where $P4_{2}22$ is the symbol for SG 93 and $T_{x}$ is lattice translation in the $x$ direction.  Because the addition of perpendicular lattice translation (here $T_{x}$) does not change the minimal insulating filling~\cite{WPVZ,WiederLayers}, then we can use the minimal insulating filling of SG 93, previously determined in~\cite{WPVZ} to be $4\mathbb{Z}$, to infer that $(p4_{2}22)_{RG}$ also exhibits a minimal insulating filling of:
\begin{equation}
\nu_{chain}'\in 4\mathbb{Z}.
\label{eq:chainLessMinimal}
\end{equation}
We could have also equivalently derived $\nu_{chain}'$ by recognizing that the $4_{2}$ screw symmetry in $(p4_{2}22)_{RG}$ [Eq.~(\ref{eq:42screw})] divides each chain unit cell into $s_{4_{2}}$-related halves, which, along with $\mathcal{T}$ symmetry, enforces a minimal band connectivity of $4$ in the single-particle limit~\cite{WiederLayers,QuantumChemistry,Bandrep1,Bandrep2,Bandrep3}.

Next, we compare $\nu_{chain}'$ to the electronic filing.  Because all of the (formerly) valence atomic orbitals of the iodine ions are occupied by electrons, we take \emph{all} of the iodine ion electrons to be (closed-shell) core electrons~\cite{mcQuarriePchem}.  Therefore, from the electronic configuration of an iodine ion:
\begin{equation}
\text{I}^{-} \equiv [\text{Kr}]4d^{10}5s^{2}5p^{6},
\end{equation}
we obtain:
\begin{equation}
N_{I^{-}}^{c} = 54,\ N_{I^{-}}^{v} = 0.
\label{eq:electronicCountingIodine}
\end{equation}
For a TaSe$_4$ chain with two iodine atoms [which exhibits a total chemical formula of (TaSe$_4$)$_{4}$I$_{2}$ per 1D unit cell], this implies that:
\begin{equation}
N_{T}^{c'} = N_{T}^{c} +  2N_{I^{-}}^{c} = 828,\ N_{T}^{v'} = \tilde{N}_{T}^{v} = 114.
\label{eq:fullIodineChainE}
\end{equation}
Because $N_{T}^{c'}\text{ mod }4=0$ [Eq.~(\ref{eq:chainLessMinimal})], then we can again take the core electrons and atomic orbitals to be separated from the valence electrons and orbitals by an energy gap, and can restrict focus to the valence states, as we did previously in the text following Eq.~(\ref{eq:chainTotalEs}).  Crucially, for the valence electrons, $N_{T}^{v'}\text{ mod }4 \neq 0$, implying that a TaSe$_4$ chain, when it is doped with two iodine atoms, remains a filling-enforced semimetal.  Specifically, if all of the bands in an isolated (TaSe$_4$)$_4$I$_2$ chain are minimally connected, then $N_{T}^{v'}\text{ mod }4 =2$ implies the presence of a half-filled set of four connected, spinful bands (two connected bands per spin in the limit of vanishing SOC) at the Fermi level of a single TaSe$_4$ chain doped with two iodine atoms.  Because we have previously shown in Fig.~\ref{fig:feSem}(a) that the filling-enforced gaplessness of a TaSe$4$ chain that is missing two electrons manifests as a $4_{2}$-screw-symmetry-enforced nodal degeneracy at $k_{z}=\pi/c$ in the absence of SOC (and in a time-reversed pair of hourglass nodal points along $k_{z}$ when SOC is incorporated), and because a TaSe$_4$ chain with two iodine atoms [\emph{i.e.} (TaSe$_4$)$_{4}$I$_{2}$] still respects $4_{2}$ screw symmetry [Eq.~(\ref{eq:42screw})], then we conclude that the filling-enforced gaplessness of a single chain of (TaSe$_4$)$_{4}$I$_{2}$ also manifests as a filling-enforced nodal degeneracy near $k_{z}=\pi/c$.

For our final step towards constructing the 3D unit cell of (TaSe$_4$)$_2$I, we simply superpose (but do not yet couple) two copies of the previous (TaSe$_4$)$_{4}$I$_{2}$ chain.  We perform this intermediate step because the conventional cell of (TaSe$_4$)$_2$I consists of two TaSe$_4$ chains and four iodine atoms, for a total chemical formula of (TaSe$_4$)$_{8}$I$_{4}$ [Fig.~1(a,b) of the main text].  Though this superposition does not change the symmetry, in the limit that the two chains are decoupled, it does introduce an additional (sublattice) symmetry that relates the chains (we will shortly see that in 3D (TaSe$_4$)$_2$I crystals, the ``sublattice'' symmetry relating the two chains is the fractional lattice translation $T_{BCT} = \{E|\frac{a}{2}\frac{a}{2}\frac{c}{2}\}$ of the conventional cell [Fig.~1(a,b) of the main text]).  After superposing the two chains, the rod group remains $(p4_{2}22)_{RG}$, but with the additional sublattice symmetry relating the two chains.  The presence of the sublattice symmetry does not change the minimal insulating filling, which remains $\nu_{chain}'\in 4\mathbb{Z}$ [Eq.~(\ref{eq:chainLessMinimal})].  However, superposing two chains doubles the total number of electronic states, as well as the total number of electrons.  We schematically depict the band structure of two decoupled (TaSe$_4$)$_{4}$I$_{2}$ chains in the absence of SOC in Fig.~\ref{fig:feSem}(b).  Before introducing any coupling between the chains, in the limit of perturbatively weak SOC employed in this section, the bands in Fig.~\ref{fig:feSem}(b) away from $k_{z}=0,\pi/c$ are \emph{fourfold degenerate} (one band per spin per chain), and there is now an \emph{eightfold} nodal degeneracy at $k_{z}=\pi/c$, which is \emph{not} filling-enforced.  We can understand the stability of this eightfold nodal degeneracy from two perspectives.  First, the symmetries of rod group $(p4_{2}22)_{RG}$ cannot stabilize an eight-dimensional corepresentation~\cite{BigBook,DDP,NewFermions} or spinful eightfold band connectivity~\cite{QuantumChemistry,Bandrep1}; therefore, the eightfold fermion will split in the presence of SOC.  However, in the absence of SOC and interchain coupling, the eightfold fermion \emph{cannot} directly gap without lowering the system symmetry (which includes the chain-exchanging ``sublattice'' symmetry).  Second, from the perspective of minimal insulating filling, the total number of core and valence electrons in the doubled chain is:
\begin{equation}
N_{2}^{c'} = 2N_{T}^{c'} = 1656,\ N_{2}^{v'} = 2N_{T}^{v'}  = 228,
\label{eq:fullIodineChainTimesTwo}
\end{equation}
where $N_{T}^{c',v'}$ are defined in Eq.~(\ref{eq:fullIodineChainE}).  Because $N_{2}^{c'}\text{ mod }4 = N_{2}^{v'}\text{ mod }4=0$, then the eightfold fermion is not filling-enforced in the doubled chain.  In this sense, it bears a resemblance to the double Dirac fermions in SG 135 ($P4_{2}/mbc$) introduced in~\cite{DDP}.  Specifically, like the tetragonal eightfold fermions in~\cite{DDP}, the eightfold degeneracy in Fig.~\ref{fig:feSem}(b) is not a consequence of filling-enforced gaplessness, but rather occurs because of symmetry constraints on the single-particle Hamiltonian.

Finally, having established the electronic filling of two superposed and decoupled TaSe$_4$ chains doped with two iodine atoms each, we can construct the 3D unit cell of (TaSe$_4$)$_2$I, which contains the same atoms and exhibits the same (as well as additional) symmetries.  We begin by placing an array of $z$-directed TaSe$_4$ chains -- separated by a distance $a$ in the $x$ and $y$ directions -- at $(x,y) = (a/2,0)$ in each unit cell.  We then, as previously, remove two electrons from each chain, such that the chains now each have an overall chemical formula [(TaSe$_{4}$)$_{4}]^{2+}$.  Next, we place a second [(TaSe$_{4}$)$_{4}]^{2+}$ chain at $(x,y)=(0,a/2)$ in each unit cell that is related by:
\begin{equation}
C_{4z} = \{C_{4z}|000\},
\label{eq:realC4}
\end{equation}
to the chain at $(a/2,0)$.  To restore charge neutrality, we then place four I$^{-}$ ions in each unit cell at:
\begin{equation}
(x,y,z) = (0,0,0.15c),\ (0,0,0.85c),\ (0.5a,0.5a,0.35c),\ (0.5a,0.5a,0.65c).
\label{eq:IodineLocations}
\end{equation}
We note that the first (last) pair of iodine atoms in Eq.~(\ref{eq:IodineLocations}) could be shifted to lie together at $z=0$ ($z=c/2$) [\emph{i.e.}, to their $z$-coordinates in the superposed charge-neutral chains discussed in the text surrounding Eq.~(\ref{eq:fullIodineChainTimesTwo})] while preserving $C_{2x}$, $s_{4_{2}}$, and $C_{4z}$ symmetries as respectively defined in Eqs.~(\ref{eq:normal4Rod}) and~(\ref{eq:realC4}).  The resulting crystal of [(TaSe$_{4}$)$_{4}]^{2+}$ chains and I$^{-}$ ions is identical to the structure shown in Fig.~1(a,b) of the main text.  While containing the same number of states and electrons [Eq.~(\ref{eq:fullIodineChainTimesTwo})] and respecting the same symmetries [Eq.~(\ref{eq:normal4Rod})] as the two superposed (TaSe$_4$)$_{4}$I$_{2}$ chains discussed in the text surrounding Eq.~(\ref{eq:fullIodineChainTimesTwo}), the 3D crystal also respects \emph{additional} symmetries, which include $C_{4z}$ about $x=y=0$ [Eq.~(\ref{eq:realC4})] and translations in the $x$ and $y$ directions:
\begin{equation}
T_{x} = \{E|a00\},\ T_{y} = \{E|0a0\}.
\label{eq:conventionalSyms}
\end{equation}
Furthermore, because the chains are not centered at $x=y=0$, then the previous generating symmetries of rod group $(p4_{2}22)_{RG}$ [Eq.~(\ref{eq:normal4Rod})] now additionally contain translations in the $xy$-plane when enforced in the 3D crystal.  Specifically, while the previous twofold rotation about the $x$-axis can be expressed in 3D without additional translations:
\begin{equation}
C_{2x} = \{C_{2x}|000\},
\label{eq:realC2}
\end{equation}
the previous $4_{2}$ screw symmetry from Eq.~(\ref{eq:42screw}) contains additional translations when enforced in 3D about $(x,y)=(0,a/2)$:
\begin{equation}
s_{4_{2}} = \bigg\{C_{4z}\bigg|\frac{a}{2}\frac{a}{2}\frac{c}{2}\bigg\}.
\label{eq:fakeScrew}
\end{equation}
Crucially, by combining the (inverse of the) new $C_{4z}$ symmetry [Eq.~(\ref{eq:realC4})] and the previous $4_{2}$ screw symmetry [Eq.~(\ref{eq:fakeScrew})], we realize a new, purely translational symmetry:
\begin{equation}
T_{BCT} = \bigg\{E\bigg|\frac{a}{2}\frac{a}{2}\frac{c}{2}\bigg\},
\label{eq:BCtranslation}
\end{equation}
which we recognize as one of the primitive lattice vectors of a body-centered tetragonal SG~\cite{BigBook}.  Specifically, we recognize Eqs.~(\ref{eq:realC4}),~(\ref{eq:realC2}), and~(\ref{eq:BCtranslation}), as the generating elements of SG 97 ($I422$), the SG of (TaSe$_4$)$_2$I~\cite{Ta2Se8IPrepare,gressier1984characterization,gressier1984electronic}.  Using the~\href{http://www.cryst.ehu.es/cryst/minsup.html}{MINSUP} tool on the Bilbao Crystallographic Server (BCS)~\cite{BCS1,BCS2,BCSSuper}, we confirm that SG 97 ($I422$) is indeed a $k$-type index-2 supergroup of SG 93 ($P4_{2}22$), the space supergroup of the rod group of decoupled TaSe$_4$ chains and iodine atoms [Eq.~(\ref{eq:spaceSuperGroup})], and is specifically generated by:
\begin{equation}
I422 \equiv E(P4_{2}22) \cup T_{BCT}(P4_{2}22),
\end{equation}
where $T_{BCT}$ is defined in Eq.~(\ref{eq:BCtranslation}).

Though the body-centered structure generated by Eqs.~(\ref{eq:realC4}),~(\ref{eq:realC2}), and~(\ref{eq:BCtranslation}) contains all of the same symmetries as the 3D primitive tetragonal crystal [SG 93 ($P4_{2}22$)] that we previously constructed from two [(TaSe$_{4}$)$_{4}]^{2+}$ chains and four I$^{-}$ ions [text surrounding Eq.~(\ref{eq:realC4})], it contains half as many atoms (states) per unit cell as the previous structure in SG 93.  Furthermore, because the generating translations in the body-centered structure [Eq.~(\ref{eq:BCtranslation}) and its $C_{2x}$ and $C_{4z}$ conjugates] are shorter than the previous lattice translation $T_{z} = \{E|c\}$ of isolated TaSe$_4$ chains [Eq.~(\ref{eq:normal4Rod})], then the TRIM points in the larger BZ of the body-centered structure (SG 97) do not lie in the same locations that they did previously in the smaller BZ of the larger (conventional) cell (SG 93).  This has the effect of unfolding the weak-SOC eightfold degeneracy at $k_{z}=\pi/c$ in Fig.~\ref{fig:feSem}(b) into a pair of fourfold nodal planes at $k_{z}=\pm \pi/c$ in the BZ of the body-centered structure.  In the presence of interchain coupling (but the absence of SOC), the fourfold nodal planes split at all crystal momenta \emph{except} the $P$ points and $k_{z}=\pm \pi/c$ along $\Gamma Z$ (further details available at~\url{https://topologicalquantumchemistry.org/#/detail/35190}~\cite{QuantumChemistry,AndreiMaterials,BCS1,BCS2}).  When the effects of SOC and interchain coupling in the $xy$-plane are incorporated, the nodal degeneracies at $k_{z}=\pm\pi/c$ further split into 48 Fermi-surface WPs away from $k_{x,y}=0$ [Table~\ref{tbwps}], as well as four time-reversal pairs of $C_{4z}$-symmetry-enforced chiral fermions (WPs) below the Fermi energy [Fig.~1(d) of the main text and Fig.~\ref{fig:feSem}(c)].  The $C_{4z}$-enforced chiral fermions below $E_{F}$ represent the ``enforced'' semimetallic crossings along $\Gamma Z$ predicted from the Topological Quantum Chemistry and band connectivity~\cite{QuantumChemistry,Bandrep1,AshvinIndicators,ChenTCI,SlagerPRX} analyses of the electronic structure of (TaSe$_4$)$_2$I performed in~\cite{AndreiMaterials,ChenMaterials,AshvinMaterials}.

We can also understand the band structure of (TaSe$_4$)$_2$I from the perspective of Topological Quantum Chemistry~\cite{QuantumChemistry} by analyzing the orbitals closest to $E_{F}$.  In (TaSe$_4$)$_2$I, the bands nearest $E_{F}$ arise from the $d_{z^{2}}$ orbitals of the Ta atoms.  In the primitive cell [Fig.~1(a,b)], there are two inequivalent pairs of Ta atoms -- one pair occupies the $4c$ position [$\left(0,\frac{a}{2},0\right)$, $\left(\frac{a}{2},0,0\right)$] and one pair occupies the $4d$ position [$\left(0,\frac{a}{2},\frac{c}{2}\right)$, $\left(\frac{a}{2},0,\frac{c}{2}\right)$] in the notation of the BCS (Wyckoff multiplicities on the BCS are given with respect to the conventional cell, such that a multiplicity-2 position in a body-centered SG is labeled with the number $4$)~\cite{BCS1,BCS2}.  Specifically, by projecting the bands along $\Gamma Z$ onto the atomic orbitals of (TaSe$_4$)$_2$I, we find that the eight bands at $E_{F}$ (two bands below $E_{F}$ and six bands above) originate from linear combinations of the Ta $d_{{z}^{2}}$ orbitals occupying the intermediate $8f$ position [$\left(0,\frac{a}{2},z\right)$, $\left(\frac{a}{2},0,z\right)$, $\left(0,\frac{a}{2},-z\right)$, $\left(\frac{a}{2},0,-z\right)$].  As discussed earlier in this section [text surrounding Eq.~(\ref{eq:chainDoped})], the eight bands are quarter-filled because the two I$^{-}$ ions in each primitive cell combine to remove two total electrons from the valence shells of the four Ta atoms.  Using~\href{http://www.cryst.ehu.es/cgi-bin/cryst/programs/bandrep.pl}{BANDREP} on the BCS~\cite{BCS1,BCS2,QuantumChemistry,Bandrep1,Bandrep2,Bandrep3}, we determine that the eight bands are formed from a sum of two four-dimensional, \emph{disconnected} elementary band representations (EBRs) from $4c$ and $4d$.  Specifically, the two valence and conduction bands closest to $E_{F}$ represent two pieces of a single four-dimensional EBR that can be disconnected (along high-symmetry lines) without breaking a symmetry.  This implies that the eight $C_{4z}$-enforced chiral fermions along $k_{x}=k_{y}=0$ in (TaSe$_4$)$_2$I could in principle be removed through a (very large) band inversion, leaving behind either non-minimally connected WPs in the BZ interior, or a topological gap~\cite{QuantumChemistry}.  Crucially, this also implies that the CDW in (TaSe$_4$)$_2$I accesses a gap between the pieces of a disconnected EBR, which emphasizes the topological nature of the low-temperature CDW phase~\cite{QuantumChemistry,Bandrep1,Bandrep2,Bandrep3}.  Furthermore, while the momentum separation of the chiral fermions at $k_{z}=\pm \pi/c$ is large in the BZ of 3D (TaSe$_4$)$_2$I in body-centered SG 97 ($I422$), it is still small in the limit of decoupled chains, because $k_{z}=\pm \pi/c$ are related by $(2\pi/c){\bf \hat{z}}$, which \emph{is} a reciprocal lattice vector in the primitive tetragonal structure of decoupled chains [SG 93 ($P4_{2}22$), generated by Eqs.~(\ref{eq:normal4Rod}) and~(\ref{eq:conventionalSyms})].

As discussed in the main text, though the $C_{4z}$-enforced WPs along $k_{x}=k_{y}=0$ in (TaSe$_4$)$_2$I lie below the Fermi level and sit in narrowly-separated groupings with compensating chiral charges [Fig.~\ref{fig:feSem}(c)], and are thus unlikely to contribute in experiment to transport or exhibit observable Fermi-arc surface states, we have shown that they still play a key role in understanding the symmetry-enforced band connectivity of (TaSe$_4$)$_2$I.  Specifically, although the Fermi pockets and $C_{4z}$-enforced WPs are not pinned to $k_{z} = \pm \pi/c$ in SG 97, they still appear localized close to $k_{z}=\pm\pi/c$ in the electronic structure of (TaSe$_4$)$_2$I in its high temperature phase [Fig.~1(d) of the main text].  This occurs precisely because interchain coupling is weak in (TaSe$_4$)$_2$I~\cite{Ta2Se8IPrepare,gressier1984characterization,gressier1984electronic}, such that the Fermi pockets and WPs remain (perturbatively) close to their original, filling-enforced location(s) ($k_{z}=\pi/c$) in the electronic structure of decoupled TaSe$_4$ chains [Fig.~\ref{fig:feSem}(a)].

In summary, in this section, we have shown that the Fermi surface of the high-temperature (Weyl semimetal) phase of (TaSe$_4$)$_2$I is localized in the vicinity of $k_{z}=\pm\pi/c$ because (TaSe$_4$)$_2$I derives from weakly coupled TaSe$_4$ chains, which exhibit filling-enforced nodal degeneracies close to $k_{z}=\pi/c$ [Fig.~\ref{fig:feSem}(a)].  This provides additional insight on the CDW phase in (TaSe$_4$)$_2$I.  Specifically, while we have shown in Table~I of the main text and in SI~H.1 that the CDW gap in (TaSe$_4$)$_2$I originates from coupling 3D WPs with opposite chiral charges, the symmetry and filling analysis performed in this section suggests that the CDW phase could in principle also be constructed by weakly coupling an array of interacting, filling-enforced semimetallic wires with nonsymmorphic rod group $(p4_{2}22)_{RG}$.  In this construction, each wire would then become gapped by electron-electron interactions, either through symmetry-lowering in the mean-field, or through another, more exotic mechanism that has not yet been elucidated~\cite{WPVZ,DDP,DDPMott2}.  To conclude, the presence of a CDW in (TaSe$_4$)$_2$I, which opens a symmetry- and filling-enforced gap in the limit of decoupled TaSe$_4$ chains, is also consistent with the recent recognition that filling-enforced semimetals; such as SrIrO$_3$~\cite{OriginalIridate,WiederLayers,IridateInstability}, CuBi$_2$O$_4$~\cite{DDPMott1,DDPMott2,NewFermions}, and organic stable radicals~\cite{BalatskyInstability}; are often susceptible in experiment to interacting instabilities.

\clearpage
\onecolumngrid

\end{document}